\documentclass[aps,pra,twocolumn,showpacs,superscriptaddress]{revtex4-2}
\usepackage{amsmath,amssymb,amsfonts,mathtools}
\usepackage{graphicx}
\usepackage{textcomp}
\usepackage{braket}
\usepackage{physics}
\usepackage{tabularx}
\usepackage{xcolor}
\usepackage{multirow}
\usepackage{hyperref}
\usepackage{rotating}
\usepackage{dirtytalk}

\usepackage{algorithm}
\usepackage[noend]{algpseudocode}

\usepackage{comment}
\usepackage[T1]{fontenc}

\newcolumntype{Y}{>{\centering\arraybackslash}X}
\def\BibTeX{{\rm B\kern-.05em{\sc i\kern-.025em b}\kern-.08em
    T\kern-.1667em\lower.7ex\hbox{E}\kern-.125emX}}
\begin{document}

\title{Modular Architectures and Entanglement Schemes for Error-Corrected Distributed Quantum Computation}

\author{Siddhant Singh}
\altaffiliation{These authors contributed equally to this work}
%\email{siddhant.singh@tudelft.nl}
\affiliation{QuTech, Delft University of Technology, Lorentzweg 1, 2628 CJ Delft, The Netherlands}
\affiliation{Networked Quantum Devices Unit, Okinawa Institute of Science and Technology Graduate University, Okinawa, Japan}

\author{Fenglei Gu}
\altaffiliation{These authors contributed equally to this work}
%\email{f.gu@tudelft.nl}
\affiliation{QuTech, Delft University of Technology, Lorentzweg 1, 2628 CJ Delft, The Netherlands}

\author{Sébastian de Bone}
\affiliation{QuTech, Delft University of Technology, Lorentzweg 1, 2628 CJ Delft, The Netherlands}
\affiliation{QuSoft, CWI, Science Park 123, 1098 XG Amsterdam, The Netherlands}

\author{Eduardo Villase\~nor}
\affiliation{CSIRO, Research Way, Clayton, 3168 Victoria, Australia}

\author{David Elkouss}
\email{david.elkouss@oist.jp}
\affiliation{QuTech, Delft University of Technology, Lorentzweg 1, 2628 CJ Delft, The Netherlands}
\affiliation{Networked Quantum Devices Unit, Okinawa Institute of Science and Technology Graduate University, Okinawa, Japan}

\author{Johannes Borregaard}
\email{borregaard@fas.harvard.edu}
\affiliation{QuTech, Delft University of Technology, Lorentzweg 1, 2628 CJ Delft, The Netherlands}
\affiliation{Department of Physics, Harvard University, Cambridge, Massachusetts 02138, USA}

\date{\today}

\begin{abstract}
Connecting multiple smaller qubit modules by generating high-fidelity entanglement is a promising path for scaling quantum computing hardware. The performance of such a modular quantum computer depends on the quality and rate of entanglement generation. However, identifying optimal architectures and entanglement generation protocols remains an open question. How can modular quantum architectures be designed to achieve fault tolerance while requiring only feasible entanglement rates and hardware? Focusing on solid-state quantum hardware, we investigate the threshold and logical failure rate of a fully distributed surface code.  We consider both emission-based and scattering-based entanglement schemes between the modules to link the performance to the physical hardware and identify the regime for fault tolerance. We compare architectures with one or two data qubits per module. For some entanglement schemes, thresholds nearing the thresholds of non-distributed implementations ($\sim0.4 \%$) appear feasible with future parameters minimizing the performance gap between modular and monolithic quantum processors.
\end{abstract}

\maketitle

\section{\label{sec:introduction}Introduction}
%In the current pursuit of large-scale quantum computation, two major obstacles are encountered: hardware noise~\cite{Nielsen_Chuang_2010,Preskill_2018,Clerk2014,Hornberger2009} and scalability limitations~\cite{reilly2019challenges,Acharya2023,Gyongyosi2021}. %The goal for scalability is to construct as large quantum processors as possible, with numerous high-quality qubits \cite{patra2023efficient}. 
Error-corrected~\cite{Devitt_2013,Roffe_2019,Bluvstein2024} modular quantum computers ~\cite{caleffi2022distributed,AghaeeRad2025,bombin2021interleaving,VANMETER2010,Main2025,PhysRevA.59.4249},
%~\cite{FigueiredoRoque2021,Evered2023} 
offer a promising route to overcome the noise and scalability challenges of large-scale fault-tolerant quantum computing. A modular quantum computer consists of individual computing modules with separate control, which are linked together by means of quantum entanglement~\cite{VANMETER2010,PhysRevA.59.4249,caleffi2022distributed,10.1145/2494568,10214316}. 
%This simplifies the engineering challenge of scaling the number of qubits and offers more flexibility in the connectivity of the device~\cite{awschalom2021}. 
Modular architectures have been proposed for several types of quantum computing hardware, such as superconducting qubits~\cite{Nickerson2013, bravyi2022}, neutral atoms~\cite{Ramette2022,li2024highrate}, and color centers~\cite{Li2024,afzal2024,10.1116/5.0200190}. However, several fundamental questions remain: What are the optimal ways to generate high-fidelity entanglement between modules? Can modular architectures achieve fault-tolerant thresholds comparable to those of monolithic systems?
%However, several outstanding questions on the operation and performance of distributed quantum error-correction (QEC) codes across a modular quantum computer remain unanswered. 
Previous work has hinted that the performance of a distributed QEC code heavily depends on the degree of modularity~\cite{ramette2023faulttolerant} and the specific choice of quantum entanglement generation protocol~\cite{10.1116/5.0200190}. The many suitable protocols for entanglement generation~\cite{Beukers2024}, as well as specific choice of modular architecture and QEC codes~\cite{Terhal2015,Breuckmann2021}, amount to a vast and largely unexplored design space. How do different entanglement protocols compare in terms of performance for real hardware? What hardware constraints limit their applicability?

In this article, we explore this space for quantum hardware with efficient spin-photon interfaces such as color centers in diamond or silicon, which is a promising hardware for fully connected, modular quantum computers~\cite{Abobeih2022,Bradley2022,doi:10.1126/science.aah6875, PhysRevLett.119.223602,PhysRevB.100.165428, PhysRevLett.123.183602, stas2022,knaut2024,PhysRevLett.119.253601, PhysRevX.11.031021, PhysRevX.11.041041,Simmons2024}. We investigate the performance of a specific surface code, the toric code~\cite{KITAEV20032,bravyi1998quantum,10.1063/1.1499754,PhysRevA.86.032324} across two types of modularized architectures, considering both emission-~\cite{Bernien2013} and scattering-based~\cite {knaut2024,Welte2018} entanglement generation schemes.
%and demonstrate significant differences in code performance for the two depending on the system parameters. 

\begin{figure*}
\centering
\includegraphics[width=\textwidth]{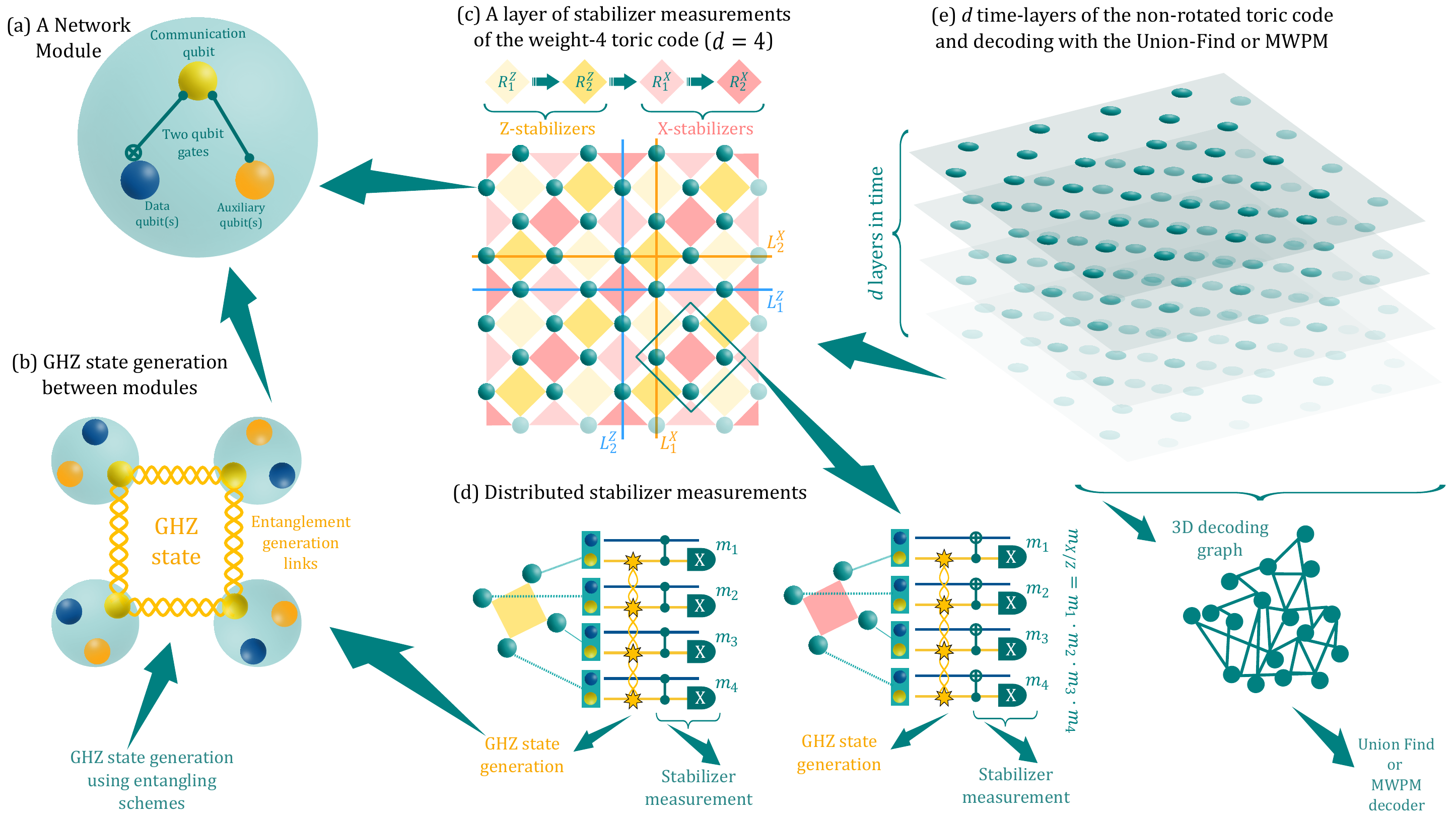}
\caption{Weight-4 toric code architecture with one data qubit per module. (a) The structure of a module. It consists of a single communication qubit that allows for optical connection to the other modules and some memory qubits acting as either data qubits of the code or auxiliary qubits for the stabilizer measurements. (b) Multipartite GHZ states used for measuring the stabilizers that spread among modules. (c) One round of $Z$ and $X$ type stabilizer measurements, each divided into two subroutines. The four colors of the squares in the toric-code lattice correspond to the four subroutines as shown in the sequence. The toric-code lattice encodes two qubits with logical operators ($L_{1,2}^{X, Z}$). The corresponding blue and orange lines indicate the qubit involved in implementing the logical operators. (d) Quantum circuits for implementing the distributed $Z$ and $X$ stabilizer measurement.  A GHZ state is generated upon demand (star signs) and used to measure the joint parity $ZZZZ$ or $XXXX$ on the stabilizer data qubits via the application of local controlled-$Z/X$ gates. The outcome of the stabilizer measurement is the joint parity outcome from the four measurements, i.e., $m_{X/Z}=m_1\cdot m_2\cdot m_3 \cdot m_4$. (e) Time layers of the stabilizer measurement. The stabilizer measurements are repeated $d$ times, the same as the distance of the square toric code. These $d$ layers constitute one full QEC cycle, and all syndrome data is then sent to a decoder. Thereafter, suitable corrections are applied to the qubits.}
\label{fig:weight_4_architecture}
\end{figure*}

We go beyond the generic circuit-level noise model and develop hardware-tailored noise models that allow us to link the modular QEC performance directly to the physical parameters of the quantum hardware. From this investigation, we specify the requirements of key parameters such as qubit coherence times, photonic link efficiency, and quality of the spin-photon interface for the existence of a circuit-level noise code threshold and identify the break-even point where the QEC suppresses errors beyond the physical error rates. We perform a comprehensive comparison of multiple GHZ-generation protocols for stabilizer measurements, including several direct GHZ-generation protocols. Crucially, we demonstrate that by avoiding the slow Bell pair combination and distillation process of emission-based (EM) protocols as considered in Ref.~\cite{10.1116/5.0200190}, one can achieve substantially higher error thresholds and reduce the required coherence times by over an order of magnitude. While we find that near-term experimental parameters are sufficient to realize fault-tolerant quantum computation, modest improvements in hardware performance result in QEC thresholds as high as $\sim0.4\%$, which is comparable to the thresholds found for monolithic architectures of about half a percent under circuit-level noise~\cite{PhysRevA.86.032324}. Our findings provide a road map for designing scalable, fault-tolerant modular quantum computers and answer key questions about their feasibility with near-term and future quantum technology.

%The article is organized as follows.  In Sec. \ref{sec:architectures}, we describe the modular architectures inspired by the toric surface code. Next, in Sec. \ref{sec:entangling schemes}, we describe the theory and analytical model of the entanglement generation schemes, including emission-based and scattering-based schemes. We construct detailed physical models for the hardware performance in Sec. \ref{sec:physical_model}. Finally, we describe the results on distributed code error thresholds and architecture implementation in Sec. \ref{sec:results}.

\section{\label{sec:results}Results}
\subsection{\label{subsec:results_modular_architectures}Modular architectures}
We consider a setup where each module of the computing architecture consists of a single communication qubit (CQ) and a few memory qubits, as shown in Fig.~\ref{fig:weight_4_architecture}(a). The modules are arranged on a square lattice, and the communication qubits can establish entanglement with the nearest-neighboring modules through optical links. We will assign one or two of the memory qubits in a module as data qubits of the code, while additional memory qubits are assigned as auxiliary qubits that can be utilized for stabilizer measurements~\cite{Nickerson2013, PhysRevX.4.041041,10.1116/5.0200190}, as shown in Fig.~\ref{fig:weight_4_architecture}(a).

Two-qubit interactions are allowed only between the communication qubit and a memory qubit within the same node, and only the communication qubit can be directly measured.  This model is tailored to a broad range of quantum hardware based on color centers such as Nitrogen-Vacancy (NV)~\cite{PhysRevApplied.15.024049}, Silicon-Vacancy (SiV)~\cite{doi:10.1126/science.aah6875, PhysRevLett.119.223602,PhysRevB.100.165428, PhysRevLett.123.183602, knaut2024}, Tin-Vacancy (SnV)~\cite{PhysRevLett.119.253601, PhysRevX.11.031021, PhysRevX.11.041041}, and Silicon defect centers~\cite{afzal2024} where the electronic spin states of the defect are optically addressable and can couple through a dipolar interaction to nearby nuclear memory spins.

We consider the topological non-rotated toric surface code~\cite{Dennis_2002,PhysRevA.86.032324,Nickerson2013}, which requires measurement of four-body stabilizers between neighboring data qubits. The stabilizers are measured by creating multi-qubit GHZ states between the modules, either through direct GHZ state generation or fusion of Bell pairs. For the architecture with one data qubit per module, we require the generation of 4-qubit GHZ states, while 3-qubit GHZ states are sufficient when 2 data qubits are hosted per module. We will refer to these as the weight-4 (WT4) ~\cite{10.1116/5.0200190} and weight-3 (WT3) architectures, respectively. WT4 architecture is a direct translation of monolithic surface code architecture to our modular design, which has been explored in ~\cite{Nickerson2013,PhysRevX.4.041041,10.1116/5.0200190}. We propose WT3 architecture in this work to lower the stringent entanglement generation requirements.

\begin{figure}[hbtp]
\centering
\includegraphics[width=0.5\textwidth]{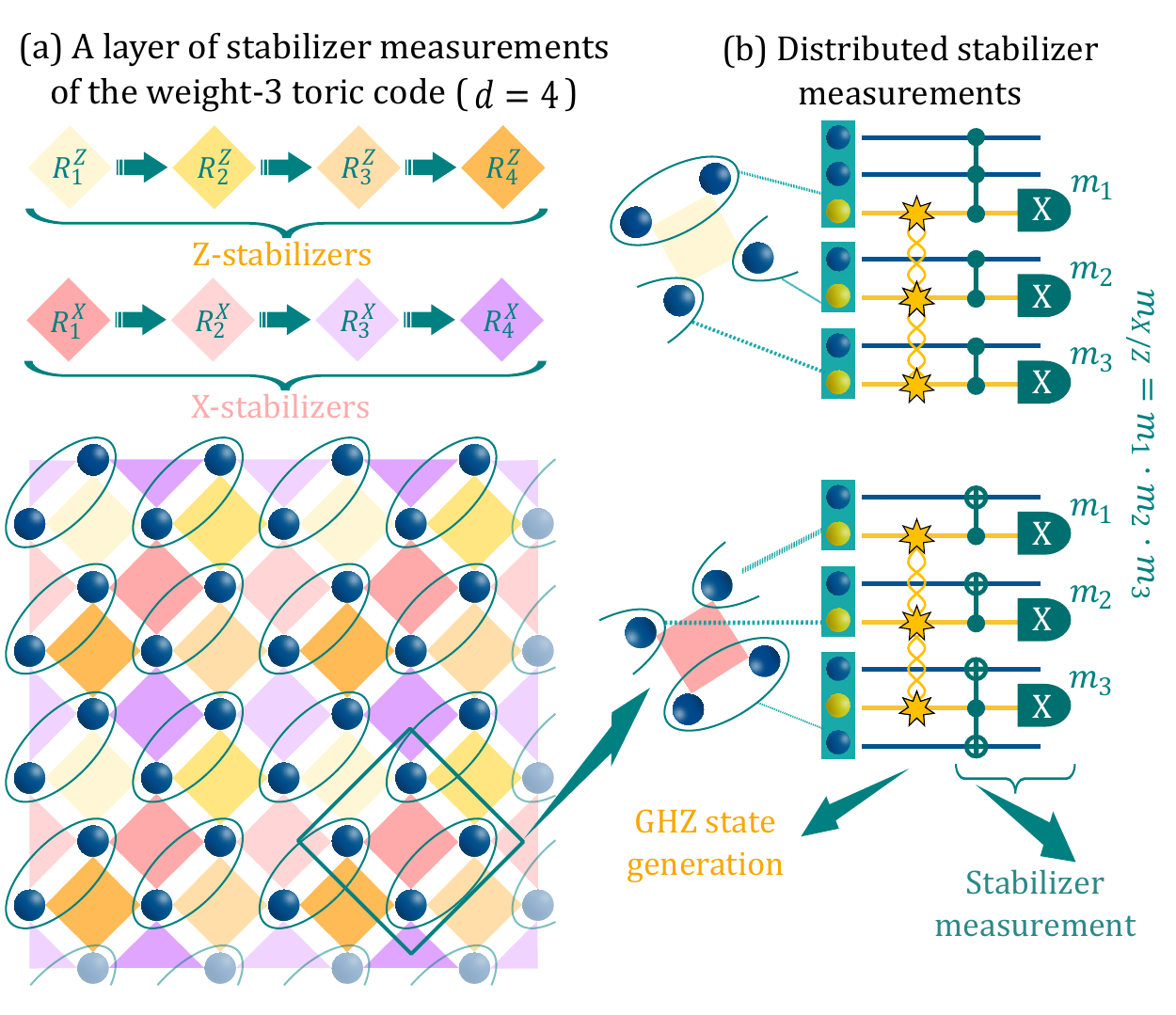}
\caption{Weight-3 toric code architecture of code distance $d=4$ with two data qubits per module. (a) One round of stabilizer measurements, consisting of measurements for $X$ and $Z$ type stabilizers, each containing four subroutines. The logical qubits of the code are the same as for the WT4 architecture, as shown in Fig.~\ref{fig:weight_4_architecture}(c). (b) Quantum circuits for the stabilizer measurement. The notation is the same as in Fig.~\ref{fig:weight_4_architecture}
}
\label{fig:weight_3_architecture}
\end{figure}

%\subsection*{\label{subsubsec:wt4stabs_measure}Stabilizer measurements}
Since each module only contains one communication qubit, it is not possible to measure all the stabilizers simultaneously. This constraint results in a checkerboard pattern of stabilizer measurement sub-rounds (each of X and Z types), which differs for the WT4 and WT3 architectures. 

For the WT4 architecture, each X and Z-type stabilizer is divided into two sub-rounds, resulting in a four-sequence QEC cycle (see Fig. \ref{fig:weight_4_architecture}(c)). 
For the WT3 architecture, the modules are arranged obliquely on the code background array, such that one module contributes two data qubits on the code and exactly three modules span each stabilizer. We consider the even-distance (for each logical operator) WT3 toric code, as this allows all the modules to have exactly two data qubits.
The topology of the WT3 architecture means that four sub-rounds for each stabilizer type (X, Z) are required, resulting in an 8-sequence QEC cycle (see Fig. \ref{fig:weight_3_architecture}).
%The non-simultaneous sub-round stabilizer measurement sequence also leads to additional errors between these sub-rounds for both architectures. This is explained in more detail in the App. \ref{app:architectures}.

%\subsection*{\label{subsubsec:wt4cutoff}Cut-off times}

\begin{figure*}[hbtp]
\centering
\includegraphics[width=0.7\linewidth]{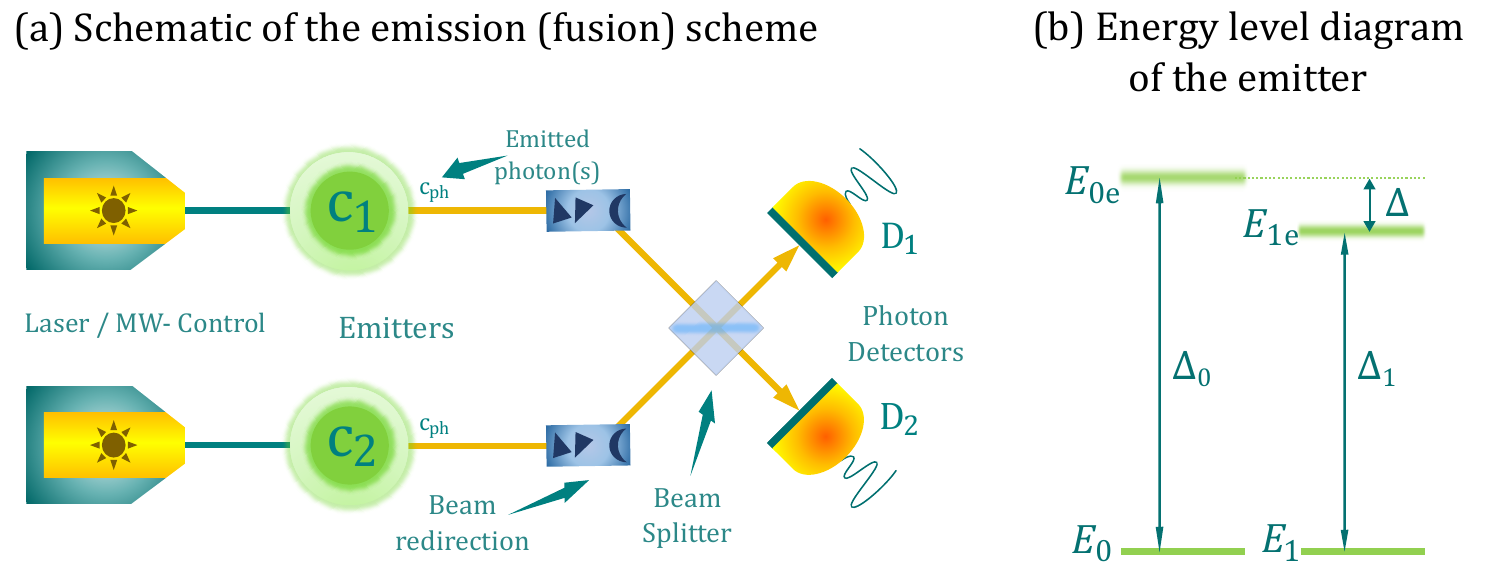}
\caption{Emission-based (EM) scheme. (a) An optical setup with two emitters as communication qubits can be initialized in a CQ-photon entangled state. Then, the emitted photons ($\text{c}_\text{ph}$) are sent to a middle station where an optical Bell measurement is performed using a balanced beam splitter and single photon detection. The detection of an early and a late photon at the detectors successfully heralds a Bell pair between the two CQs ($c_i$). (b) The energy levels of the photon-CQ system. The CQ has two sets of independent transitions: $E_{0}\leftrightarrow E_{{\rm 0e}}$ and $E_{1}\leftrightarrow E_{{\rm 1e}}$ with transition frequencies $\Delta_0$ and $\Delta_1$, respectively. Here, $E_1$ refers to the energy level of the bright-state $|1\rangle$ that emits a photon upon excitation. 
%(c) An example of a GHZ state fusion protocol with distillation used to create a high-fidelity WT3 GHZ state presented as a tree graph. Events in yellow indicate Bell pair generation between the respective modules, while the teal-colored arrows labeled with a multi-Pauli operator indicate a distillation event by measuring that operator. Modules with a teal color indicate the distilled entangled state.
}
\label{fig:emission based}
\end{figure*}

%\subsection*{\label{subsubsec:decoherence_architectures} Decoherence model for QEC cycles}
%In addition to the circuit-level noise, and hardware noise in GHZ state generation, 
%we adopt a continuous time decoherence model of all qubits. The probability to decohere (modeled as an effective depolarizing channel) is $1-\text{exp}(-t/T^{\text{dec}})$ where $t$ measures time and $T^{\text{dec}}$ is the characteristic decoherence time. 
%Importantly, we consider two different decoherence times depending on whether a module is active in an entanglement generation attempt or not. 
%During an entanglement generation attempt, the memory qubits have a decoherence time of $T_\text{link}^\text{dec}$. When there is no active entanglement generation attempts in a module, all qubits (including the communication qubit) have a decoherence time of $T_\text{idle}^\text{dec}$. In general, $T_\text{idle}^\text{dec}\geq T_\text{link}^\text{dec}$ for the solid-state hardware considered in this work~\cite{Reilly2015,Gold2021,Crawford2023}.Sec. \ref{sec:physical_model} provides more details on this noise model.

\subsection{\label{subsec:results_GHZ_generation}GHZ-state generation:}
%Since distributed surface code architectures rely on GHZ states, these GHZ states must be created with high fidelity and probability of success for high code performance. 
%In this section, we describe the emission—and scattering-based entanglement generation schemes we consider for generating high-fidelity GHZ states between the modules. Each scheme can be adapted to generate either 4-qubit or 3-qubit GHZ states.
We consider emission and scattering-based schemes for generating GHZ states between the modules. Each scheme can be adapted to generate either 4-qubit or 3-qubit GHZ states. 

The EM scheme was first proposed in 1999~\cite{PhysRevA.59.1025} and has been well developed to realize long-distance entangled Bell pairs~\cite{stolk2024metropolitan}. However, for realizing the multipartite GHZ states, sequential generations of Bell pairs, intra-modular two-qubit gates~\cite{Childress2013} on the memory qubits, and multiple-qubit readouts are required. These requirements present a significant time overhead in the QEC cycle with corresponding memory requirements on the data qubits.

Nevertheless, scattering-based entangling schemes~\cite{evans2018photon} harness spin-dependent reflection/transmission, which is several orders of magnitude faster than the gates based on spin-spin interactions. Based on whether the reflected or transmitted photon is used to generate the heralding signal, we classify the scattering-based schemes into the reflection (RFL) scheme and the carving (CAR) scheme. The RFL scheme is inspired by previous studies of cavity-mediated spin-photon entanglement gates~\cite{spin_pho3_reiserer2014quantum, spin_pho4_kalb2015heralded} used for Bell-state realizations~\cite{spin_pho1_bhaskar2020experimental, spin_pho2_wei2024universal}. On the other hand, the CAR scheme is closely related to carving schemes for Bell-state generation~\cite{carving2017Welte}. 
%The gate can be realized with a qubit-coupled waveguide/cavity system with Group-IV defect centers in diamond~\cite{thiering2018ab} or silicon-based defect centers~\cite{castelletto2020silicon, radulaski2017scalable}.
While the RFL scheme requires optical circulators and a delay line to guide and manipulate the photons, which raises challenges in on-chip integrated photonics, the CAR scheme circumvents these problems. However, the probabilistic-success nature lower its entanglement generation rates. The principles of these schemes are illustrated below.

\begin{figure*}
\centering
\includegraphics[width=\textwidth]{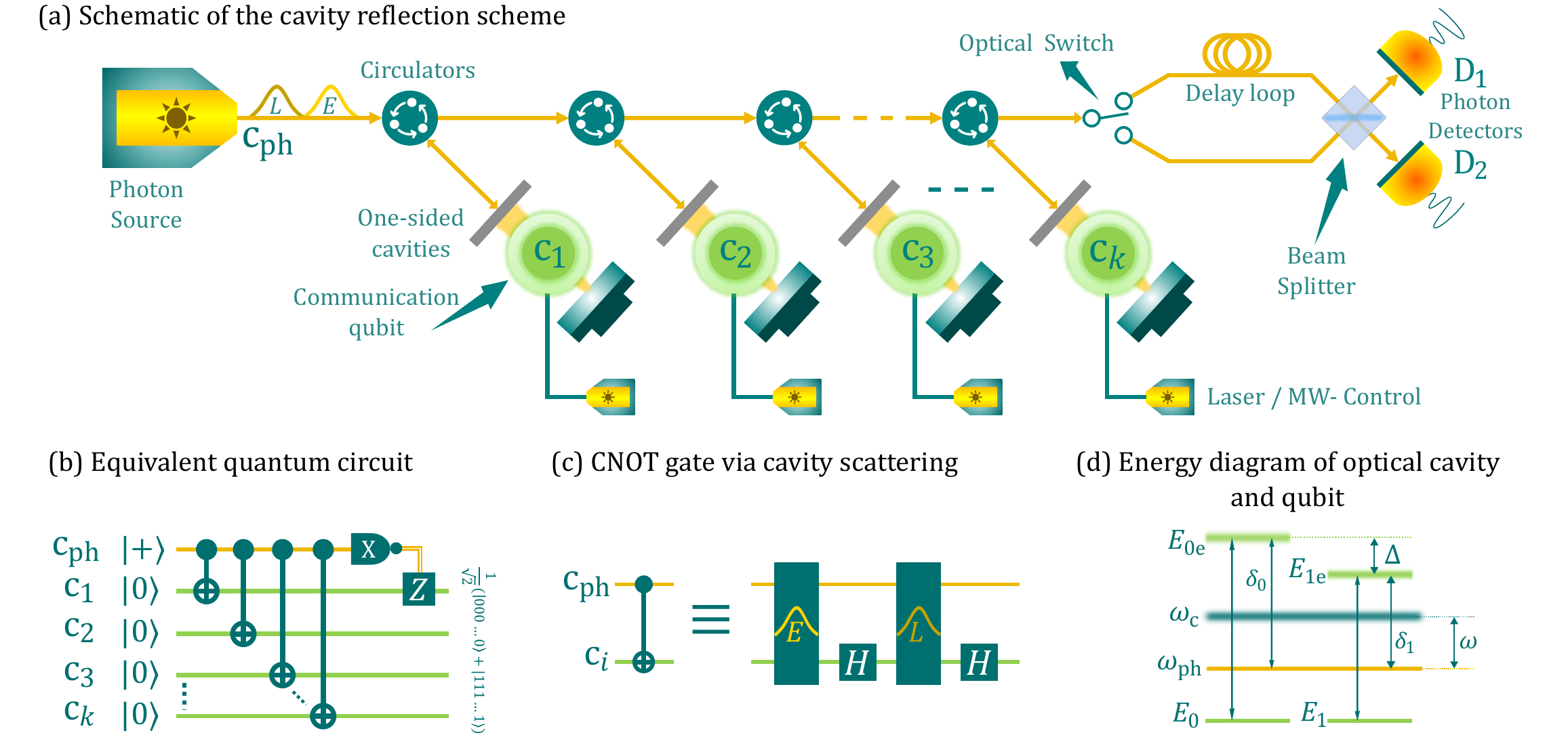}
\caption{Principles of the reflection (RFL) scheme. (a) System setup of the RFL scheme. It contains a series of atoms, each embedded within a single-side cavity. The fibers and circulators guide the photon to interact with each atom-cavity system successively. The delay loop, beam splitter, and photon detectors are used to detect the photonic qubit in the X basis. (b) Quantum circuit that the reflecting protocol implements. It contains \textsc{cnot} gates and a photon measurement in its X basis. (c) Physical implementation for the \textsc{cnot} gate. It contains four steps, including scattering the photon of the two time bins and implementing Hadamard gates on the atom. (d) The energy levels of the photon-atom-cavity system. The spin has two sets of independent transitions coupled to the input photon: $E_{0}\leftrightarrow E_{{\rm 0e}}$ and $E_{1}\leftrightarrow E_{{\rm 1e}}$. $\omega_{\rm c}$: cavity resonance frequency; $\omega_{\rm ph}$: input photon frequency.}
\label{fig:reflection}
\end{figure*}

\textbf{Emission-based (EM) scheme:} 
In this scheme, GHZ states are realized by the fusion of multiple Bell pairs~\cite{10.1116/5.0200190}. For creating the Bell pairs, we consider both the single-click~\cite{PhysRevA.59.1025} and double-click (Barrett-Kok~\cite{PhysRevA.71.060310}) protocols, as they are suitable for the near-term parameters and boosted future parameters, respectively.

The single-click protocol works as follows. In each module, the CQ has a four-level structure with two ground states, as shown in Fig.~\ref{fig:emission based}(b). Ideally, one ground state can be selectively excited to emit a photon. Through a controlled operation, a photon is created and entangled with the CQ. The two photons from two modules are then sent and meet at a balanced beam splitter in a middle station. Then, after a single photon detection, the successful creation of an entangled state between the two CQs will be heralded, as shown in Fig.~\ref{fig:emission based}(a). In the double-click protocol,  a second round of photon emission is used to boost the fidelity by eliminating terms that do not contribute to single photon emission after the first round, by applying local operations before the second emission. For the single-click (double-click) protocol, the overall success probability of the protocols depends linearly (quadratically) on the effective photon detection probability $\eta_\text{ph}$.

 Due to imperfections of both the Bell states and the local gate operations, it is necessary to employ distillation protocols to boost the final fidelity of the GHZ state. Similar to the method illustrated in Ref.~\cite{10.1116/5.0200190}, we optimize over various distillation protocols where additional Bell pairs are used to increase the final GHZ state fidelity. Notably, we only consider distillation protocols where, at most, two auxiliary memory qubits per module are available to accommodate realistic hardware constraints.  Full analytical details of the noisy EM schemes and the distillation protocols can be found in the supplementary material.

\begin{figure*}
\centering
\includegraphics[width=\textwidth]{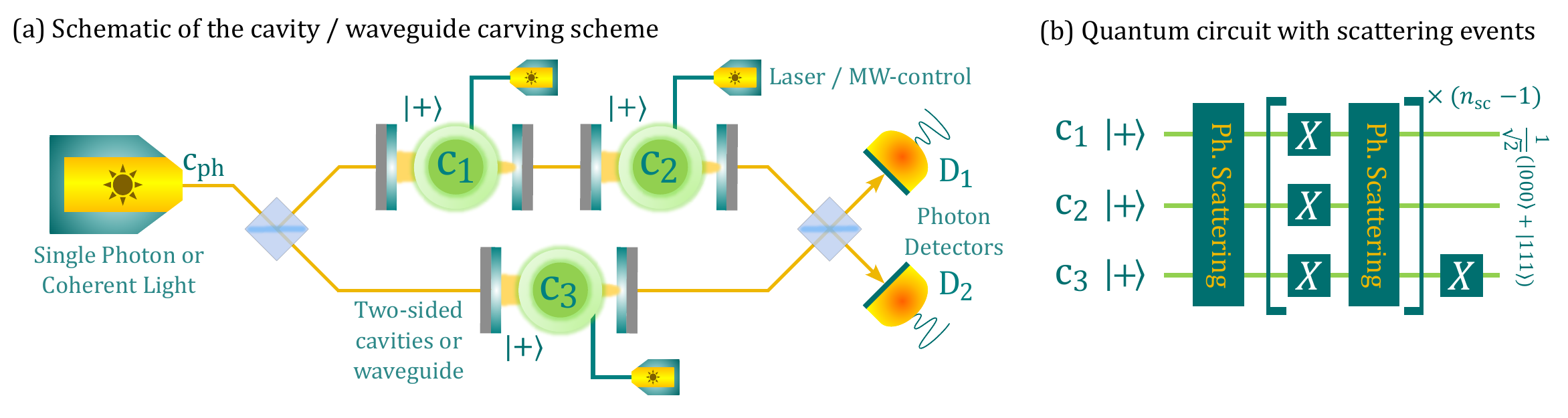}
\caption{Principle of the carving (CAR) scheme. (a) System setup of the CAR protocol. It consists of spins embedded inside a two-sided cavity (or waveguide), a light source, beam-splitters, and photon detectors. (b) The operating sequence of the CAR protocol. All the spins are initialized in $|+\rangle$ states. The operating sequence consists of scatterings of single photons through the optical circuit and \textsc{not} gates implemented on spins between every two sequential scattering rounds. }
\label{fig:carving}
\end{figure*}

%\subsection{Reflection scheme}

\textbf{Reflection (RFL) scheme:} In the RFL scheme, the GHZ state is realized by employing a photon as the flying auxiliary qubit, interacting sequentially with all the CQs involved, and finally being detected, as shown in Fig.~\ref{fig:reflection}(a). This procedure realized a quantum circuit as shown in Fig.~\ref{fig:reflection}(b). The photon-spin interaction realizes a \textsc{cnot} gate~\cite{Bhaskar2020}, where the photon encodes a qubit in its early and late time bins, and two qubit rotations are applied on the CQ, with one at the time between the scattering of the two time bins and the other after the late time-bin scattering, as shown in Fig.~\ref{fig:reflection}(c).

When examining the performance, we include various hardware imperfections in our modeling of the RFL scheme, such as photon loss, finite cavity coupling, and unwanted optical couplings, considering the CQ level structure and photon and cavity frequencies, as shown in Fig.~\ref{fig:reflection}(d). We consider both a near-perfect single-photon source and an attenuated coherent light source for the photonic qubit. The details of our model are provided in the supplementary material. 

%\subsection{Carving scheme}
\textbf{Carving (CAR) scheme:} 
The CAR scheme exploits the spin-dependent reflection or transmission mechanism of a qubit-coupled two-sided cavity or waveguide to realize the GHZ state. For an ideal mechanism, an incoming photon will be reflected (transmitted) when the CQ is in state $|1\rangle (|0\rangle)$. The GHZ state can be probabilistically carved out by means of a single photon scattering at the cavities or waveguides arranged in two routes, and finally getting detected, as shown in Fig.~\ref{fig:carving}(a). The protocol can be repeated with \textsc{not} gates between every two successive rounds, as shown in Fig.~\ref{fig:carving}(b), to increase the fidelity of the output GHZ state. Note that the CAR scheme is inherently probabilistic, with a probability of success that decreases exponentially with the number of CQs in the target GHZ state. The maximum success probability for a 3-CQ (4-CQ) GHZ state is 1/16 (1/32). 
%The actual success probability will be further lowered by photon loss and imperfect detection. 
For this scheme, we consider both a cavity (CAV) implementation and a waveguide (WG) implementation, where the cavity (waveguide) is of a finite (infinite) bandwidth. The level structure and the frequencies of the incoming photon and cavity are the same as those with the EM and RFL protocols. We include hardware imperfections, such as finite coupling strengths and unwanted transmission/reflection, in our modeling. The performance analysis for the CAR scheme with a perfect single photon source (SPS) and an attenuated coherent light (COH) source is presented in the supplementary material.

\subsection{\label{subsec:results_hardware_to_QEC}Hardware parameters to quantum error-correction performance}
%We investigate the surface code architectures' code thresholds and sub-threshold behavior for each entangling scheme for circuit-level noise. 
We consider the physical hardware parameter sets as described in our methods section. For each scheme, we have two sets of parameters: 1) the near-term parameters (NTP), which are based on experimental demonstrations and within the reach of present quantum technology, and 2) the future parameters (FP), where we boost some key parameters in the experiment. Three sets (Set-1, Set-2, and Set-3) of operational and coherence times are also considered, in the increasing order of the ratio between the coherence times and gate times (see the Methods section for details). We simulate a square toric surface code of code distance and number of time layers both equal to an even-number $d$, as shown in Fig.~\ref{fig:weight_4_architecture}(e). Since entanglement generation is probabilistic, we consider a maximum time allowed (GHZ cut-off time $t_\text{cut}$) for GHZ generation for each stabilizer measurement to maintain synchronicity of the error correction sub-rounds.

The code thresholds are calculated for different physical error possibilities $p$, which account for both the gate and measurement errors, different coherence-time sets, and the NTP and FP scenarios. For each case, $t_\text{cut}$ was swept to find the highest threshold $p_\text{th}$ via a binary search. We report the optimized code threshold for each case using the weighted-growth version of the Union-Find decoder~\cite{Delfosse2021almostlineartime} is used to identify the qubit error.

\begin{figure*}
\centering
\includegraphics[width=0.85 \textwidth]{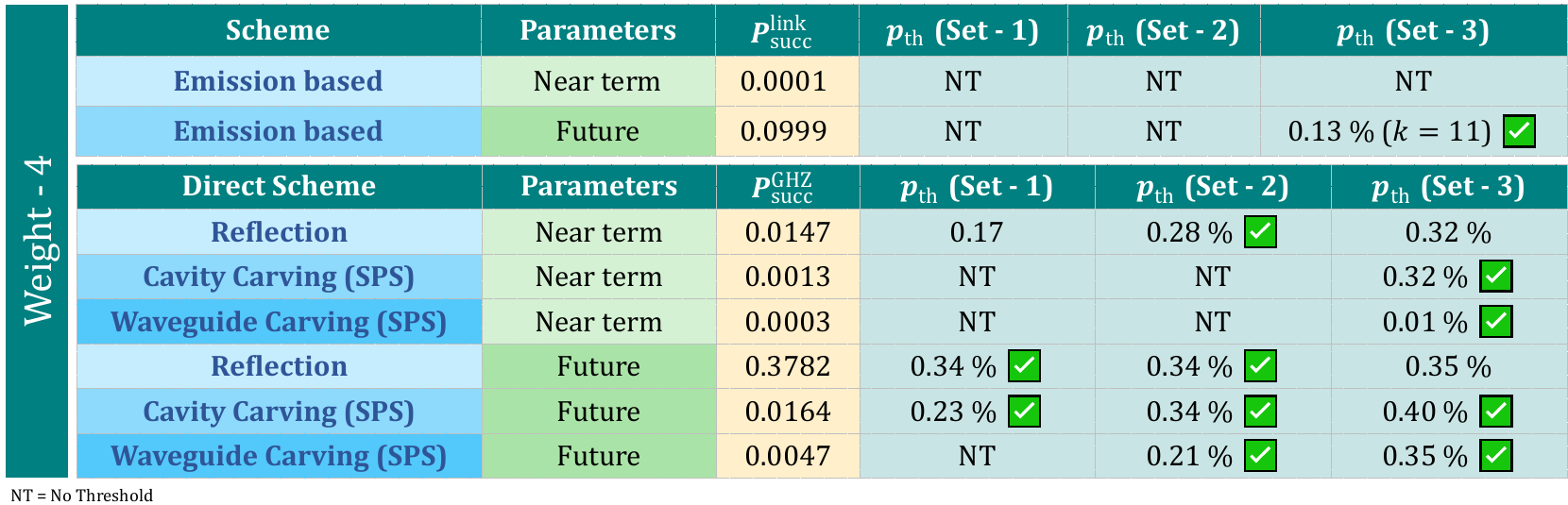}
\caption{Weight-4 toric code thresholds. For every scheme and hardware parameter set, we report the code threshold for each coherence time set. We also report the optimal GHZ generation fusion-based protocol for the EM scheme via the number of Bell pairs ($k$) used in that protocol. The reported code thresholds ($p_\text{th}$), followed by a check mark, are the ones that are higher than the corresponding WT3 toric code threshold. NT represents ``no threshold exist''.}
\label{tab:wt-4_thresholds}
\end{figure*}

\begin{figure*}
\centering
\includegraphics[width= 0.85 \textwidth]{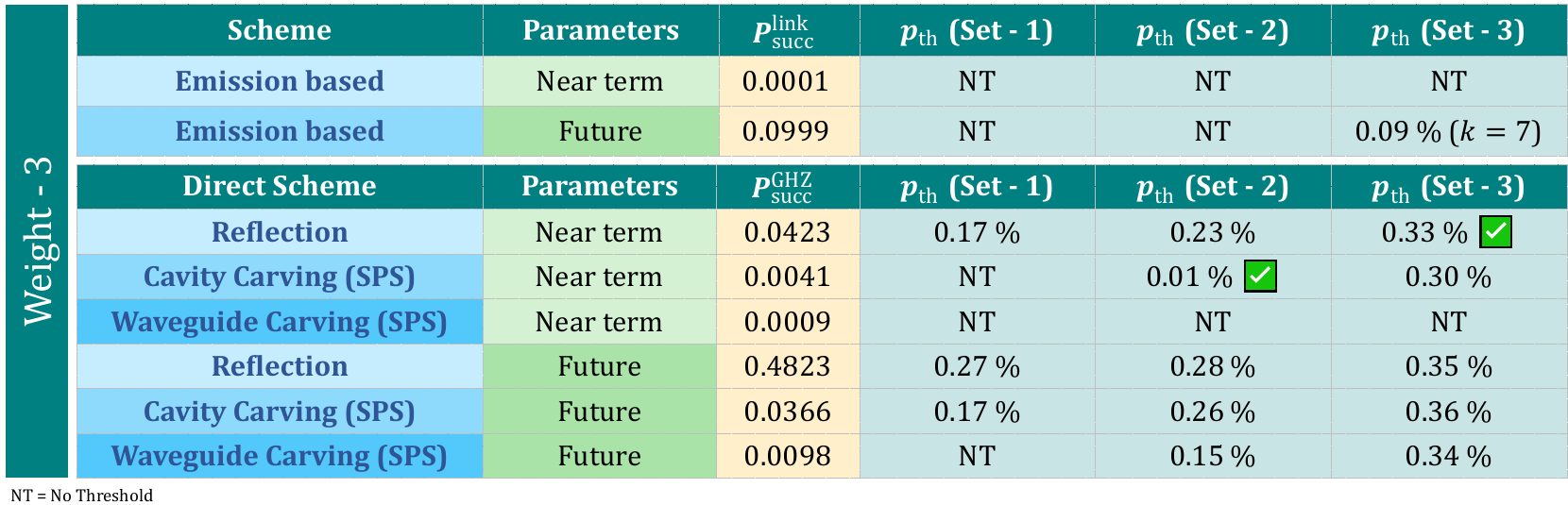}
\caption{Weight-3 toric code thresholds. Refer to the caption of Fig.~\ref{tab:wt-4_thresholds} for the explanation.}
\label{tab:wt-3_thresholds}
\end{figure*}

Based on the numerical investigation with different parameter sets, we, in general, required $P_\text{succ}^\text{link}> 10^{-4}$ and a GHZ fidelity above $99\%$, for a threshold $p_\text{th}$ to exist. Coherence and operational times of Set-1 also serve as the lower bound for the existence of a threshold. Together, these bounds give a rough guide to determine the necessary hardware requirements.

We report that we do not find any code thresholds using a COH photon source for the CAR scheme. As similar performance for the RFL scheme is expected, we skip the simulation of the RFL scheme with a COH source, considering only a deterministic SPS instead. Note that the performance for a non-deterministic SPS can be estimated from this by absorbing the inefficiency of the photon source into the overall detection efficiency.

%\subsection{\label{subsec:wt4_thresholds} Weight-4 toric code thresholds}

\textbf{Thresholds for modular architectures:} 
We present the WT4 architecture threshold results in Fig.~\ref{tab:wt-4_thresholds} and WT3 thresholds in Fig.~\ref{tab:wt-3_thresholds}, where thresholds up to only two decimal places precision are shown. More accurate values can be found in the supplementary material. We found no threshold for the EM scheme for the NTP, which we believe is primarily due to the low success probability ($P_\text{succ}^\text{link}$) of Bell pair generation of $10^{-4}$.  However, even for future parameters (with $P_\text{succ}^\text{link}=0.0999$), we get a threshold for only the best coherence times of Set-3. This threshold was found for the protocol with $k=11(7)$ Bell pairs for the WT4 (WT3) architecture.

%Refer to App.~\ref{appsubsec:EM_protocols} for details on EM fusion and distillation protocols.

For the scattering-based (RFL and CAR) schemes with NTP parameters, we observe promising thresholds ($\sim$0.3\%) for some parameter sets. The RFL scheme supports thresholds for all the examined parameter sets. However, for the WT4 architecture, we only get a threshold value for the highest coherence times (Set-3) for the CAV-CAR and WG-CAR schemes with an SPS. For the WT3 architecture, we find no thresholds for WG-CAR scheme. This is associated with the low probability of success for the CAR scheme, around an order of magnitude lower than the RFL scheme. This result suggests that the decoherence noise becomes the dominant error source in this regime due to long waiting times.

In contrast, we see higher thresholds for FP. The RFL scheme has the highest success possibilities, $P_\text{succ}^\text{GHZ}=0.3782 \;(0.4823)$ for the WT4 (WT3) architecture, boosting the thresholds. The thresholds for the RFL scheme saturate around $\sim 0.35\%$ for WT4, meaning further improvements on the coherence parameters do not increase the threshold due to other noise sources present. However, for the WT3 case, due to the longer QEC cycles, the thresholds improve when the coherence parameters improve. Similarly, for the CAV-CAR protocol, we see a boost in the thresholds, with the highest value of $0.40\%$ for WT4 with Set-3. This happens because the CAV-CAR scheme has higher GHZ state fidelity than the RFL scheme, and with Set-3, the coherence times are high enough for both schemes that the GHZ fidelity becomes the dominant factor in drawing the difference in the performance. Similar trend is seen for WT3. Finally, we get an even lower success probability for the WG-CAR scheme, with no thresholds for coherence times of Set-1. However, rapidly increasing thresholds for Set-2 and Set-3 indicate that the threshold is very sensitive to the coherence times in this regime for both WT4 and WT3 architectures.

\textbf{Perspective on modular architectures and entangling schemes:} The WT4 and WT3 architectures both give rise to competitive code thresholds for all the examined physical parameter sets. 
%First, we must note that the WT3 architecture has half the number of modules compared to the WT4 architecture. However, the WT3 architecture has twice the data qubits per module. For the modules, we considered that the module can host multiple memory qubits that can be used as code data qubits.
When the number of modules is a bottleneck for the experiment, the WT3 architecture can be chosen because it has half the number of modules that are required for the same code distance. 
%Another way of looking at the construction would be to choose WT3 architecture when we want the largest code distance for the same number of available network modules.
However, the QEC cycle times of the WT3 architecture are twice as long for the same distance of the code. %WT4 would be preferred over WT3 architecture if many modules are available, without compromising the logical clock speed. If the physical measurement error rates are lower, one could afford to do fewer time layers for the code, bringing the runtime of the WT3 closer to that of the WT4 architecture.
The bare WT3 architecture also has an adverse effect due to error propagation within each node, as it has two local data qubits per module that can have hook errors due to local operations within the module.
%The locally controlled operations can propagate errors within each node. These error propagations can be tracked using flag qubits \cite{PRXQuantum.1.010302,PhysRevX.10.011022} for improved performance, but only the bare stabilizer measurements are considered for this work. This is discussed in more detail in App.~\ref{appsubsec:wt3_fault_tolerance}. 
%Also, due to a sub-round structure of stabilizer measurements, errors can occur on the data qubits between two rounds, affecting the decoding. This leaves a heavier impact on the WT3 architecture as it has twice the sub-rounds. See App.~\ref{appsubsec:simultaneous_vs_sub-round} for details on these errors. We also report that the WT4 architecture generally has about $5\%$ lower logical error rates, estimated numerically. For the reasons above, we associate this observation with the less-than-optimal decoding of the WT3 toric code.
Errors occurring between the stabilizer sub-rounds also affect the decoding of the two modular architectures. Numerically, we find that the WT4 architecture generally has about $5\%$ lower logical error rates than the WT3 architecture.

\begin{figure*}
\centering
\includegraphics[width= \textwidth]{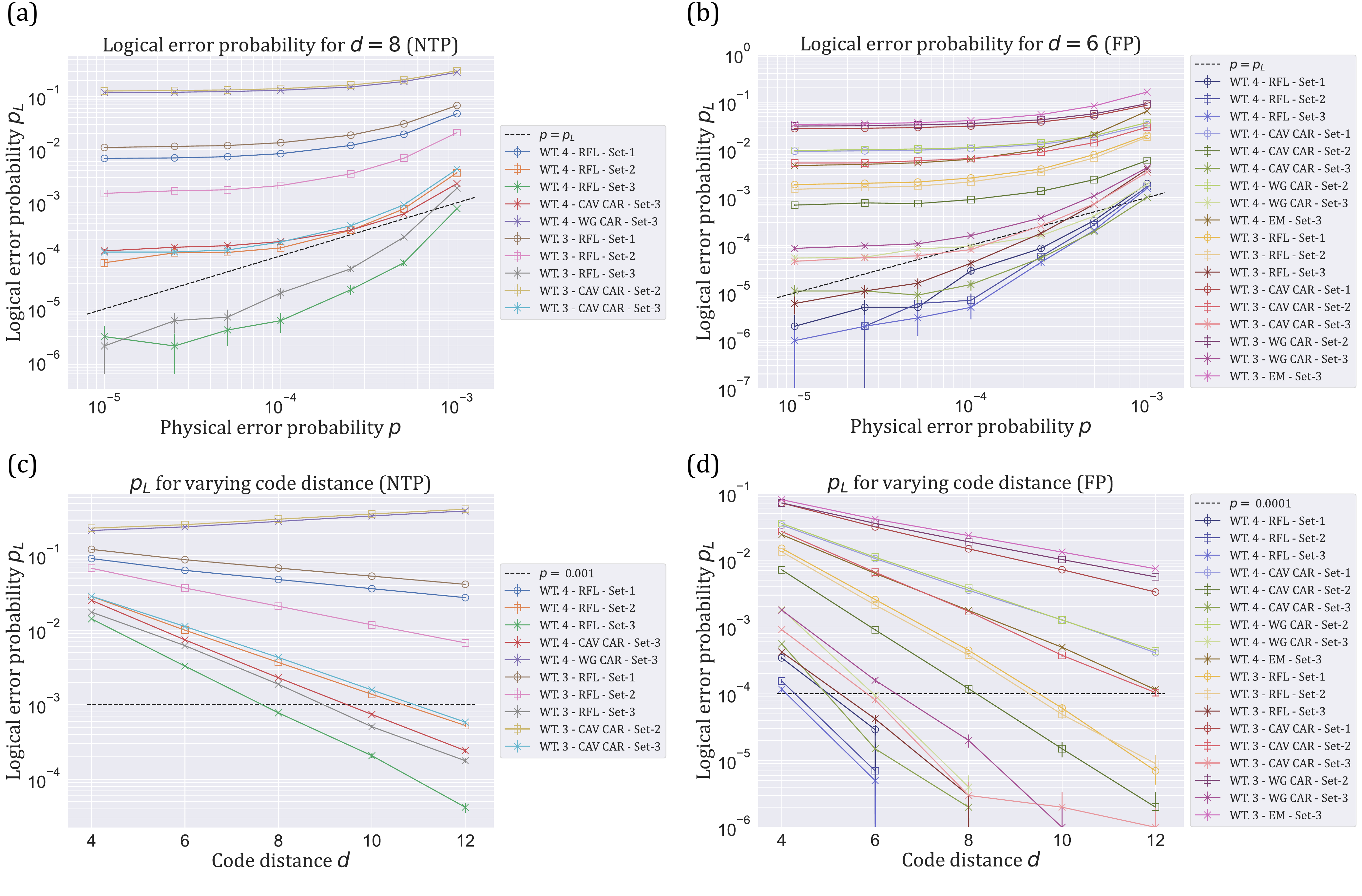}
\caption{Sub-threshold behavior of architectures that have an error threshold. For clarity, Set-1, Set-2, and Set-3 are shown with circle, square, and cross markers. All curves are checked for monotonicity, including the error bars. (a) Logical error probability ($p_L$) for NTP for the smallest even-sized code distance $d=8$ shows better than break-even performance. (b) $p_L$ with FP, for the smallest even-sized code distance $d=6$, which shows better than break-even performance. (c) $p_L$ for NTP, against varying code distance, with constant $p=10^{-3}$. For WT4 WG-CAR Set-3 and WT3 CAV-CAR Set-2, we see an increase in $p_L$ for an increase in $p$ because the threshold for these architectures is less than $10^{-3}$, which is about $0.01\%$. (d) $p_L$ with FP, against varying code distance, with constant $p=10^{-4}$. Some curves are discontinued after a certain code distance because no logical failure was recorded beyond this point, even with $10^6$ iterations per data point. More iterations are required for accuracy, but it was beyond our computational resources' ability to estimate.}
\label{fig:sub_threshold}
\end{figure*}

%\subsection{\label{subsec:schemes_comparison}Comparison between entangling schemes}
%\textbf{Comparison between entangling schemes:} 
The threshold results show that the scattering-based schemes outperform the EM schemes. Two factors are responsible for this. One is that the EM schemes require two-qubit gates involving the memory qubits for Bell pair fusion operations, while the scattering-based schemes do not. The other factor is the high success possibilities $P_\text{succ}^\text{GHZ}$ for scattering-based schemes, which reduces the $t_\text{cut}$ for the same ratio of GHZ generation rate. This thereby reduces the effect of decoherence noise on the memory qubits.

We conducted a separate investigation to verify the noise impact via the fusion process in the EM scheme. We considered a fictitious set of parameters for WT4 EM Set-3 (FP). We choose the boosted parameters $P_\text{succ}^\text{link}=0.3782$ and Bell-pair fidelity $F_\text{link}=99.75\%$ so that it matches the effective GHZ state fidelity and success probability to the WT4 RFL Set-3 (FP) (with $p_\text{th}=0.35\%$). The new threshold only increased from $0.13\%$ to $\approx 0.16\%$, and saturated around this value despite any further parameter improvements. This lack of improvement in the EM threshold can be explained via the noisy fusion process, which suppresses the performance of the EM scheme.

%\subsection{\label{subsec:sub-threshold_behavior} Sub-threshold performance of the architectures}
\textbf{Sub-threshold performance of the architectures:} Most reported thresholds are near $p=10^{-3}$. We simulated the logical error probability ($p_L$) for physical error probability as low as $p=10^{-5}$ and found the smallest code distance for NTP and FP, which gives break-even performance.

The logical error rates for code distance $d=8$ lattice size are shown in Fig. \ref{fig:sub_threshold}(a) for NTP. 
%We see the logical error rates for all the coherence-time sets (1,2,3) and schemes. 
It shows that $d=8$ is the smallest code distance for which we get $p_L\ll p$ for at least one combination of architecture and scheme. We show the logical error rates in the FP case in Fig. \ref{fig:sub_threshold}(b). We do this for distance $d=6$, for which we get better than break-even performance. For the smallest $d=4$, WT4 RFL Set-3 shows just the break-even performance (with $p_L\approx p$) near $p=10^{-4}$. A preferred scheme and architecture combination can be chosen for the minimal logical error rates. The plot shows that the logical error rates saturate and do not decrease when going near $10^{-5}$. This is because the system has residual noise beyond the circuit-level noise, which originates from the hardware that remains constant in the plots.
%mainly due to the noisy GHZ state preparation. We still have all the noise due to the noisy hardware implementation and decoherence on the data qubits, which are not varied in these plots.

Additionally, we increase the code distance to see how the logical error rates decrease. Shown in Fig. \ref{fig:sub_threshold}(c) for NTP parameters with $p=10^{-3}$ and Fig. \ref{fig:sub_threshold}(d) for NF parameters with $p=10^{-4}$, respectively. The logical error rates fall exponentially upon increasing the distance below the code threshold.  For the NTP parameters, we see break-even performance for WT4 RFL Set-3 with only $d=8$, and a few other schemes catch up with $d=10$. We see improved sub-threshold performance for the FP case with a break-even starting lattice size of just $d=6$. Most of the FP schemes demonstrate $p_L\leq p$ for distance 12 onward.

\section{\label{sec:discussion}Discussion}
Our work has outlined several feasible approaches for modular error-corrected quantum computations with solid-state quantum hardware. Our results demonstrate that architectures with either one or two data qubits per module exhibit comparable code thresholds, which are strongly influenced by the choice of GHZ state generation schemes. Besides, we found that scattering-based schemes result in higher thresholds due to the faster and higher quality GHZ state generation for the distributed stabilizer measurements. This is because they circumvent the noisy and slow two-qubit gates and measurements required in the emission-based schemes. Furthermore, we have also provided bounds on the coherence times vs. operation times for the threshold performance of the modular architectures. Architectures with $T_\text{link}^\text{dec}/t_\text{link}<10^4$ will not be a reliable candidate for distributed modular architectures as defined in our framework. A GHZ state success probability $P^\text{GHZ}_\text{succ}>10^{-4}$ and GHZ fidelity $F_\text{GHZ}>99\%$ are required to achieve a code threshold.

While we investigated the surface code due to its well-studied performance, it will be interesting to consider other codes that might better exploit the GHZ-state-enabled nonlocal stabilizer measurements. In particular, codes that could offer reduced qubit overhead, such as the recently introduced quantum low-density parity check (qLDPC) family known as the Bivariate Bicycle (BB) codes, by IBM \cite{Bravyi2024}. Furthermore, extending the simulations to investigate the performance of logical gates and computation~\cite{Gidney2021howtofactorbit}, even for the surface code, would be an interesting future direction. 

\section{\label{sec:methods}Methods}
\subsection{Cut-off times:} The GHZ state generation is probabilistic for the distributed architecture due to the nature of GHZ generation protocols and due to photon loss and imperfect photon measurements. To maintain the stabilizer synchronicity on the code lattice, we consider a cut-off time ($t_\text{cut}$) that caps the maximum time duration allowed for GHZ state generation among the modules before stabilizers are measured. A low $t_\text{cut}$ will lead to less syndrome measurement information and affect the detection of errors, and a larger $t_\text{cut}$ will lead to more noise accumulation due to decoherence of the qubits while idle. Therefore, we optimize $t_\text{cut}$ to find the highest error-correcting threshold of the code.

\subsection{Operation and coherence times:}
\begin{figure}[hbtp]
\centering
\includegraphics[width=0.38\textwidth]{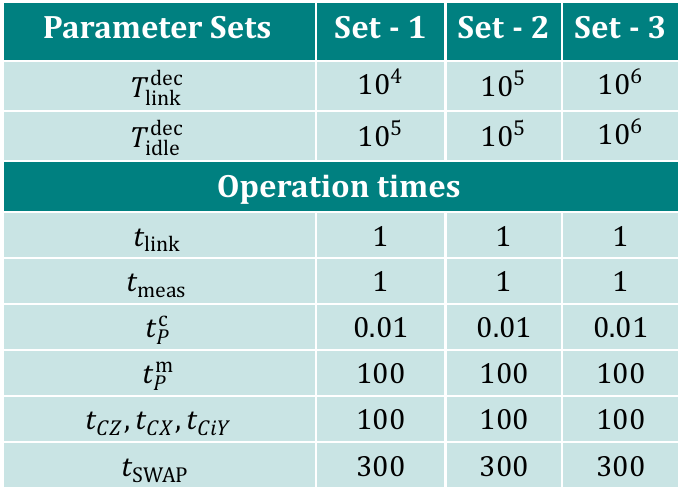}
\caption{Coherence times and operation times. All values are expressed w.r.t. $t_\text{link}$, which is assumed to take one unit of time.}
\label{tab:coherence_parameters}
\end{figure}

%\section{\label{sec:physical_model}Simulation parameters}
%In this section, we expand on the hardware parameters for the surface code architectures and the entanglement generation schemes utilized in their stabilizer measurements.
%\johs{Would this section fit better in Methods? To cut words, you can also expand the table captions and remove the text discussion. E.g. you can cut the discussion of the different sets and the choice of T1=T2.}
We consider a broad set of coherence and operation time parameters as shown in Fig. \ref{tab:coherence_parameters}. There, $T_\text{link}^\text{dec}$ and $T_\text{idle}^\text{dec}$ describe the coherence times applicable to the CQs and memory qubits when entanglement generation is attempted and during idling, respectively. Next, we describe the operational times with $t_\text{link}$ as the time for one entanglement generation attempt (either a Bell pair generation or direct GHZ state generation), and $t_\text{meas}$ meaning the measurement time duration for the communication qubit. Further,  $t_{P}^\text{c/m}$ is the time duration of a native single-qubit $P$ gate with $P\in \{X, Y, Z, H\}$ acting on the communication (c)/memory (m) qubit. Lastly, we have the two-qubit gate times for controlled-$Z, X, iY$ rotations $t_{CZ}, t_{CX}, t_{CiY}$ and \textsc{swap} operation $t_\text{\textsc{SWAP}}$. 
%The two-qubit operations only apply between the communication qubit (control) and the memory qubits (target).

We express all the parameters with respect to $t_\text{link}$ duration times. We consider three parameter sets with varying coherence times for link generation and idling. The first set has the lowest coherence times among the ones for which we find code thresholds. Set-2 has increased link coherence times by one order of magnitude, and finally, we consider Set-3, which has coherence times of $10^6$.

Although our simulation model allows for distinct $T_1$ and $T_2$ times, for each communication and memory qubits, we considered $T_\text{1, link}^\text{dec} = T_\text{2, link}^\text{dec}$ and $T_\text{1, idle}^\text{dec} = T_\text{2, idle}^\text{dec}$, where the subscripts 1, 2 denote the $T_1$ and $T_2$ decay times. This model describes an effective depolarizing noise on the qubits. We choose this noise model against the optimistic case of pure dephasing where  $T_\text{1, link}^\text{dec} = \infty$ and $T_\text{1, idle}^\text{dec} = \infty$. 
%Moreover, our surface code architecture is not biased for $X$ or $Z$ type of errors; therefore, it is a natural choice for the noise model due to decoherence effects.

\subsection{Emission-based scheme parameters:}
\begin{figure}[hbtp]
\centering
\includegraphics[width=0.25\textwidth]{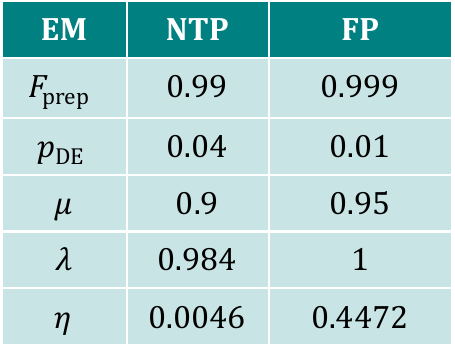}
\caption{Parameter sets for the emission-based scheme. }
\label{tab:emission_parameters}
\end{figure}

We describe these parameters for the EM scheme in Fig.~\ref{tab:emission_parameters}. 
Here, $F_\text{prep}$~\cite{Hensen2015,PhysRevA.99.052330} describes the fidelity of emitter-photon state preparation, $p_\text{DE}$~\cite{Humphreys2018} is the probability of the double excitation error on the emitter caused during emitter-photon state preparation, $\mu$ is the Hong-Ou-Mandel visibility of the photon interference~\cite{Humphreys2018}\cite{Pompili2021}, $\lambda$ is the standard deviation in the path difference for the interference setup~\cite{Humphreys2018}\cite{10.1145/3341302.3342070}, and the effective photon detection probability is $\eta_\text{ph}$~\cite{Coopmans2021}. Furthermore, we have the probability of successful entanglement generation between two nodes $p^\text{link}_\text{succ}$ which depends on $\eta_\text{ph}$ and has the approximate value of $p^\text{link}_\text{succ}\approx 10^{-4}$ and $p^\text{link}_\text{succ}\approx 0.099$ for NTP and FP, respectively. The FP sets are inspired by potential improvements in the experiments.

\subsection{Reflection (RFL) scheme parameters:}
\begin{figure}[hbtp]
\centering
\includegraphics[width=0.48\textwidth]{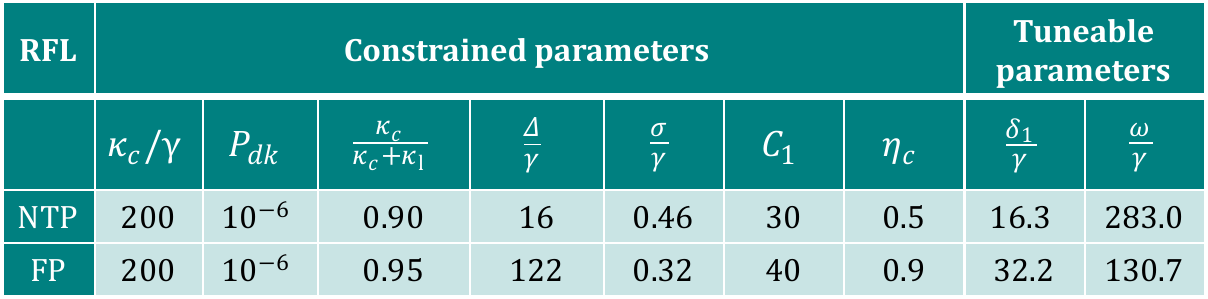}
\caption{\label{tab_para_rfl}System parameters of the reflection scheme (RFL). }
\end{figure}

We simulate the performance of the RFL schemes based on the parameters of the diamond's silicon-vacancy (SiV) color center. The relevant parameters of the physical setup for the RFL scheme are shown in Fig.~\ref{tab_para_rfl}. The constrained parameters are based on NTP and FP, and the tunable parameters are based on maximizing the final GHZ state fidelity. The near-term parameters are modeled from recent experimental demonstrations in Refs.~\cite{Bhaskar2020,stas2022,knaut2024}. $\kappa_{\rm c}$ describes the cavity loss rate from the cavity to the coupling fiber. $P_{{\rm dk}}$ is the dark count rate of the photon detectors, assuming the dark count happens at 1 Hz and the photon detection window is 1 $\mu$s. $\gamma$ denotes the natural linewidth of the spin, $\kappa_{\rm l}$ is the cavity loss rate from the cavity to the environment. Here, $\Delta=\delta_0-\delta_1$, with $\delta_0=E_{\rm 0e}-E_{\rm 0}-\omega_{\rm ph}$ and $\delta_1=E_{\rm 1e}-E_{\rm 1}-\omega_{\rm ph}$. $\sigma$ means the calibration error in $\delta_1$ and $\delta_0$. The cavity cooperativity, $C_{1}$, between the spin and the one-side cavity, is defined as $C_1=\frac{g_{1}^{2}}{\gamma\left(\kappa_{{\rm c}}+\kappa_{{\rm l}}\right)}$, with $g_{1}$ the coupling strength of the spin and the one-side cavity. Finally, $\eta_{{\rm c}}$ is the circulator efficiency of preserving a photon and $\omega$ is defined as $\omega=\omega_{\rm ph}-\omega_{\rm c}$.

\subsection{Cavity-carving (CAV-CAR) scheme parameters:}
\begin{figure}[hbtp]
\centering
\includegraphics[width=0.48\textwidth]{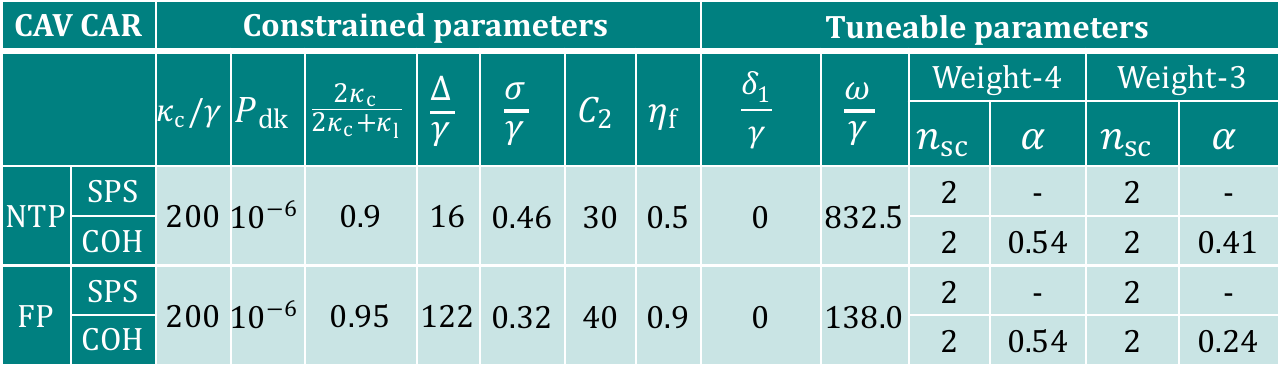}
\caption{\label{tab_para_car_cav}Hardware parameters of the cavity carving scheme (CAV-CAR).}
\end{figure}

The parameters for the CAV-CAR scheme are presented in Fig.~\ref{tab_para_car_cav}, for both single-photon source (SPS) and coherent photon source (COH).
Like the RFL scheme, NTP for the CAV-CAR scheme is also modeled from recent experimental demonstrations in Refs.~\cite{Bhaskar2020,stas2022,knaut2024}. Here, $C_{2}$ is the cooperativity between the spin and the two-side cavity defined as $C_{2}=\frac{g_{2}^{2}}{\gamma\left(2\kappa_{{\rm c}}+\kappa_{{\rm l}}\right)}$, with $g_{2}$ the coupling strength of the spin and the two-side cavity. $\eta_{{\rm f}}$ captures the fiber efficiency of preserving a photon. $n_{\rm sc}$ indicates the number of total scattering times and $\alpha$ is the parameter of the coherent state $\ket{\alpha}$ (not to be confused with $\alpha$ as the bright-state parameter in the EM scheme). Other hardware parameters are similar to the RFL scheme. The detailed methods for calculating the final GHZ state with the SPS and COH are shown in the supplementary material.

\begin{figure}[hbtp]
\centering
\includegraphics[width=0.48\textwidth]{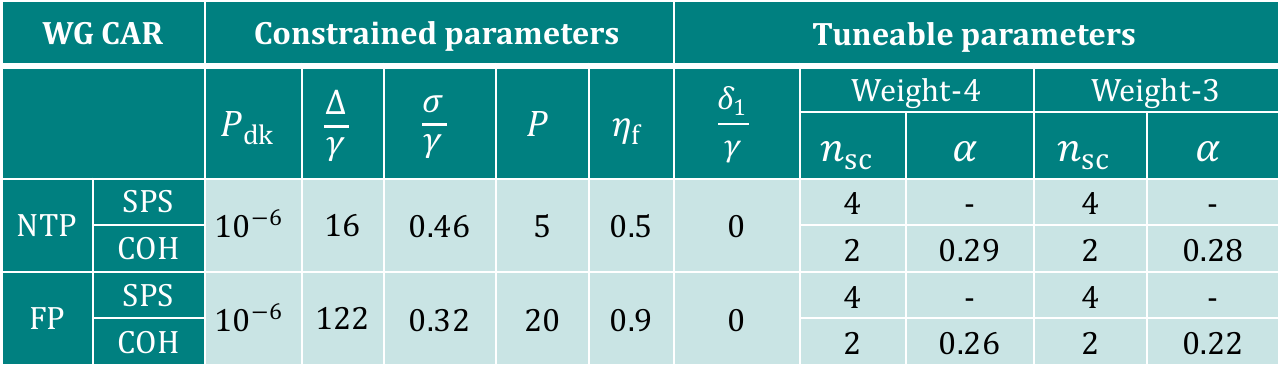}
\caption{\label{tab_para_car_wg}Hardware parameters of waveguide-carving (WG-CAR) scheme.}
\end{figure}

\textbf{Waveguide-carving (WG-CAR) scheme parameters:}
The parameters for the CAV-CAR scheme are shown in Fig.~\ref{tab_para_car_wg}, respectively. Here, $P=\Gamma/\gamma$, with $\Gamma$ being the intensity of the light emitted from the spin and going into the waveguide. Other variables are the same as the CAV-CAR scheme. The NTP sets are inspired from the same recent experiments as the RFL scheme parameters.

\subsection{Logical error rates and thresholds:} 
To calculate the logical error rates, we start with the noisy GHZ state (on the communication qubits) generated by the entangling schemes parameterized by all the input hardware parameters for that architecture. We simulate the X and Z stabilizer unit cell for the architecture's lattice using our source code \href{https://github.com/Poeloe/oop_surface_code/tree/auto_generated_GHZ_prots}{\textsc{CircuitSimulator}}. This is done by executing circuits in Fig.~\ref{fig:weight_4_architecture}(d) and Fig.~\ref{fig:weight_3_architecture}(b). It results in the density matrix of all the qubits involved in the stabilizer measurement, including all communication and memory qubits. All the qubits, except the data qubits of the code, are measured at the end. This projects the data qubits in the desired stabilizer state. We find this effective projector acting on the data qubits of the code, similar to Ref.~\cite{Nickerson2013}. The Pauli twirled representation of this superoperator is
\begin{equation}
        \Pi_M^\text{eff}(\rho_D) =\sum_e ( a_e^{M}E_e\Pi_M ^\text{ideal}(\rho_D)E_e^\dagger
        +  b_e^{\Bar{M}}E_e\Pi_{\Bar{M}} ^\text{ideal}(\rho_D)E_e^\dagger  )
\end{equation}
where $\rho_D$ is the data qubits' density matrix, and $\Pi_M$ denotes the projector due to the measurement, acting on the state. $E_e$ are all the possible Pauli errors acting on the data qubits in the unit cell of the code lattice, and the bar denotes a measurement error. Coefficients $a_e$ and $b_e$ are estimated by calculating the overlap of a given Pauli error string with the superoperator. This superoperator object is then represented as a table with all possible error configurations. Errors are then sampled from the superoperator onto the unit cells of a given code lattice of distance $d$. This task is done via our software package \href{https://github.com/siddhantphy/qsurface}{\textsc{Qsurface}} that samples errors from the superoperator and then decodes for the errors using the UnionFind decoder to calculate the logical error rates. The logical error rates are then fitted to the curve $p_L=A+B\gamma+C\gamma^2$, where $p_L$ is the logical error rate, $\gamma=(p-p_\text{th})L^{1/\nu_0}$, where $L$ is the lattice size of the code, $p$ is the physical error probability and $A, B, C, p_\text{th}, \nu_0$ are the fitting parameters. We estimate the threshold $p_\text{th}$ using this fit.

\section*{Data Availability}
Simulation data for this research's findings are openly available in 4TU.ResearchData~\cite{project_data}.

\section*{Code Availability}
The source code is openly available on 4TU.ResearchData~\cite{project_data}.

\section*{Acknowledgments}
The authors thank Tim Taminiau, Ronald Hanson, Erwin van Zwet, and Barbara Terhal for helpful discussions and feedback. We gratefully acknowledge support from the joint research program ``Modular quantum computers'' by Fujitsu Limited and Delft University of Technology, co-funded by the Netherlands Enterprise Agency under project number PPS2007. D.E. was partially supported by the JST Moonshot R\&D program under Grant JPMJMS226C. J.B. acknowledges support from The AWS Quantum Discovery Fund at the Harvard Quantum Initiative. We thank SURF (www.surf.nl) for the support in using the National Supercomputer Snellius.

\section*{Author Contributions}
S.S. developed the theory for modular architectures and EM schemes under the guidance of D.E. and performed numerical simulations for the superoperators and threshold calculations across all architectures and schemes. F.G. developed the theory and performed numerical simulations for the RFL and CAR schemes under the guidance of J.B., and authored the corresponding sections of the manuscript. S.D.B. contributed to the development of the former source code for threshold simulations for WT4 architecture. E.V. conceptualized the initial idea for the WT3 architecture. S.S. wrote the overall manuscript. The project was jointly supervised by D.E. and J.B. All authors reviewed and contributed to the final manuscript.

\section*{Competing Interests}
The authors declare no competing financial or non-financial interests.

\clearpage
\appendix
\onecolumngrid

\section{Noise model for stabilizer measurements}
\label{app:noise_model}
In this appendix, we describe the noise modeling for the stabilizer circuits. As discussed before, the stabilizer circuit can be decomposed in two steps. The first step is the GHZ state generation using an entanglement generation scheme, and the second step is the application of locally controlled gates and measurement of the communication qubits. The detailed noise modeling of the entanglement generation is described in the respective appendices. We apply circuit-level noise and decoherence noise to all the qubits involved in the stabilizer circuit at all times. We describe these noise channels in detail here.

\subsection{\label{appsubsec:circuit_level_noise}Circuit-level noise}
We consider the gate set \{$X^\text{c/m},Y^\text{c/m},Z^\text{c/m},H^\text{c/m},CZ, CiY,\textsc{CNOT},\text{SWAP}$\} as the native gate set for the module, where c/m denotes whether the single qubit gate is applied to the communication qubit or memory qubit. Circuit-level noise~\cite{PhysRevX.13.031007} implies that we apply noise during state preparation, single-qubit gates, two-qubit gates, and measurement operations. We model this as a depolarizing channel characterized by error probability $p_g$, which we call gate error. The measurement error, denoted as $p_m$, is modeled as the probability of a flipped measurement outcome. State preparation, idling, and single-qubit gate noise are modeled as the ideal operation followed by a single qubit depolarizing channel as~\cite{Nielsen_Chuang_2010}:
\begin{equation}
    \mathcal{N}_\text{gate}^\text{single-qubit}(\rho)=(1-p_g)\rho+\frac{p_g}{3}\sum_{P_j} P_j\rho P_j
\end{equation}
where $P_j\in \{X, Y, Z\}$ are the Pauli matrices for single qubit. While the two-qubit gate noise (here $\rho$ is now a two-qubit density matrix) is modeled via the two-qubit depolarizing channel:
\begin{equation}
    \mathcal{N}_\text{gate}^\text{two-qubit}(\rho)=(1-p_g)\rho + \frac{p_g}{15}\sum_{(P_j,P_k)\notin (\mathbb{I},\mathbb{I})}(P_j\otimes P_k)\rho(P_j\otimes P_k)^\dagger 
\end{equation}
This channel is applied after the noiseless application of any two-qubit gates involving a communication and memory qubit. We choose $p_g=p_m=p$ as the physical error rate for the simulations and sweep $p$ in our simulations to parameterize all the circuit-level noise.

\subsection{\label{appsubsec:decoherence}Decoherence noise}
\textbf{Decoherence noise:}
We adopt a continuous time decoherence model of all qubits. The probability to decohere (modeled as an effective depolarizing channel) is $1-\text{exp}(-t/T^{\text{dec}})$ where $t$ measures time and $T^{\text{dec}}$ is the characteristic decoherence time. 
%Importantly, we consider two different decoherence times depending on whether a module is active in an entanglement generation attempt or not. 
During an entanglement generation attempt, the memory qubits have a decoherence time of $T_\text{link}^\text{dec}$. When there is no active entanglement generation attempts in a module, all qubits (including the communication qubit) have a decoherence time of $T_\text{idle}^\text{dec}$. In general, $T_\text{idle}^\text{dec}\geq T_\text{link}^\text{dec}$ for the solid-state hardware considered in this work~\cite{Reilly2015,Gold2021,Crawford2023}.
We model it as a continuous function of time with relevant coherence times as stated in Fig. \ref{tab:coherence_parameters}, with all operations and gates tagged with their time duration. We implement the decoherence channel in terms of its Kraus operator sum representation as~\cite{Nielsen_Chuang_2010}:
\begin{equation}
\label{eqn:kraus_definition}
    \mathcal{N}^\text{dec}(\rho)=\sum_{j=1}^{\kappa}K_j\rho K_j^\dagger
\end{equation}
where $\rho$ is any general single qubit density matrix and $K_j$ are the Kraus operator elements, which satisfy $\sum_{j=1}^\kappa K_jK_j^\dagger=\mathbb{I}$.

The $T_1$ time decay is described by the generalized amplitude damping (GAD) noise channel with Kraus operators:
\begin{equation}
\begin{split}
K^\text{GAD}_1&=\frac{1}{\sqrt{2}}
\begin{bmatrix}
1 & 0\\
0 & \sqrt{1-\gamma_1}
\end{bmatrix}\\
K^\text{GAD}_2&=\frac{1}{\sqrt{2}}
\begin{bmatrix}
0 & \sqrt{\gamma_1}\\
0 & 0
\end{bmatrix}\\
K^\text{GAD}_3&=\frac{1}{\sqrt{2}}
\begin{bmatrix}
\sqrt{1-\gamma_1} & 0\\
0 & 1
\end{bmatrix}\\
K^\text{GAD}_4&=\frac{1}{\sqrt{2}}
\begin{bmatrix}
0 & 0\\
\sqrt{\gamma_1} & 0
\end{bmatrix}
\end{split}
\end{equation}
where $\gamma_1=1-e^{-t/T_1}$ is the decay factor with $t$ the time duration of the operation and $T_1$ the coherence time applicable.

$T_2$ time decay is the pure dephasing noise channel modeled via phase damping (PD) with coherence times $T_2$. This has the Kraus operators:
\begin{equation}
K^\text{PD}_1=
\begin{bmatrix}
1 & 0\\
0 & \sqrt{1-\gamma_2}
\end{bmatrix},\;
K^\text{PD}_2=
\begin{bmatrix}
0 & 0\\
0 & \sqrt{\gamma_2}
\end{bmatrix}.
\end{equation}
Again with $\gamma_2=1-e^{-t/T_2}$.
These decoherence noise channels are applied for given operations or gate times, including link generation times, idling, and measurement times. When both these channels are applied together, they give rise to an effective depolarizing channel.

\section{\label{app:superoperator}Superoperator approach to QEC simulation}
This appendix describes the core of our QEC simulations using the superoperator approach for the noisy stabilizer measurement circuits. This approach is adopted from the former work on fully distributed architectures from the references~\cite{Nickerson2013, PhysRevX.4.041041}. We adopted this method to simulate our WT4 and WT3 architectures' noisy stabilizer measurements. To simulate noise beyond the circuit-level noise model, we need to simulate the full-density matrix of the noisy stabilizer circuit. For an $n$ qubit system, we would need $2^{2n}$ density matrix elements to be stored and would need very large matrix multiplications for applying any unitary operator. The problem becomes rapidly intractable as the number of qubits grows. This is where the superoperator representation is efficient. The idea is to estimate the noisy superoperator that describes the noisy stabilizer measurements on a unit cell of the surface code architecture and then use that noisy superoperator to simulate the other many unit cells of the code, depending upon its distance in each logical operator. Here, the underlying assumption is that errors are not correlated between distant modules of the distributed code. This is true because all modules are independent and only interact during the entanglement generation protocols via their communication qubits.

As we will describe, the noise applied to the code via the superoperator application is done by sampling Pauli errors to align with the stabilizer simulations. This is computationally easy to decode, in contrast to full-density matrix simulations of all the code data qubits. This is equivalent to Pauli twirling the more general noisy channel. Since the stabilizer measurement operators then incorporate only Pauli errors, they give deterministic measurement outcomes that can be fed to the decoder of choice. Now, we describe the analytical treatment of the superoperator approach.

\subsection{Superoperator analytical model}
\label{appsubsec:superoeprator_analytical}
Represent the Hilbert space of all the qubits involved in the stabilizer measurement circuit as $\mathcal{H}$. This involves all the data qubits of the stabilizer and the communication qubits. We can see this total Hilbert space as the direct product of the subspace of data qubits, which undergo the unitary transformation during the stabilizer measurement, and the communication qubits, which are measured out at the end of this transformation, as $\mathcal{H}=\mathcal{H}_\text{D}\otimes \mathcal{H}_\text{c}$. Our stabilizer circuit involves a GHZ creation over the desired communication qubits of the nodes, followed by the application of controlled gates and measurements. The density matrix of the GHZ state created via a chosen scheme is the input to this circuit. The action of the noisy stabilizer circuit $\mathcal{S}$ is to perform the transformation:
\begin{equation}
\mathcal{S}(\rho_\text{D}\otimes \rho_\text{c})=\rho_\text{D}'
\end{equation}
where $\rho_\text{D}$ is the state of the data qubits, $\rho_c$ state of the communication qubits. And $\rho_\text{D}'$ is the reduced density matrix of the data qubits after the communication qubits are measured out. Suppose we figure out a Pauli decomposition of the effective transformation using Kraus operators on the post-measurement state and their corresponding probabilities. In that case, we can use these to sample errors on the code lattice for the decoding. Choi-Jamiolkowskii isomorphism~\cite{Jamiolkowski1972} can be exploited to find this desired decomposition of the stabilizer circuit action. Consider each of the incoming data qubits to be one qubit of the Bell pair $|\Phi_+\rangle=\frac{1}{\sqrt{2}}(|00\rangle+|11\rangle)$. The map $\mathcal{S}$ acts only the data qubits, while their entangled counterparts remain unchanged. Then, the initial joint state to the stabilizer circuit can be expressed as:
\begin{equation}
    |\rho\rangle=\frac{1}{\sqrt{2^{n_D}}}\sum_{i=0}^{2^{n_D}-1}|i\rangle |i\rangle 
\end{equation}
Where $n_\text{D}$ is the total number of data qubits in a stabilizer unit cell. Here, $|i\rangle |i\rangle$ denote joint qubit computational basis states over the stabilizer unit-cell data qubits and the auxiliary system which together form Bell-pairs as described above and can be expanded and also be written in the binary form. For instance, if $n_D=4$, then $\rho=\frac{1}{4}|\Phi_+\rangle^{\otimes 4}=\frac{1}{4}(|0000\rangle|0000\rangle+\hdots +|1111\rangle|1111\rangle)$. The noisy stabilizer circuit acts as:
\begin{equation}
    \rho_\mathcal{S}=(\mathcal{S}\otimes \mathbb{I})|\rho\rangle \langle \rho |= \sum _j p_j(K_j \otimes \mathbb{I})|\rho\rangle \langle \rho |(K_j \otimes \mathbb{I})^\dagger
\end{equation}
where each Kraus operator $K_j$ can be expressed in the computational basis as $K_j=\sum _{x,y}[K_j]_{xy}|x\rangle \langle y| $. Then, the action of each Kraus operator element yields an interesting fact:
\begin{equation}
\begin{split}
    [K_j]|\rho\rangle &= \frac{1}{\sqrt{2^{n_D}}}\sum _{x,y}[K_j]_{xy}|x\rangle \langle y|\sum_i |i\rangle |i\rangle \\
    &= \frac{1}{\sqrt{2^{n_D}}}\sum_{x,y}\sum_i [K_j]_{xy}\delta_{yi}|x\rangle |i\rangle \\
    &=\frac{1}{\sqrt{2^{n_D}}}\sum_{x,y}[K_j]_{xy}|x\rangle |y\rangle 
\end{split}
\end{equation}
This equation reveals the relationship between the Kraus operators $[K_j]$ and their action on a computational basis. The output state density matrix can be diagonalized to reveal the Kraus operators.

\subsection{\label{appsubsec:using_superoperator}Using the superoperator decomposition}
The idea in App. \ref{appsubsec:superoeprator_analytical} can be readily applied to the noisy parity projections for the stabilizer measurements. If $\rho_D$ is the state of incoming data qubits to the stabilizer measurement circuit, and $\Pi_M$ is the associated projector of the measurement, then the output state is $\rho_D'=\Pi_M(\rho_D)/\text{Tr}[\Pi_M(\rho_D)]$. However, because the stabilizer measurement is noisy, we also get erroneous outcomes, which leads to the effective projector:
\begin{equation}
        \Pi_M^\text{eff}(\rho_D) =\sum_e ( a_e^{M}E_e\Pi_M ^\text{ideal}(\rho_D)E_e^\dagger
        +  b_e^{\Bar{M}}E_e\Pi_{\Bar{M}} ^\text{ideal}(\rho_D)E_e^\dagger  )
\end{equation}
where $\Bar{M}$ denotes a faulty measurement, and $e$ runs over all the possible Pauli errors on the data qubits. The errors on the data qubits $E_e=P_1P_2\hdots P_w$ with $P_j\in \{I,X,Y,Z\}$ are the Pauli errors on the data qubits involved in the stabilizer measurement ($w=4$ for WT4 and $w=8$ for WT3 architecture, respectively; note that it is not the weight of the stabilizer but the number of data qubits involved in one unit cell of the architecture's lattice). Here, we assumed the Pauli basis decomposition for the Kraus operators in the spirit of the previous argument. However, instead of diagonalizing the resulting density matrix, we can instead find the coefficients $a_e$ and $b_e$ for known error permutation $e$ on the data qubits. Assuming the form $[K_j]=(P_1P_2\hdots P_w)\Pi_{M/\Bar{M}} ^\text{ideal}$ and get the output state:
\begin{equation}
    \rho_j=(K_j\otimes \mathbb{I})|\rho\rangle \langle \rho|(K_j\otimes \mathbb{I})^\dagger
\end{equation}
Using this pre-computed output state, we can directly calculate the overlap with the noisy output state of the stabilizer circuit to get the error probabilities as:
\begin{equation}
    p_j=F(\rho_\mathcal{S},\rho_j) = \text{Tr}\left[ \sqrt{\sqrt{\rho_j}\rho_\mathcal{S}\sqrt{\rho_j}} \right]
\end{equation}
This method fully decomposes data qubit, measurement errors, and their corresponding error probabilities. We use the software stack developed by us for these superoperator calculations, called \href{https://github.com/Poeloe/oop_surface_code/tree/auto_generated_GHZ_prots}{\textsc{CircuitSimulator}}. We can randomly sample from this error probability distribution to apply errors on the code lattice and decode those errors using a decoder of choice~\cite{Nickerson2013,10.1116/5.0200190}. This part of the simulation is handled by our QEC simulator for surface codes, called \href{https://github.com/siddhantphy/qsurface}{\textsc{Qsurface}}, which can handle monolithic or distributed surface code simulations for an input superoperator. We describe this part of the simulations in the next subsection.

\section{\label{app:architectures}Distributed surface code simulations}

This appendix describes the QEC simulations for the WT4 and WT3 architecture. The input to these simulations is the superoperator in a tabular format using methods in the App. \ref{app:superoperator} to decompose the superoperator. Error probability for each Pauli string Kraus operator is calculated and arranged in a tabular form, along with measurement errors, either when there is GHZ success within the cutoff time $t_\text{cut}$ or a GHZ state generation failure. For all these possible scenarios, it has a total number of rows equal to $4^4\times 2\times 2=1024$. Once we sample GHZ success, data qubit, and measurement errors from this error-probability distribution based on the weights, the effective joint measurement outcome is deterministic, either $\pm 1$, for each sampled error. These stabilizer outcomes are fed to the decoder thereafter. The decoder module in our simulation workflow does not take into account specific error mechanisms and is independent of the architecture and entangling scheme considered.

In comparison to a monolithic architecture, the fully distributed implementation can be achieved in a non-fault-tolerant way through a single auxiliary qubit interacting sequentially with the four data qubits via gate teleportation~\cite{yimsiriwattana2004generalizedghzstatesdistributed}. Such a scheme is, however, sub-optimal in our case since it requires repeated entanglement distribution between a single module and its neighbors while keeping the auxiliary qubit coherent, which is challenging in solid-state hardware~\cite{Reilly2015,Gold2021,Crawford2023}. Which is why we consider a fully distributed architecture. We also briefly mention the modeling of cut-off times imposed on the stabilizer measurements. Suppose some stabilizers' GHZ state generation is unsuccessful within this time window. In that case, we abandon the stabilizer measurement and substitute the value from the latest successful stabilizer measurement before carrying on with the next sub-round of the stabilizer measurements. Now, we describe how these Monte Carlo simulations are executed for each of our architectures, with examples.

\subsection{QEC simulations for WT4 architecture}
\label{appsubsec:qec_wt4}

\begin{figure}[hbtp]
\centering
\includegraphics[width=0.48\textwidth]{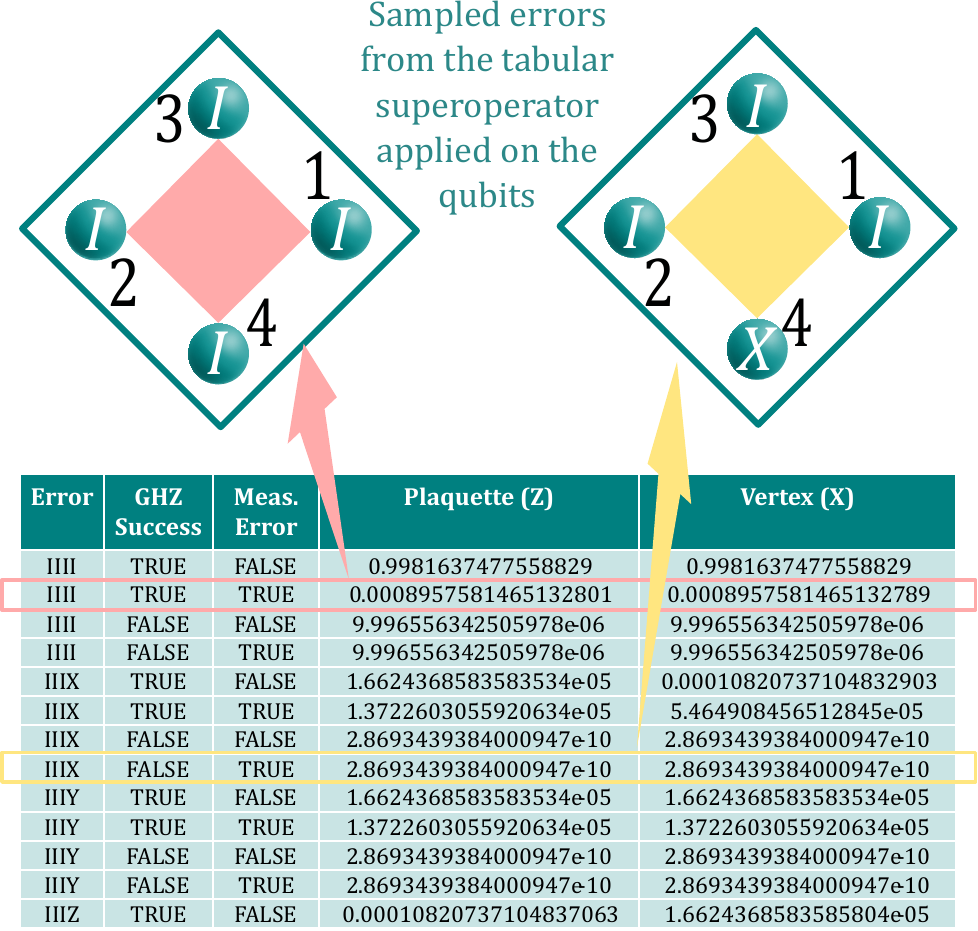}
\caption{Application of data qubit errors on the code WT4 lattice, using the tabular superoperator calculated from Sec. \ref{appsubsec:using_superoperator}. For each plaquette and vertex stabilizer, we sample errors from the superoperator based on the given error probability and apply the respected errors in the right order.}
\label{fig:wt_4_sup_errors}
\end{figure}
The compact tabular format of the WT4 superoperator is shown in Fig. \ref{fig:wt_4_sup_errors}. The "Error" column describes the data qubit errors on the stabilizer data qubits, the "GHZ Success" column tells whether the GHZ generation was successful or not, and the "Meas. Error" column states whether there is a measurement error on the communication qubits during the stabilizer measurement, and the "Plaquette (Z)" and "Vertex (X)" columns describe the probability of these error combinations for each stabilizer type. This is a compact form of the tabular superoperator constructed in Ref.~\cite{10.1116/5.0200190}. For a QEC cycle, we start with $R_1^Z$ stabilizers. For each stabilizer, we sample for error probabilities from the "Plaquette (Z)" column and output one row. Say we sampled \{IIII, TRUE, TRUE\} (shown in light-red). This means there is no data qubit error on the stabilizer data qubits; the GHZ state was successfully generated, but there is a measurement error on the communication qubits. Therefore, we apply no data qubit error on the data qubits. We will flip the outcome when we measure the stabilizer due to the measurement error. Then, we move on to the next stabilizer and sample from the superoperator until the sub-round is completed.

Another instance can be understood from the X stabilizer measurements (shown in yellow); after sampling, we randomly output \{IIIX, FALSE, TRUE\}. This means the fourth data qubit has an $X$ error. However, there was also a GHZ failure, so we cannot perform the stabilizer measurement. Therefore, we will copy the previous stabilizer outcome value to this current stabilizer, but we apply the $X$ error on the data qubit caused due to noise during the stabilizer attempt. The measurement error does not matter when GHZ state generation fails, as we see that both TRUE/FALSE scenarios for measurement error have the same weight. This way, we apply all the different error combinations to the stabilizer data qubits. After all the sub-rounds in a layer are done, we measure that layer, including the measurement errors sampled for each stabilizer. For the final layer, we do perfect measurements to have an even number of syndromes (this is a simplification for the simulation purpose). Once all the layers are completed for the QEC cycle, we send the 3D syndrome graph data composed of the stabilizer outcomes to the decoder, either the Minimum Weight Perfect Matching (MPWM) or the Union-Find (UF) (weighted growth version) decoder.

\subsection{QEC simulations for WT3 architecture}
\label{appsubsec:qec_wt3}
\begin{figure}[hbtp]
\centering
\includegraphics[width=0.48\textwidth]{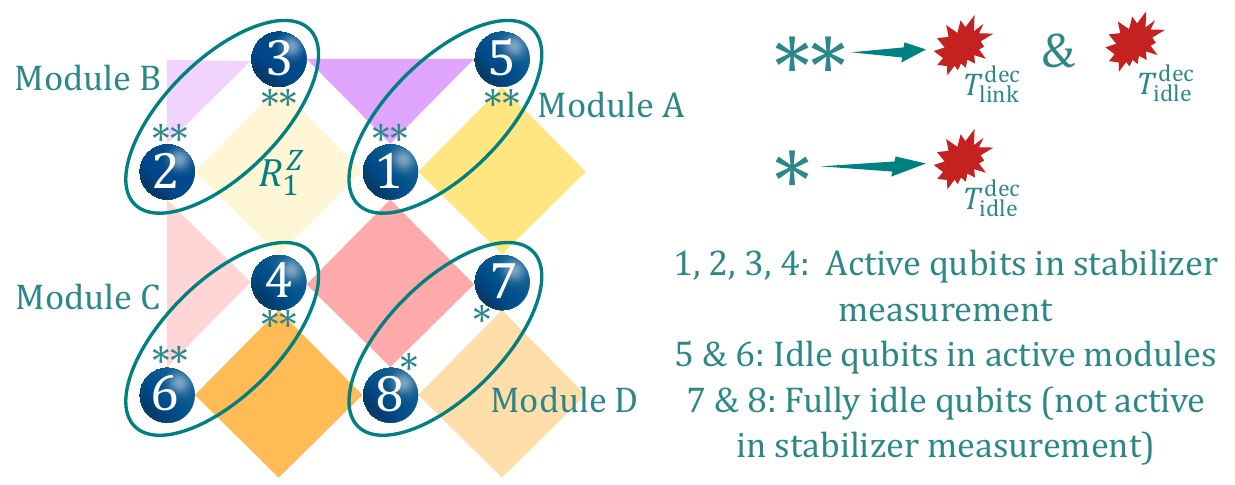}
\caption{Unit-cell for WT3 surface code layout, for a $R_1^Z$ stabilizer. There are 8 data qubits involved in the current stabilizer measurement, out of which 4 are active stabilizer qubits and 4 are idling qubits. These qubits undergo different decoherence rates depending on GHZ success and failure of the current stabilizer.}
\label{fig:wt_3_unit_cell}
\end{figure}

The WT3 superoperator is a little more complicated. It helps to first look at the WT3 unit cell structure and identify different data qubits involved, as shown in Fig. \ref{fig:wt_3_unit_cell}.

Let us look at one unit cell of the WT3 toric surface code, consisting of modules A, B, C, \& D with the corresponding labeled data qubits.
We consider the same noise model as the WT4 surface code. However, additional idling is involved on data qubits for WT3 surface code.  Consider the current stabilizer measurement on the light-yellow, round-1 Z-type stabilizer. Modules A, B, and C are active in the stabilizer measurement. But module D is entirely idle during this sub-round. Here, we first point out the noise action on various qubits involved.

During the generation of the GHZ state by a chosen entanglement generation scheme, modules A, B, and C attempt entanglement generation; during this attempt, all the memory qubits (consisting of data qubits 1,2,3,4,5,6) are subjected to the decoherence with coherence times $T_\text{link}^\text{dec}$ (labeled with **). The attempt could be a Bell pair generation (EM scheme) or direct GHZ state (RFL and CAR schemes). We refer to data qubits 5 and 6 as active-idle qubits because they are idle in the active modules. While module D does not attempt any entanglement generation during this stabilizer sub-round (entirely idle module), and the memory qubits 7 and 8 are subjected to the decoherence $T_\text{idle}^\text{dec}$ (labeled with *) at all times. We refer to these qubits as fully idle data qubits. While certain neighboring round-1 Z stabilizers attempt GHZ generation, the considered $R_1^Z$ stabilizer might succeed in GHZ generation. In which case, the stabilizer will be immediately measured out, and there will be some idling on all the eight data qubits, again with the coherence times $T_\text{idle}^\text{dec}$.

\begin{figure}[hbtp]
\centering
\includegraphics[width=0.48\textwidth]{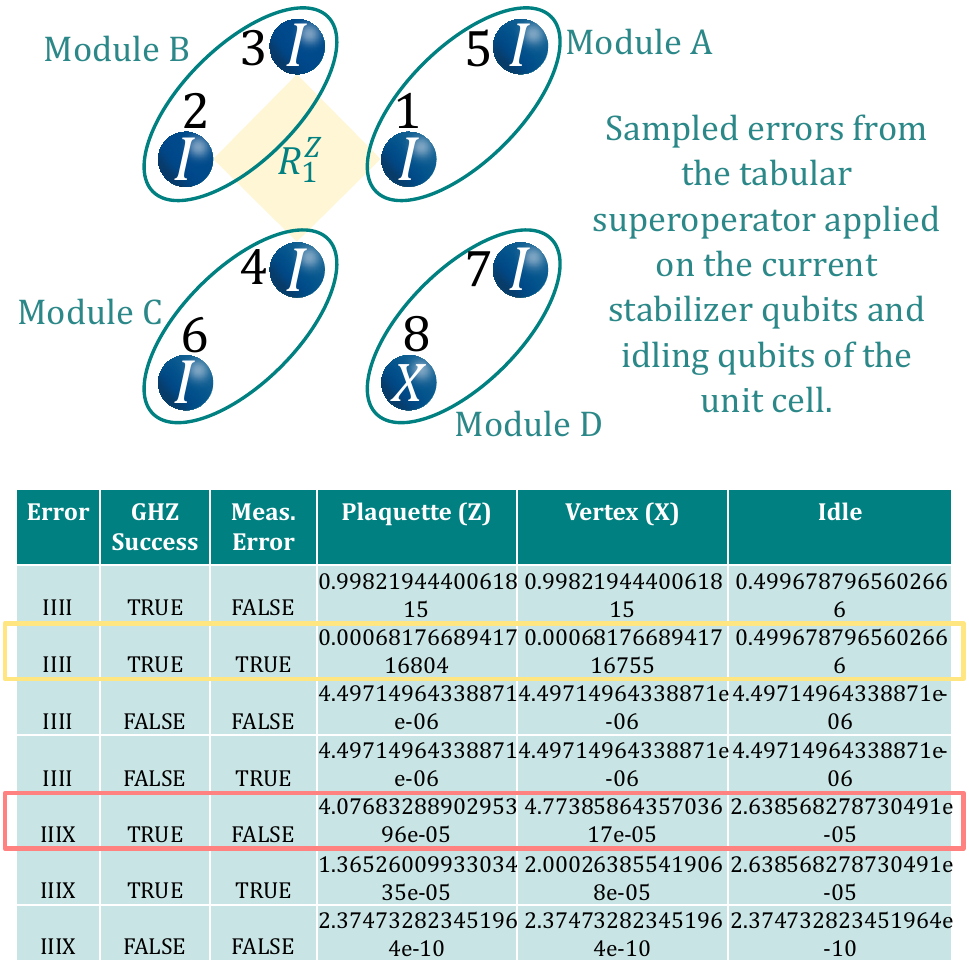}
\caption{Application of data qubit errors on the WT3 code lattice, using the tabular superoperator calculated from Sec. \ref{appsubsec:using_superoperator}. We show a $R_1^Z$ stabilizer unit cell as an example where we apply the errors on data qubits of the current $R_1^Z$ stabilizer and idling errors on other data qubits in the unit cell.}
\label{fig:wt_3_sup_errors}
\end{figure}
The WT3 superoperator's tabular form has an additional column, "Idle", that indicates the noise on the four idling qubits. Instead of creating another idling superoperator, we use the compact form shown in Fig. \ref{fig:wt_3_sup_errors}, with the same number of rows as the WT4 superoperator. The simulation works as follows. If we again consider the $R_1^Z$ stabilizer, and we sample \{IIII, TRUE, TRUE\} for the "Plaquette (Z)" (shown in light-yellow), we can apply the noise on data qubits 1,2,3,4, following the same argument as WT4 simulations. However, we must also apply the idling noise on the idling data qubits 5,6,7,8. Given that the stabilizer has a GHZ success, we now sample using the "Idle" column only from the rows that have "GHZ Success $=$ TRUE" (shown in red). Say we sampled the error configuration "IIIX", in which case we will serially apply this configuration to the idling data qubits (in the order 5,6,7,8). We loop over all the stabilizer unit cells within a sub-round to apply all the data qubit errors, and we repeat this for all the sub-rounds for each type of stabilizer. This completed one layer of measurements in the QEC cycle. After applying all the data qubit errors, we measure the layer, including the measurement errors that were previously recorded for each stabilizer. Finally, again, we do perfect measurements in the final layer and send the 3D syndrome graph to the decoder.

\subsection{Flag fault-tolerance of the WT3 architecture}
\label{appsubsec:wt3_fault_tolerance}
\begin{figure}[hbtp]
\centering
\includegraphics[width=0.68\textwidth]{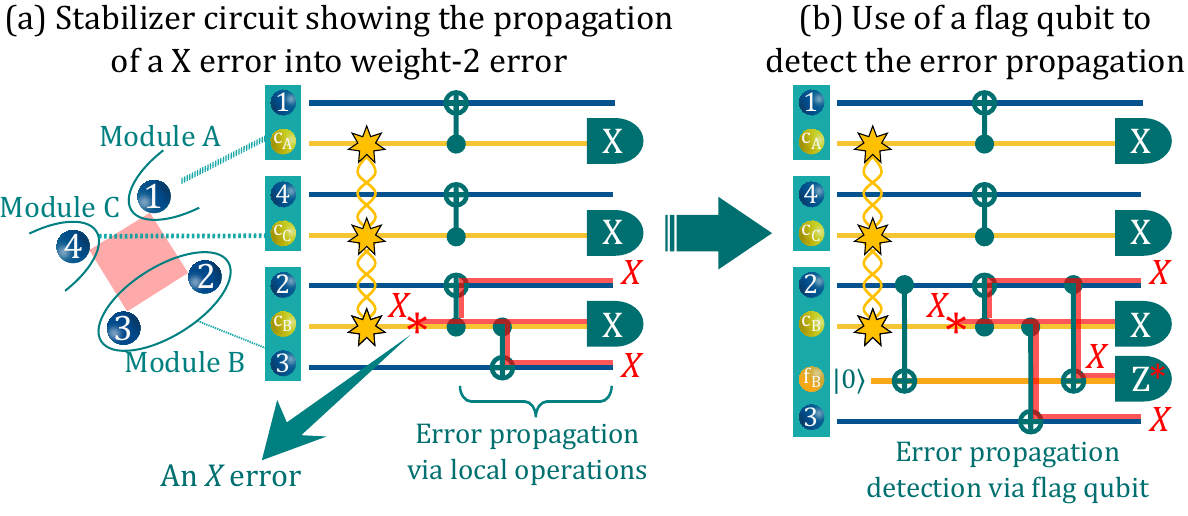}
\caption{Detection of error propagation (hook errors) using flag qubit within one module (Module B) of WT3 architecture stabilizer. (a) Shows the example of an $X$ error on a communication qubit ($\text{c}_\text{B}$) that propagates to two data qubits (2 \& 3) within one module via local \textsc{cnot} operations. (b) Depicts the usage of additional memory qubit (auxiliary) as flag qubit ($\text{f}_\text{B}$) where the error also propagates to the flag qubit (initialized in $|0\rangle$) and is detected via its $Z$ measurement flip (shown as $Z^*$).}
\label{fig:wt_3_flagged}
\end{figure}

The distributed implementation of the WT4 code is fault-tolerant by construction, given the fault-tolerant creation of the GHZ states, as there is no propagation of errors (hook errors) \cite{PhysRevA.90.062320} from one node to another. However, the bare WT3 architecture is not fault-tolerant w.r.t. hook errors, as there can be weight-2 error propagation from single qubits errors within each node due to local two-qubit controlled gates. A flagged version of the WT3 stabilizer measurement circuit could be considered to detect and correct for these hook errors~\cite{PhysRevLett.121.050502}. We describe one way to implement the flagged circuit in this appendix.
%In the main text, we briefly mentioned that the bare stabilizer circuit of the WT3 architecture is not fault-tolerant as an undetectable weight-2 (two-qubit) error propagation is possible with only a weight-1 (single qubit) error within each module. 
This error propagation is shown in Fig.~\ref{fig:wt_3_flagged}(a). An $X$ error can propagate from the communication qubit to both data qubits within the module, which goes undetectable. This error propagation can be detected using a flag qubit~\cite{PRXQuantum.1.010302}. The error is intentionally propagated to the flag qubit, where it is detected upon the measurement. 
This flagged stabilizer circuit (Fig.~\ref{fig:wt_3_flagged}(b)) can be used after the logical initialization (in the first layer of the first QEC cycle) of the code to detect such hook errors. Please note that this circuit does not provide a unique detection pattern for this error because, for instance, an $X$ error on qubit-2 after the first \textsc{cnot} would also lead to the same detection pattern (flip of $\text{f}_\text{B}$ measurement outcome). However, detecting such errors helps to decode errors better as this information can be combined with additional syndrome data to improve the decoding. For instance, in case of a hook error (as shown), another Z-stabilizer of qubit-3 will also flip, and this condition can be pre-programmed in the decoder to improve the logical error rates of the surface code. However, for the scope of this work, we only limit ourselves to simulating the unflagged version of the WT3 architecture.

\subsection{Simultaneous stabilizer measurements vs. sub-round structure}
\label{appsubsec:simultaneous_vs_sub-round}
A natural question arises: what is the impact of dividing the stabilizer measurement round into sub-rounds on decoding the QEC code? In a monolithic architecture, all the stabilizers in each layer are measured simultaneously via intertwined stabilizer circuits~\cite{PhysRevA.90.062320,McEwen2023relaxinghardware}. We can consider a small example in the case of perfect measurements. For two neighboring Z type stabilizers $Z_\text{abcd}$ and $Z_\text{befg}$, any incoming error $X_\text{b}$ will flip both the stabilizers and is trivial to correct for the decoder. However, if we first measure $Z_\text{abcd}$ in the first sub-round and $Z_\text{befg}$ in the second sub-round, there may be $X_\text{b}$ error that can occur between the two sub-rounds, triggering only the later $Z_\text{befg}$ stabilizer. This deceives the decoder to infer that a measurement error as occurred on one of the communication qubits b,e,f,g. Thereby missing the data qubit error correction via $X_\text{b}$ and increasing the $X$-logical error probability. This kind of error occurs more often in the WT3 architecture as it has twice the number of sub-rounds, which makes catching easy recovery configurations challenging for the decoder. More detailed analytical or numerical analysis of these errors is beyond the scope of this present work, and we leave this for future work.

\section{\label{app:emission-based}Emission based scheme and distillation protocols}
This section provides a full analytical treatment of the emission-based scheme and its performance evaluation in the presence of circuit-level noise. We first describe the entanglement generation through a Bell-pair between two modules via the single-click protocol~\cite{PhysRevA.59.1025}.

\subsection{\label{appsubsec:single_click_protocol}Single-click protocol}
Consider two modules as identical emitter setups, A and B, where each emitter (the communication qubit of each module acts as the emitter qubit) can be initialized in the arbitrary state
\begin{equation}
    |\Psi\rangle_\text{init}^\text{A/B}= \sqrt{1-\alpha}|0\rangle +\sqrt{\alpha}|1\rangle
\end{equation}
where $\alpha$ is the bright state population parameter. This state can be initialized for each emitter with fidelity $F_\text{prep}$. This gives the density matrix
\begin{equation}
\rho_\text{init}^\text{A/B}=(1-\alpha)|0\rangle \langle 0|+(2 F_\text{prep}-1)\sqrt{\alpha(1-\alpha)}|0\rangle \langle 1| +(2 F_\text{prep}-1)\sqrt{\alpha(1-\alpha)}|1\rangle \langle 0|+\alpha |1\rangle \langle 1|
\end{equation}
where the preparation infidelity noise is modelled as a dephasing channel over the state. Each of the emitters is now excited via a $\pi$-pulse, which causes the bright state $|1\rangle$ to emit a photon ($|1\rangle \rightarrow |11_\text{ph}\rangle$) while the dark state $|0\rangle$ does not lead to any emission ($|0\rangle \rightarrow |00_\text{ph}\rangle$), where the second ket is for the photon number, denoted with subscript \lq ph\rq. The resulting density matrix for each emitter-photon is
\begin{equation}
\rho_\text{emit}^\text{A/B}=(1-\alpha)|00_\text{ph}\rangle \langle 00_\text{ph}|+(2 F_\text{prep}-1)\sqrt{\alpha(1-\alpha)}
|00_\text{ph}\rangle \langle 11_\text{ph}| +(2 F_\text{prep}-1)\sqrt{\alpha(1-\alpha)}|11_\text{ph}\rangle \langle 00_\text{ph}| +\alpha |11_\text{ph}\rangle \langle 11_\text{ph}|
\end{equation}
However, there could be an excitation error due to the application of this excitation pulse, which can occur due to an extra photon being emitted or unintentionally exciting the dark state. We describe this as excitation error probability $p_\text{EE}$, and this phenomenon is modeled as a dephasing channel applied to either the emitter or the emitted photon state. The resulting density matrix is
\begin{equation}
\begin{split}
\rho_\text{emit}^\text{A/B}&=(1-\alpha)|00_\text{ph}\rangle \langle 00_\text{ph}|+(2 F_\text{prep}-1)(1-2p_\text{EE}) \sqrt{\alpha(1-\alpha)}|00_\text{ph}\rangle \langle 11_\text{ph}| +(2 F_\text{prep}-1)(1-2p_\text{EE})\\ 
&\sqrt{\alpha(1-\alpha)}|11_\text{ph}\rangle \langle 00_\text{ph}|+\alpha |11_\text{ph}\rangle \langle 11_\text{ph}|
\end{split}
\end{equation}
for each emitter-photon setup. After this step, the flying photons are sent to a middle station where both photons are interfered with via a beam splitter. However, there is noise in the fiber, which carries the photons.

We consider the phase uncertainty $\lambda$ due to the path length difference between the interferometer's two arms. This can be realized as a dephasing channel with some error probability $\lambda$. This is calculated as:
\begin{equation}
\lambda=\frac{1}{2}\bigg(1+\frac{I_1(\sigma(\varphi)^{-2})}{I_0(\sigma(\varphi)^{-2})}\bigg)
\end{equation}
Where $I_0$ and $I_1$ are modified Bessel functions of the zeroth and first order, and $\sigma(\varphi)$ is the standard deviation of the phase instability. Typically, this value is around $0.984$~\cite{Humphreys2018,PhysRevA.99.052330,10.1116/5.0200190} for the relevant phase uncertainty in the recent experiments. This dephasing channel is applied to one of the photons (the one from emitter A).
Then we obtain 
\begin{equation}
\begin{split}
\rho_\text{em-ph}^\text{A}&=(1-\alpha)|00_\text{ph}\rangle \langle 00_\text{ph}|+(2 F_\text{prep}-1)(1-2p_\text{EE})
(2\lambda-1)  \sqrt{\alpha(1-\alpha)}|00_\text{ph}\rangle \langle 11_\text{ph}| +(2 F_\text{prep}-1)\\ &
(1-2p_\text{EE})(1-2\lambda)\sqrt{\alpha(1-\alpha)}|11_\text{ph}\rangle \langle 00_\text{ph}|
+\alpha |11_\text{ph}\rangle \langle 11_\text{ph}|
\end{split}
\end{equation}

Moreover, as the fiber is also noisy, we have photon loss through the fiber. We combine this effect with also the photon detector efficiency. Each excitation's total photon detection probability is parameterized by $\eta_\text{ph}$. This is a product of the total transmissivity between the emitter and the photon detector (accounts for photon loss), the detector efficiency, and the probability that the desired photon is emitted during the detection time window and within the zero-phonon line of the emitter.
We model this overall effect as an amplitude-damping channel on both the emitter photons. The Kraus operators for the amplitude damping (AD) channel are:
\begin{equation}
K^\text{AD}_1=
\begin{bmatrix}
1 & 0\\
0 & \sqrt{\eta_\text{ph}}
\end{bmatrix},\;
K^\text{AD}_2=
\begin{bmatrix}
0 & \sqrt{1-\eta_\text{ph}}\\
0 & 0
\end{bmatrix}
\end{equation}
According to Eqn.~\ref{eqn:kraus_definition}, operating with the amplitude-damping channel on the emitted photons from both the modules, we get the output states before the photon detection as:
\begin{equation}
\begin{split}
\rho_\text{em-ph}^\text{A}&=(1-\alpha)|00_\text{ph}\rangle \langle 00_\text{ph}|+(2 F_\text{prep}-1)(1-2p_\text{EE})\sqrt{\eta_\text{ph}}
(2\lambda-1)  \sqrt{\alpha(1-\alpha)}|00_\text{ph}\rangle \langle 11_\text{ph}| +\alpha(1-\eta_\text{ph})\\ &|10_\text{ph}\rangle \langle 10_\text{ph}|
+(2 F_\text{prep}-1)(1-2p_\text{EE})\sqrt{\eta_\text{ph}}(1-2\lambda)\sqrt{\alpha(1-\alpha)}|11_\text{ph}\rangle \langle 00_\text{ph}| +\alpha \eta_\text{ph} |11_\text{ph}\rangle \langle 11_\text{ph}|
\end{split}
\end{equation}
\begin{equation}
\begin{split}
\rho_\text{em-ph}^\text{B}&=(1-\alpha)|00_\text{ph}\rangle \langle 00_\text{ph}|+(2 F_\text{prep}-1)(1-2p_\text{EE})\sqrt{\eta_\text{ph}}  \sqrt{\alpha(1-\alpha)}|00_\text{ph}\rangle \langle 11_\text{ph}| +\alpha(1-\eta_\text{ph})|10_\text{ph}\rangle \langle 10_\text{ph}|\\ &
+(2 F_\text{prep}-1)(1-2p_\text{EE})\sqrt{\eta_\text{ph}}\sqrt{\alpha(1-\alpha)}|11_\text{ph}\rangle \langle 00_\text{ph}| +\alpha \eta_\text{ph} |11_\text{ph}\rangle \langle 11_\text{ph}|
\end{split}
\end{equation}
For the photon detector, we assume the regime with negligible dark count probability, similar to other schemes, about $10^{-6}$, and we neglect it. We consider the photon indistinguishability $\mu$ for the joint Bell-pair measurement for the perfect interference. The measurement projector for the two-photon measurement can be described with the following POVMs acting only on the photonic degrees of freedom~\cite{10.1145/3341302.3342070}.
\begin{equation}
    \begin{split}
        E_{00}&=|0_\text{ph}0_\text{ph}\rangle \langle 0_\text{ph}0_\text{ph}|\\
        E_{01}&=\frac{1}{2}\bigg( \frac{\sqrt{1+\sqrt{\mu}}+\sqrt{1-\sqrt{\mu}}}{\sqrt{2}}\big(|0_\text{ph}1_\text{ph}\rangle \langle 0_\text{ph}1_\text{ph}|+|1_\text{ph}0_\text{ph}\rangle \langle 1_\text{ph}0_\text{ph}|\big)+ \frac{\sqrt{1-\sqrt{\mu}}-\sqrt{1+\sqrt{\mu}}}{\sqrt{2}}\\ &\big(|0_\text{ph}1_\text{ph}\rangle \langle 1_\text{ph}0_\text{ph}|+|1_\text{ph}0_\text{ph}\rangle \langle 0_\text{ph}1_\text{ph}|\big)+ \sqrt{1+\mu} |1_\text{ph}1_\text{ph}\rangle \langle 1_\text{ph}1_\text{ph}|\bigg)\\
        E_{10}&=\frac{1}{2}\bigg( \frac{\sqrt{1+\sqrt{\mu}}+\sqrt{1-\sqrt{\mu}}}{\sqrt{2}}\big(|0_\text{ph}1_\text{ph}\rangle \langle 0_\text{ph}1_\text{ph}|+|1_\text{ph}0_\text{ph}\rangle \langle 1_\text{ph}0_\text{ph}|\big)+ \frac{\sqrt{1+\sqrt{\mu}}-\sqrt{1-\sqrt{\mu}}}{\sqrt{2}}\\ &\big(|0_\text{ph}1_\text{ph}\rangle \langle 1_\text{ph}0_\text{ph}|+|1_\text{ph}0_\text{ph}\rangle \langle 0_\text{ph}1_\text{ph}|\big) + \sqrt{1+\mu} |1_\text{ph}1_\text{ph}\rangle \langle 1_\text{ph}1_\text{ph}|\bigg)\\
        E_{11}&=\frac{\sqrt{1-\mu}}{\sqrt{2}}|1_\text{ph}1_\text{ph}\rangle \langle 1_\text{ph}1_\text{ph}|
    \end{split}
\end{equation}
Here $E_{00}$ corresponds to no photon detection. $E_{01}$ and $E_{10}$ correspond to one photon detected in the right and left detector, respectively. And $E_{11}$ means two photons being detected, one at each detector. For the successful entanglement generation, we herald on a single click, that is, one photon being detected at either the right or left detector.
For a left/right detector click (via $E_{10}/E_{01}$), the output states would be 
\begin{equation}
    \rho_\text{out}^\text{left/right}=\frac{(\mathbb{I}\otimes E_{10/01})(\rho_\text{em-ph}^\text{A}\otimes \rho_\text{em-ph}^\text{B})(\mathbb{I}\otimes E_{10/01})^\dagger}{\text{Tr}[(\mathbb{I}\otimes E_{10/01})(\rho_\text{em-ph}^\text{A}\otimes \rho_\text{em-ph}^\text{B})(\mathbb{I}\otimes E_{10/01})^\dagger]}
\end{equation}
where the tensor product $\rho_\text{em-ph}^\text{A}\otimes \rho_\text{em-ph}^\text{B}$ follows the order $|\text{emitter-1}\rangle|\text{emitter-2}\rangle|\text{photon-1}\rangle|\text{photon-2}\rangle$. And we identify the denominator as the success probability $P_\text{single}^\text{left/right}$ of the respective single-click.
We get the raw output entangled state after performing a partial trace over the photons:
\begin{equation}
\begin{split}
    &\rho_\text{out}^\text{left}=\frac{1}{4+\alpha \eta_\text{ph}(\mu-3)}\Big(2(1-\alpha)|01\rangle\langle01|+ 2(2 F_\text{prep}-1)^2(1-2p_\text{EE})^2(1-\alpha)(2\lambda-1)\sqrt{\mu} (|01\rangle \langle 10|\\
    &+|10\rangle \langle 01|)+ 2(1-\alpha)|10\rangle \langle 10|+\alpha(4+\eta_\text{ph}(\mu-3))|11\rangle \langle 11|\Big)
    \label{eqn:left_single}
\end{split}
\end{equation}

\begin{equation}
\begin{split}
    &\rho_\text{out}^\text{right}=\frac{\alpha \eta_\text{ph}}{2}\Big( (1-\alpha)|01\rangle \langle 01|+ (2 F_\text{prep}-1)^2(1-2p_\text{EE})^2(1-\alpha)(1-2\lambda)\sqrt{\mu} (|01\rangle \langle 10|+|10\rangle \langle 01|)\\
    &+(1-\alpha)|10\rangle \langle 10|+\frac{\alpha}{2}(4+\eta_\text{ph}(\mu-3))|11\rangle \langle 11|\Big)
    \label{eqn:right_single}
\end{split}
\end{equation}
To convert these states into the desired Bell-state $|\Psi^+\rangle$ we apply an $X^\text{c}$ gate to one of the emitters and apply a phase correction on one of the emitters if the right detector clicks. Both these gates are noisy and characterized by depolarization noise with gate error $p_g$. We include gate error for the post-correction gates, in contrast, to Ref.~\cite{10.1116/5.0200190}. This gives the corrected output density matrices $\rho^\text{left}_\text{final}$ and $\rho^\text{right}_\text{final}$. 
\begin{equation}
\begin{split}
       & \rho^\text{left}_\text{final}=(X^\text{c}\otimes \mathbb{I})\rho_\text{out}^\text{left}(X^\text{c}\otimes \mathbb{I})\\
        & \rho^\text{right}_\text{final}=(X^\text{c}\otimes Z^\text{c})\rho_\text{out}^\text{right}(X^\text{c}\otimes Z^\text{c})
\end{split}
\end{equation}

The average density matrix from the single click protocol is, therefore:
\begin{equation}
\begin{split}
        \rho_\text{s}&=P_\text{single}^\text{left}(\mathcal{N}_\text{gate}^\text{single-qubit}\otimes \mathbb{I})\rho^\text{left}_\text{final}(\mathcal{N}_\text{gate}^\text{single-qubit}\otimes \mathbb{I})+P_\text{single}^\text{right}(\mathcal{N}_\text{gate}^\text{single-qubit}\otimes \mathcal{N}_\text{gate}^\text{single-qubit})\rho^\text{right}_\text{final}\\&(\mathcal{N}_\text{gate}^\text{single-qubit}\otimes \mathcal{N}_\text{gate}^\text{single-qubit})
\end{split}
\end{equation}
It turns out that $P_\text{single}^\text{left}=P_\text{single}^\text{right}$ and the total success probability of the single-click protocol is the sum of both:
\begin{equation}
\label{eqn:p_succ_single}
    P_\text{single}^\text{succ}=\frac{1}{2}\alpha \eta_\text{ph}\big(4-\alpha \eta_\text{ph}(3-\mu) \big)
\end{equation}
We calculate the overall fidelity of the single-click protocol as
\begin{equation}
    F_\text{single}=\text{Tr}\Big[\sqrt{\sqrt{\Phi^+}\rho_\text{s}\sqrt{\Phi^+}} \Big]^2
\end{equation}
where $\Phi^+=|\Phi^+\rangle \langle \Phi^+|$ is the target Bell-pair density matrix.
The full-analytical expression of the average fidelity is:
\begin{equation}
\label{eqn:fid_single}
        F_\text{single}=\frac{\alpha \eta_\text{ph}}{36 P^\text{succ}_\text{single}}\Big( 36 (1-\alpha)(1+\phi)+(9p_g-4p_g^2)\big(36\phi-4\alpha (8+\eta_\text{ph}(\mu-3))\big) \Big)
\end{equation}
where
\begin{equation}
    \phi=\sqrt{\mu}(2F_\text{prep}-1)^2(2\lambda-1)(1-p_\text{EE})^2
\end{equation}
captures the overall noise in the photon generation and detection.

\subsection{\label{appsubsec:double_click}Double click protocol}
The raw output state ($\rho_\text{out}^\text{left/right}$) from the single-click protocol has a contribution from the pure bright-state term $|11\rangle \langle 11|$, which leads to an overall mixed state. See Eqn.~\ref{eqn:left_single} and Eqn.~\ref{eqn:right_single}. Double-click (Barrett-Kok) protocol eliminates this term by the application of $X^\text{c}$ gate on each emitter on the raw output state after the single click. Another round of single-click protocol is performed and heralded on the consecutive single-click.  This eliminates the $|00\rangle \langle 00|$ term (after $X^\text{c}$ gates) because it does not emit photons in the second round. The detection pattern now has four possibilities: $\text{left}\otimes \text{left}$, $\text{left}\otimes \text{right}$, $\text{right}\otimes \text{left}$, and $\text{right}\otimes \text{right}$, as the photon detector click patterns. However, for the double-click protocol~\cite{PhysRevA.71.060310}, we initialize the initial state with $\alpha=1/2$, and the phase uncertainty $\lambda$ gets eliminated due to the symmetry of the protocol.
The total success probability for the double-click protocol is:
\begin{equation}
P^\text{succ}_\text{double}=\sum_{\text{a}}\bigg( P_\text{single}^\text{a} P_\text{single}^{\text{a}\otimes \text{a}}+P_\text{single}^\text{a} P_\text{single}^{\text{a}\otimes \Bar{\text{a}}} \bigg)
\end{equation}
where $\text{a}\in \{\text{left},\text{right}\}$, and $\overline{\text{left}}=\text{right}$, and vice versa. All these terms are equal, and we get the total probability with post-correction with single-qubit gates as:
\begin{equation}
    \begin{split}
        P^\text{succ}_\text{double}=\frac{\eta_\text{ph}^2}{36}\bigg( 18+12p_g \big( 2+\eta_\text{ph}(\mu-3)\big)+p_g^2\eta_\text{ph}^2(\mu-3)^2 \bigg)
    \end{split}
\end{equation}

This expression reduces to $\eta_\text{ph}^2/2$ without gate error, as considered in Ref. \cite{10.1116/5.0200190}.
The full analytical expression for the average fidelity of the double-click protocol is:
\begin{equation}
\label{eqn:fid_double}
\begin{split}
    &F_\text{double}=\Big(3 p_g \big(-54 (1 + \eta_{\text{ph}}) + 
      p_g (96 + 36 \eta_{\text{ph}} + 
         p_g (-32 - (-9
         + 4 p_g) \eta_{\text{ph}} (-8 + 3 \eta_{\text{ph}})))\big) + 
   2 p_g \big(2 \big(1 - 2 F_{\text{prep}}\big)^2 \big(1 - 2 p_{\text{EE}}\big)^4 \\
   &(-189 + 
         4 p_g (81 + 4 p_g (-15 + 4 p_g))) + (27 - 
         2 p_g (9 
         + 2 p_g (-9 + 4 p_g))) \eta_{\text{ph}} + 
      3 p_g^2 (-9 + 4 p_g)\eta_{\text{ph}}^2 \big) \mu + (9 - 
      4 p_g)\\
      &p_g^3 \eta_{\text{ph}}^2 \mu^2 + 
   162 \big(1 + \big(1 - 2 F_{\text{prep}}\big)^2 \big(1 - 2 p_{\text{EE}}\big)^4 \mu\big)\Big)/\Big(18 \big(18 +
    p_g \big(24 + \eta_{\text{ph}} (12 + p_g \eta_{\text{ph}} (-3 + \mu)) (-3 + \mu)\big)\big)\Big)
\end{split}
\end{equation}
When no gate error is considered, $p_g=0$, then this reduces to $F_\text{double}=(1+\frac{\phi^2}{(1-2F_{\text{prep}})^2})/2$, given $\lambda=1$. This is slightly different from the limiting expression in Ref.~\cite{10.1116/5.0200190} because of the change in the definition of the preparation fidelity. In our work, the preparation fidelity refers to the state preparation of the emitter state only, initialized with the bright-state parameter. In contrast, in Ref.~\cite{10.1116/5.0200190}, it refers to the emitter-photon state preparation fidelity, and the dephasing noise channel is applied to the emitted photon (which is applied in both rounds, incurring an extra quadratic factor in the $\phi^2$ term).

\subsection{\label{appsubsec:single_vs_double}Single-click versus double-click}
The $P_\text{succ}^\text{link}$ for single-click protocol depends on bright state parameter $\alpha$, and changing this parameter has a tradeoff between fidelity and success probability. This is evident from Eqn.~\ref{eqn:p_succ_single} and Eqn.~\ref{eqn:fid_single}. For double-click, the $P_\text{succ}^\text{link}$ is quadratically dependent on the effective photon detection probability $\eta_\text{ph}$, which is very small (for the NTP) and therefore unfavorable for threshold performance. Through the single-click protocol, the $P_\text{succ}^\text{link}$ is linear in $\eta_\text{ph}$ (which is higher) and we can trade in some $P_\text{succ}^\text{link}$ to reduce the infidelity of the Bell state, and satisfy the numerical bounds on $P_\text{succ}^\text{link}$ and fidelity as discussed in the results section. However, the fidelity was still low (less than 0.90) for the existence of a threshold (inclusive of the physical gate error), even for very low values of $\alpha$. A similar exploration was also done in Ref.~\cite{10.1116/5.0200190} on single-click vs. the double-click protocol without the gate error. In this view, we use the single-click protocol for NTP and double-click for FP to optimize the logical error rates in these regimes.

\subsection{\label{appsubsec:EM_protocols}Fusion and distillation protocols}
We describe an example of a fusion and distillation protocol using elementary Bell pairs. As an example (shown in Fig.~\ref{fig:emission based}(c)), we consider how to generate a GHZ state between three modules, A, B, and C, while using auxiliary Bell pairs along the way to increase fidelity. First, a Bell pair, $|\Phi^+\rangle_\text{AB}$, is generated between the communication qubits of modules A and B and swapped to memory qubits within each module through local two-qubit gates. Another Bell pair $|\Phi^+\rangle_\text{AB}$ is generated and is used to distill the prior one by measuring the stabilizer $X_AX_B$ (distilled state denoted by *). Similarly, a Bell pair, $|\Phi^+\rangle_\text{BC}$, is generated between modules B and C and distilled. Finally, a \textsc{cnot} operation is applied between the two Bell pair qubits at module B, followed by a measurement of one of the qubits (the communication qubit), which results in a three-qubit GHZ state $|\Phi^+\rangle_\text{ABC}$ between A-B-C up to a local single qubit rotation. This GHZ state can be further distilled via a Bell pair $|\Phi^+\rangle_\text{AC}$.

\begin{figure}[hbtp]
\centering
\includegraphics[width=0.3\textwidth]{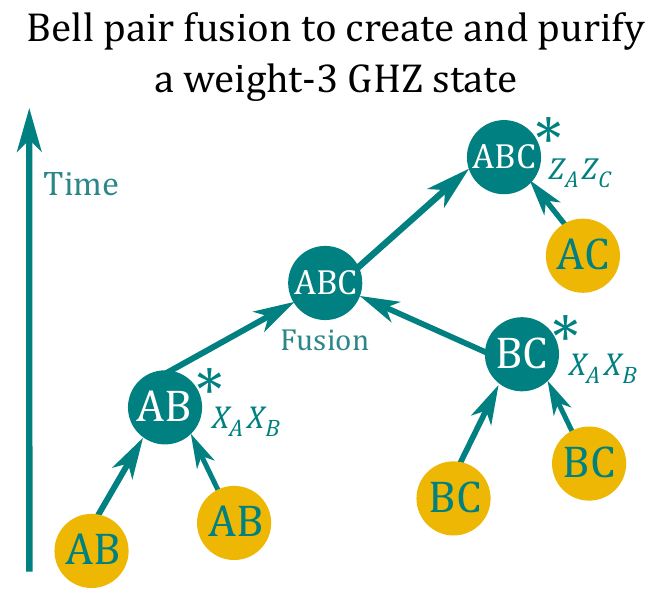}
\caption{An example of a GHZ state fusion protocol with distillation used to create a high-fidelity WT3 GHZ state presented as a tree graph. Events in yellow indicate Bell pair generation between the respective modules, while the teal-colored arrows labeled with a multi-Pauli operator indicate a distillation event by measuring that operator. Modules with a teal color indicate the distilled entangled state.}
\label{fig:emission_fusion_example}
\end{figure}

\begin{figure*}
\centering
\includegraphics[width=\textwidth]{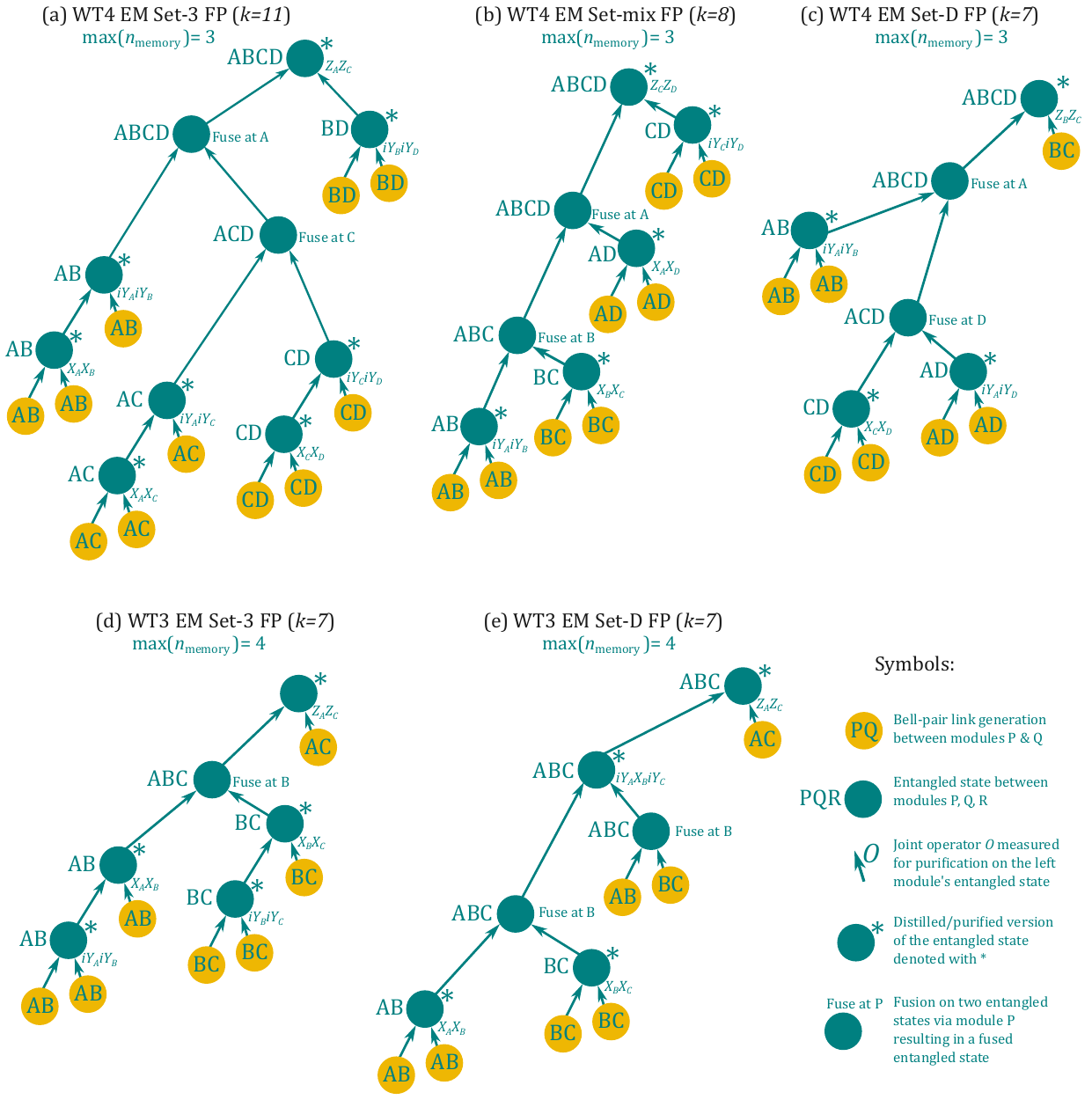}
\caption{Optimised emission-based (EM) scheme fusion and distillation protocols for various architectures and schemes, for which we obtained (highest) thresholds. We sketch these protocols as binary-tree diagrams, similar to Ref.~\cite{10.1116/5.0200190}, with time running vertically up. We also show optimized protocols for two extra coherence time sets, namely Set-mix and Set-D, introduced in App.~\ref{app:threshold_details} and App.~\ref{app:diamond_center}. (a), (b) and (c) show the protocols for WT4 architecture, while (d) and (e) show protocols for WT3 architecture. The total number of Bell-pairs used in a fusion protocol is shown as its $k$ value. These protocols consist of operations such as the Bell-pair generation (via double-click protocol for all these examples), the fusion of Bell-pair with other entangled states to increase the weight of the resulting entangled state distributed between the modules, and purification/distillation operations by measuring some joint operators (stabilizers of that entangled state). Through these protocols, the total number of memory qubit requirements can be estimated for each case. We show the maximum number of memory qubits required (including the data qubits) in a module for each reported protocol below its label.}
\label{fig:emission_protocols}
\end{figure*}

Now we give explicit examples of the EM fusion protocols for which we found a threshold. These protocols were searched using the `dynamic program' first introduced in Ref.~\cite{9292429} and 
later used in Ref.~\cite{10.1116/5.0200190}. We searched the GHZ fusion and distillation protocols pool of WT4 and WT3 GHZ states for maximum stabilizer fidelity output. Following that, a threshold calculation was run for a set of the best protocols chosen based on stabilizer fidelity. Finally, the protocol with the highest threshold and least logical error rates was chosen for each case. These are the protocols we showcase in Fig.~\ref{fig:emission_protocols}, for which we obtained thresholds. Using these binary tree diagrams, we sketch the operations involved in the GHZ creation chronologically via the EM protocol. Details of the protocol operations are provided in the caption of Fig.~\ref{fig:emission_protocols}. It consists of elementary Bell-pair generation between the modules and their distillation and fusion with other modules. For each protocol, we report the maximum number of Bell-pairs used as its $k$ value and the maximum number of memory qubits used by a node for that protocol (this includes memory qubits used as data qubits). In all these protocols, the maximum number of auxiliary qubits used is two, but WT4 architecture has one data qubit per node, while WT3 architecture has two data qubits per node. There are two extra coherence times parameter sets for which we obtained thresholds. These are not part of the main script, but we motivate these (Set-mix and Set-D) in App.~\ref{app:threshold_details} and App.~\ref{app:diamond_center}.

\section{\label{app:reflection} Reflection scheme}
The key mechanism of the RFL scheme is the photon-mediated \textsc{cnot} gate in the protocol. This direct spin-photon \textsc{cnot} operation can be achieved by reflecting a time-bin encoded single photonic qubit from a single-sided cavity with a strongly coupled communication qubit~\cite{ritter2012elementary, reiserer2015cavity}. Ideally, the photon is prepared in a state $\frac{1}{\sqrt{2}}(\ket{e}_{\text{ph}}+\ket{l}_{\text{ph}})$ while the communication qubit is prepared in $\ket{0}$. Assuming that the $\ket{0}$ state does not couple to the cavity field and that the incoming photon is on resonance with the cavity, scattering of the early photon result in the transformation $\frac{1}{\sqrt{2}}(\ket{e}_{\text{ph}}+\ket{l}_{\text{ph}})\ket{0}\to\frac{1}{\sqrt{2}}(-\ket{e}_{\text{ph}}+\ket{l}_{\text{ph}})\ket{0}$, i.e. the photon is reflected from the cavity with a $\pi$-phase shift. Before the scattering of the late photon, a Hadamard gate is applied to the communication qubit, transforming $\ket{0}$ into $\frac{1}{\sqrt{2}}(\ket{0}+\ket{1})=\ket{+}$. If the $\ket{1}$ state is strongly coupled to the cavity field, scattering of the late photon will result in the transformation, $\frac{1}{\sqrt{2}}(-\ket{e}_{\text{ph}}+\ket{l}_{\text{ph}})\ket{+}\to\frac{1}{\sqrt{2}}(\ket{e}_{\text{ph}}\ket{+}+\ket{l}_{\text{ph}}\ket{-})$ up to a global phase where $\ket{-}=\frac{1}{\sqrt{2}}(\ket{0}-\ket{1})$. This is because scattering from the $\ket{1}$ state will reflect the photon without a $\pi$-phase shift in the strongly coupled regime. A final Hadamard gate on the communication qubit taking $\ket{+}\to\ket{0}$ and $\ket{-}\to\ket{1}$ completes the \textsc{cnot} operation. By reflecting a single photonic qubit with consecutive modules, this scheme allows for the direct generation of a GHZ state as shown schematically in Fig.~\ref{fig:reflection} in the main script. The generation is heralded by a final measurement of the photonic qubit in the $X$-basis, which can be achieved with photonic switching and a fiber delay.

Here, we describe the reflection scheme step by step in mathematical detail. We consider the scattering of a photonic time-bin qubit from a single spin-coupled cavity and then generalize this to the case where the photon scatters off multiple cavities. The spin is initialized in the $\ket{0}$ state, and the photonic state is an equal superposition of the early and late time-bin states. Hence, the photon-spin state is 
$\frac{1}{\sqrt{2}} \left(|\mathrm{E}0\rangle+|\mathrm{L}0\rangle\right)$. 

Let $r_{0}$ and $r_{1}$ denote the photonic reflection coefficients corresponding to the cases when the spin is in the uncoupled state $|0\rangle$ and coupled state $|1\rangle$, respectively. After the early time-bin component of the photon scattered, the unnormalized state of the system becomes $\frac{1}{\sqrt{2}} \left(r_{0}|\mathrm{E}0\rangle+ |\mathrm{L}0\rangle\right)$. Then, a Hadamard gate is executed on the spin, which transforms the direct product state into $\frac{1}{\sqrt{2}} \left(r_{0}|\mathrm{E}+\rangle+|\mathrm{L}+\rangle\right)$. Next, the late-time-bin component of the photon is scattered, and the state of the system becomes $\frac{1}{\sqrt{2}}\left(r_{0}|\mathrm{E}+\rangle+\frac{1}{\sqrt{2}}\left(r_{0}|\mathrm{L}0\rangle+r_{1}|\mathrm{L}1\rangle\right)\right)$. Finally, a second Hadamard gate is executed on the spin, and the state of the system becomes 
\begin{equation}
    \ket{\Psi_{1}}=\frac{1}{\sqrt{2}}\left( r_{0}|\mathrm{E}0\rangle+ 
    \frac{r_{0}+r_{1}}{2}|\mathrm{L}0\rangle+\frac{r_{0}-r_{1}}{2}|\mathrm{L}1\rangle\right).
\end{equation}

For multiple cavities, the above procedure generalizes straightforwardly. Consequently, the final state of the multiple spin qubits and the photon will be

\begin{align}
    \begin{split}
        \ket{\Psi_{N}}=&\frac{1}{\sqrt{2}} \Bigg[r_{0}^{n}|\mathrm{E}\rangle |0\rangle^{\otimes n}+\\
        &\left.|\mathrm{L}\rangle\left(\frac{r_{0}+r_{1}}{2}|0\rangle+\frac{r_{0}-r_{1}}{2}|1\rangle\right)^{\otimes n}\right],
    \end{split}
\end{align}
where $n$ is the number of the spins. 

At the final step, an optical switch and a delay line are used to measure the photon in the X basis ($|\pm\rangle_{\rm ph}=\left(|\mathrm{E}\rangle\pm|\mathrm{L}\rangle\right)/\sqrt{2}$). If the photon is detected in the $\ket{+}_{\rm ph}$ state, the spins will be projected into an (unnormalized) state
\begin{equation}
    \ket{\Psi_{+}}=\frac{1}{2}\left[r_{0}^{n}|0\rangle^{\otimes n}+\left(\frac{r_{0}+r_{1}}{2}|0\rangle+\frac{r_{0}-r_{1}}{2}|1\rangle\right)^{\otimes n}\right].
    \label{eq:Psi_+}
\end{equation}
One can see from Eqn.~(\ref{eq:Psi_+}) that, when $r_{0} \approx -r_{1}$, it resembles the GHZ state $\ket{\text{GHZ}}=(|0\rangle^{\otimes n}+|1\rangle^{\otimes n})/\sqrt{2}$. If the photon is detected in the $\ket{-}$ state, the spins will be projected into a state
\begin{equation}
    \ket{\Psi_{-}}=\frac{1}{2}\left[r_{0}^{n}|0\rangle^{\otimes n}-\left(\frac{r_{0}+r_{1}}{2}|0\rangle+\frac{r_{0}-r_{1}}{2}|1\rangle\right)^{\otimes n}\right].
     \label{eq:Psi_-}
\end{equation}
In this case, when $r_{0} \approx -r_{1}$, the resulting state is close to the state $(|0\rangle^{\otimes n}-|1\rangle^{\otimes n})/\sqrt{2}$. In this case, a Z gate can be applied on the first spin qubit to recover the state $\ket{\text{GHZ}}$. 
Note that the states in Eqs.~(\ref{eq:Psi_+})-(\ref{eq:Psi_-}) are not normalized. The sum of the norms of these two states gives the success probability of the protocol in the absence of additional photon loss:
\begin{equation}
P_{\text{succ}}^{\text{GHZ}}=\langle\Psi_{+}|\Psi_{+}\rangle+\langle\Psi_{-}|\Psi_{-}\rangle.\label{eq:ref_P}
\end{equation}

The reflection coefficients, $r_0$ and $r_1$ can be expressed in terms of the different spin-photon interface parameters as~\cite{photon-cavity_int}:
\begin{equation}
r_{1,0} =1-\frac{2\frac{\kappa_{\rm c}}{\kappa_{\rm c}+\kappa_{\rm l}}}{1+2\mathrm{i}\frac{\omega}{\kappa_{\rm c}+\kappa_{\rm l}}+\frac{4C_{1}}{1+2\mathrm{i}\delta_{1,0}/\gamma}},
\label{eq:ref_coef}
\end{equation}
where $\delta_0$ and $\delta_1$ are the detunings between the corresponding spin transition and the input photon frequency and $\omega$ is the detuning between the cavity resonance frequency and the input photon. The two spin transitions have a fixed detuning $\Delta$ hence $\delta_0-\delta_1=\Delta$ (see Fig.~\ref{tab_para_rfl} in the main text).

\begin{figure}
    \centering
    \includegraphics[width= 0.48\textwidth]{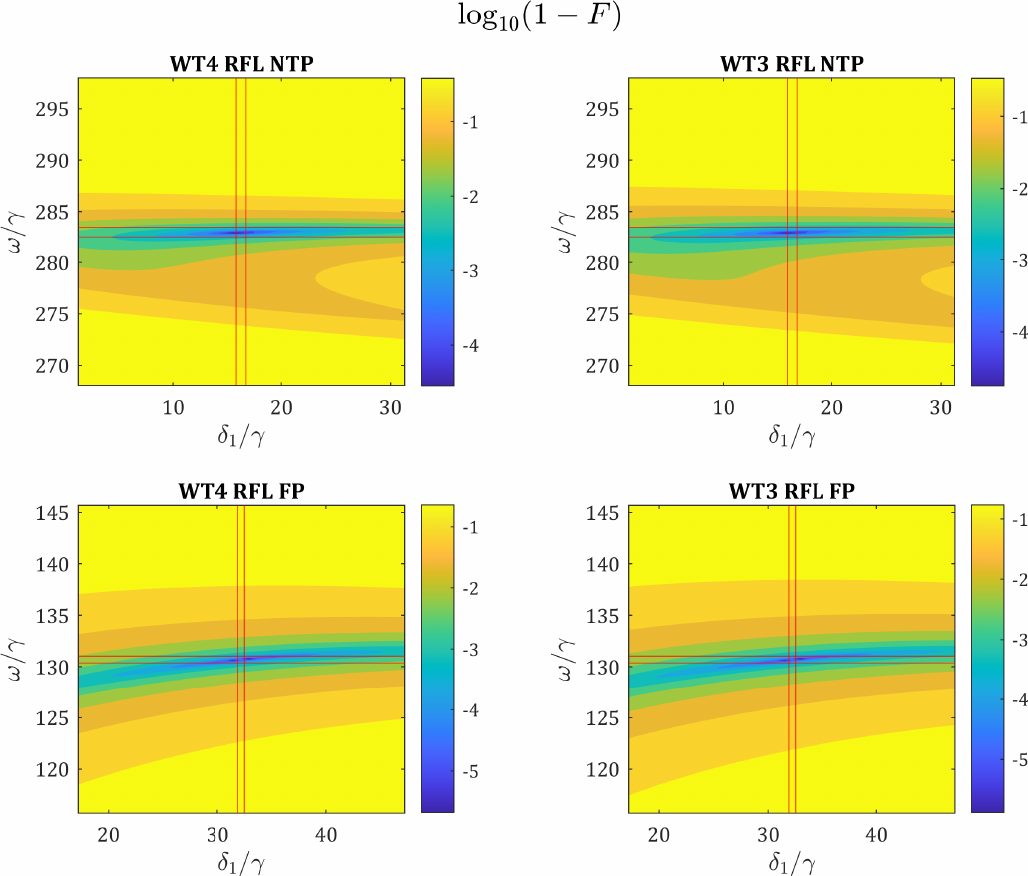}
    \caption{Scanning of the base-10 logarithm of the infidelity of the GHZ state over the variables of $\omega$ and $\delta_1$. The subtitles of the figures denote different cases of state-of-the-art parameters NTP and FP for WT4 and WT3 GHZ states. The found optimum $\omega$ and $\delta_1$ and their deviations of $\sigma/\gamma$ are indicated with the two horizontal and two vertical red lines with the same separations of $2\sigma/\gamma$. The gate error is set to 0 in this calculation.}
    \label{fig:scan_ref_fid}
\end{figure}

\begin{figure}
    \centering
    \includegraphics[width= 0.48\textwidth]{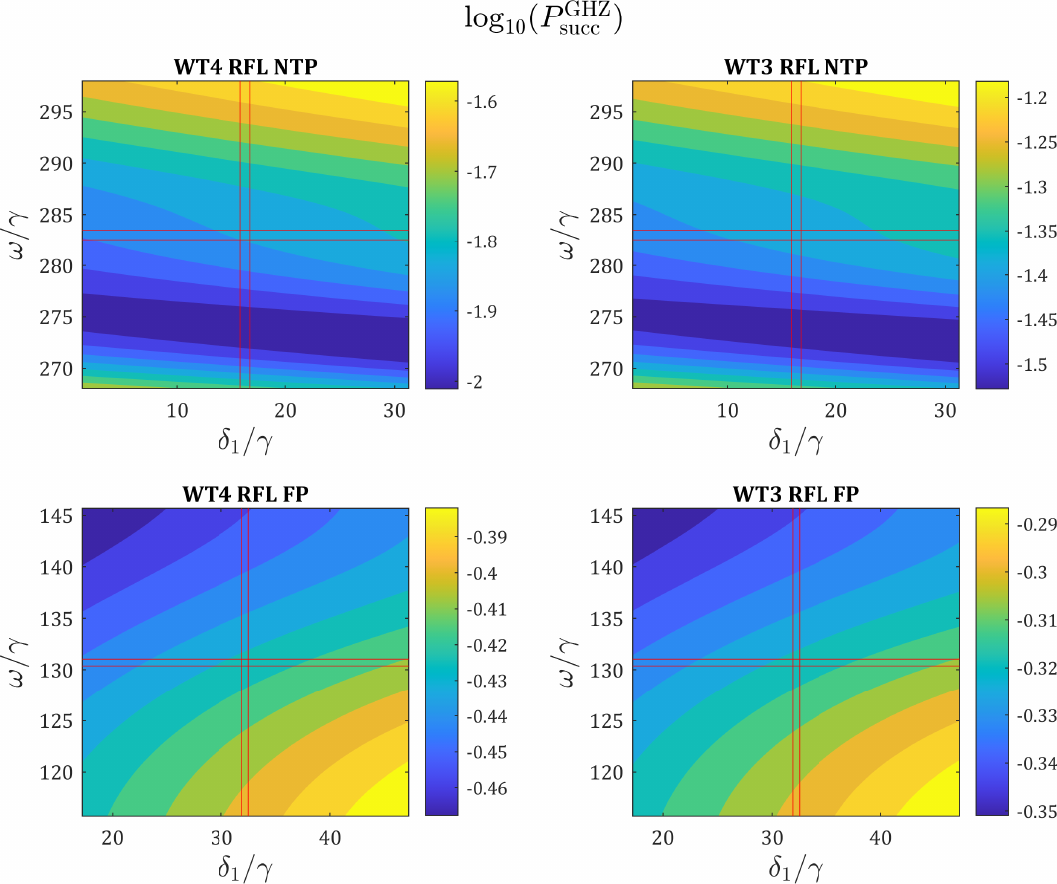}
    \caption{Scanning of the base-10 logarithm of the success rate of the GHZ state generation over the variables of $\omega$ and $\delta_1$. The plotting follows the same fashion as Fig.~\ref{fig:scan_ref_fid}.}
    \label{fig:scan_ref_rate}
\end{figure}

To determine the optimal performance, we optimize these two detunings and find a region where the fidelity of the GHZ state reaches a global maximum while the success rate is nearly flat, as shown in Fig.~\ref{fig:scan_ref_fid} and Fig.~\ref{fig:scan_ref_rate}. Hence, the location of the maximum fidelity of the GHZ state determines the optimum values of $\omega$ and $\delta$ as listed in the table in Fig.~\ref{tab_para_rfl} in the main text. For the simulation of the surface code performance, our simulation also includes faulty  Hadamard gates and Z gates as well as additional photon loss from e.g. the optical circulators. 

\section{\label{app:car_single}Carving scheme with single photon source}
First, we describe the brief principle of the CAR scheme. In the setup, the communication qubit state $\ket{0}$ is uncoupled while state $\ket{1}$ is strongly coupled to the incoming photon through the cavity/waveguide field as in the RFL scheme. Consequently, an incoming photon will be reflected if the qubit is in state $\ket{1}$ and transmitted if the qubit is in state $\ket{0}$ because of interference between the incoming and scattered fields~\cite{sorensen2003}. This can be used to project an initially un-entangled multi-qubit state to a GHZ state probabilistically using a setup as depicted in Fig.~\ref{fig:carving}(a) of the main script.

The communication qubits are initially all prepared in the state $\ket{+}$, and a single photon is sent through the setup. The first beam splitter creates a superposition of the photon propagating in the upper (subscript "u") or lower (subscript "l") arm. Notably, the photon can only be transmitted through an arm if all communication qubits in that arm are in state $\ket{0}$. The final beam splitter erases the information about which arm the photon propagated through. The subsequent detection of the photon projects all communication qubits in the lower or upper arm in state $\ket{0}$. Consequently, the detection of the photon will project the communication qubits into a state $\frac{1}{\sqrt{2}} \Big(|\,0\,0\,0\,\cdots\rangle_{\rm u }|+++\cdots\rangle_{\rm l } \pm |+++\cdots\rangle_{\rm u }|\,0\,0\,0\,\cdots\rangle_{\rm l }\Big)$, with the phase determined by which detector recorded the photon. Each qubit state is flipped by applying \textsc{not} gates, and a second photon is sent through the setup. The detection of this photon will project the communication qubits into a GHZ state up to single qubit rotations.

Here, we describe our model of the carving protocol with a perfect single-photon source in more detail. For simplicity, we first consider a setup with only two spins, one at the upper and one at the lower optical route. We will later generalize to more spins.  The initial state of the spin reads $\ket{\psi_0}=\frac{1}{2}(\ket{00}+\ket{01}+\ket{10}+\ket{11})=\frac{1}{2}\left(1, 1, 1, 1\right)^\text{T}$. When the first photon from the single-photon source passes the first beam splitter, it turns into a superposition state $\frac{1}{\sqrt{2}}\left(|{\rm U}\rangle+|{\rm D}\rangle\right)$, where $\ket{\rm U}$ is the photon state traveling along the upper route and $\ket{\rm D}$ the lower route. The unnormalized state describing the transmission of the photon through the spin-cavity systems is:
\begin{equation}
    \ket{\psi_1^{\text{pre}}}=\frac{1}{2\sqrt{2}}\left[
    \begin{pmatrix}
        t_0\\t_0\\t_1\\t_1
    \end{pmatrix}\ket{\rm U}
    +
    \begin{pmatrix}
        t_0\\t_1\\t_0\\t_1
    \end{pmatrix}\ket{\rm D}
    \right]
\end{equation}
where $t_0$ and $t_1$ are transmission coefficients for the $\ket{0}$ and $\ket{1}$ states of the spins, respectively.

When the photon is detected after the second beam splitter, the state of the spins will be projected into
\begin{equation}
    \ket{\psi_1^{+}}=\frac{1}{2}
    \begin{pmatrix}
        2t_0\\t_0+t_1\\t_0+t_1\\2t_1
    \end{pmatrix}
\end{equation}
if the photon is detected by detector $d_+$ and 
\begin{equation}
    \ket{\psi_1^{-}}=\frac{1}{2}
    \begin{pmatrix}
        0\\t_0-t_1\\t_1-t_0\\0
    \end{pmatrix}
\end{equation}
if the photon is detected by detector $d_-$.
We focus on the state $\ket{\psi_1^{+}}$ state for now but a similar procedure is applied if the outcome is state $\ket{\psi_1^{-}}$. First, we apply \textsc{not} gates on the spins transforming the state to
\begin{equation}
    \ket{\psi_1^{+\textsc{not}}}=\frac{1}{2}
    \begin{pmatrix}
        2t_1\\t_0+t_1\\t_0+t_1\\2t_0
    \end{pmatrix}.
\end{equation}
We then attempt to transmit a second photon. Assuming again that the second photon is transmitted and detected by detector $d_+$, the spin state is projected to
\begin{equation}
    \ket{\psi_2^{++}}=\frac{1}{4}
    \begin{pmatrix}
        2t_0\\t_0+t_1\\t_0+t_1\\2t_1
    \end{pmatrix}
    \odot
    \begin{pmatrix}
        2t_1\\t_0+t_1\\t_0+t_1\\2t_0
    \end{pmatrix},
\end{equation}
where $\odot$ is the symbol of the Hadamard product or element-wise product. Here one can see that if $|t_0|\gg |t_1|$, the normalized state $\ket{\psi_2^{++}}$ will resemble a Bell state $\frac{1}{\sqrt{2}}\left(0, 1, 1, 0\right)^\text{T}$. This is also true for the other detection patterns up to single-qubit gate corrections. Note that the protocol is heralded by the detection of both photons. If either the first or the second photon is not detected, the protocol is re-initialized.  

Above, we only considered the transmission and detection of two photons. However, more photons could be scattered in order to increase the fidelity of the state at the expense of a lower success probability. For the general case with $n_{\rm wt}$ spins and $n_{\rm sc}$ scattered photons, one can write an iterative relationship between the spin states after the $\left(n_{\rm sc}-1\right)$th and $n_{\rm sc}$th scatterings as
\begin{align}
    \ket{\psi_{n_{\rm sc}}}=
    \begin{cases}
        \mathbf{T}\odot \ket{\psi_{n_{\rm sc}-1}}, \quad n_{\rm sc} =1.\\
        \mathbf{T}\odot O_{\rm not}\ket{\psi_{n_{\rm sc}-1}}, \quad n_{\rm sc} \geq 2.
    \end{cases}
\end{align}
where $\ket{\psi_{n_{\rm sc}}}$ is the spin state after $n_{\rm sc}$th photon scatterings, and $O_{\rm not}$ is an operation with \textsc{not} gates on every spin. The $\mathbf{T}$ is a vector of transmission coefficients defined as
\begin{align}
    \mathbf{T}=
    \begin{cases}
        \frac{\mathbf{t}_{{\rm u}}+\mathbf{t}_{{\rm d}}}{2},\quad \text{photon detected by detector $d_+$,} \\
        \frac{\mathbf{t}_{{\rm u}}-\mathbf{t}_{{\rm d}}}{2},\quad \text{photon detected by detector $d_-$,} \\
    \end{cases}
\end{align}
where
\begin{align}
\mathbf{t}_{\rm u}=&\left(\begin{array}{c}
t_{0}\\
t_{1}
\end{array}\right)^{\otimes n_{{\rm u}}}\otimes\left(\begin{array}{c}
1\\
1
\end{array}\right)^{\otimes n_{{\rm d}}} \\
\mathbf{t}_{\rm d}=&\left(\begin{array}{c}
1\\
1
\end{array}\right)^{\otimes n_{{\rm u}}}\otimes\left(\begin{array}{c}
t_{0}\\
t_{1}
\end{array}\right)^{\otimes n_{{\rm d}}}
\end{align}
where $n_{{\rm u}}$ and $n_{{\rm d}}$ are number of spins in the upper and lower optic routes, respectively.

\begin{figure*}
    \centering
    \includegraphics[width=\textwidth]{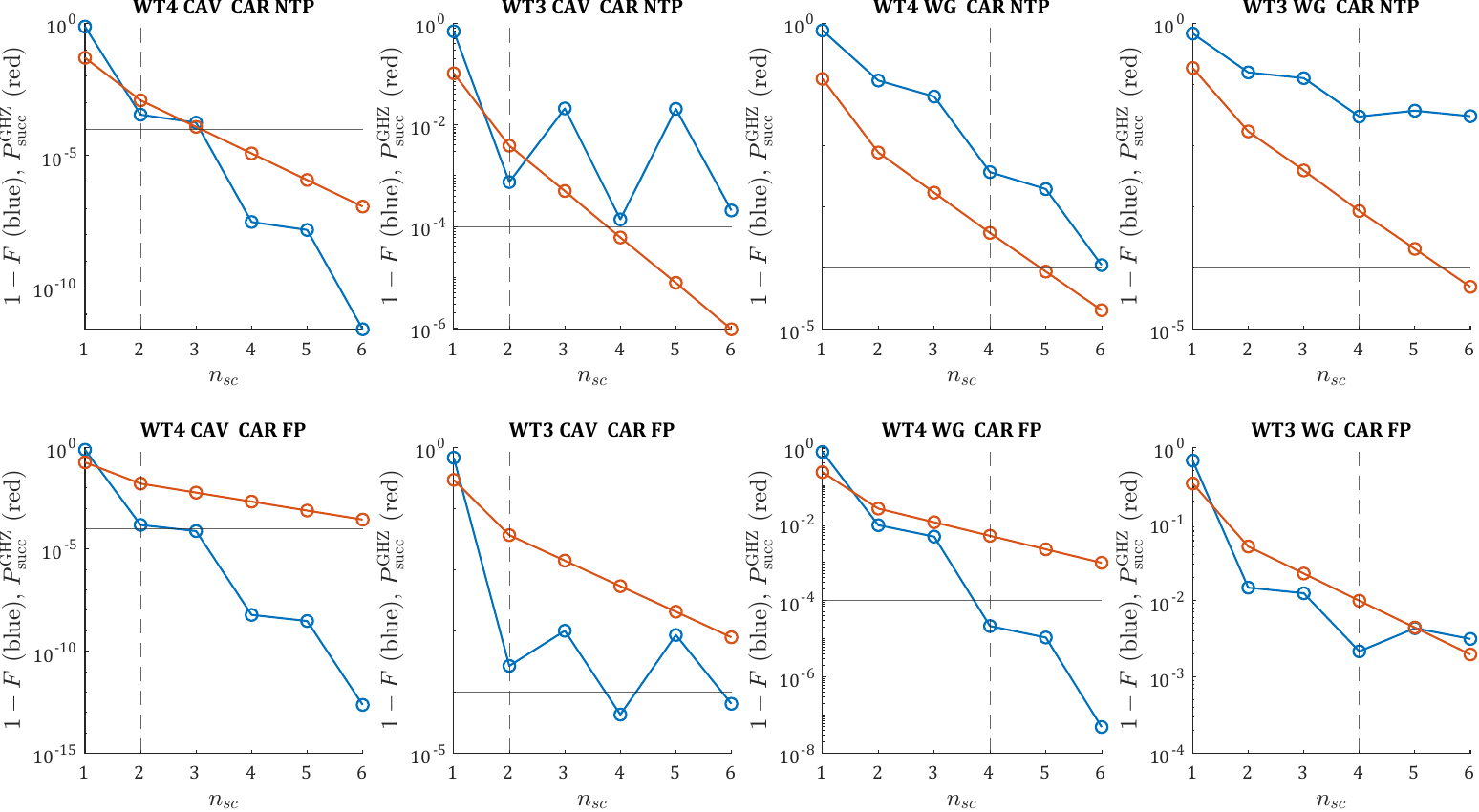}
    \caption{The scan of fidelity and success rate over the number of scattering times in the carving scheme with a single photon source. The title in each plot denotes different architectures, entangling schemes, and parameter sets. The red curves represent the success rate, and the blue curves represent the infidelity. The horizontal solid black lines denote the minimum requirement of the success rate of $10^{-4}$. The vertical dashed black lines denote the selected optimum numbers of scattering times. They are also listed in the table in Fig.~\ref{tab_para_car_cav} and \ref{tab_para_car_wg} in the main text.}
    \label{fig:scan_n_sc_single}
\end{figure*}

After at least two rounds of photon scattering, the state of the spins will resemble the state $\left(|111\cdots\rangle_{\rm u }|000\cdots\rangle_{\rm l } \pm |000\cdots\rangle_{\rm u }|111\cdots\rangle_{\rm l }\right)/\sqrt{2}$ with the $\pm$ sign depending on the parity of the number of photons detected at detector $d_-$. To recover the GHZ state $\ket{\phi_+}$, \textsc{not} gates can be applied on the spins at one of the optical routes, and a Z gate may be used on one of the spins to correct the $\pm$ sign. The corresponding gate errors will be added to the density matrix of the resulting GHZ state. The success possibility of the protocol is the trace of the unnormalized density matrix multiplied by the efficiency of the single photon detectors, single-photon source, and in/out coupling losses. 

The different transmission coefficients $t_0$ and $t_1$ correspond to two different optical transitions. Therefore, the two transitions have different detuning,  $\delta_0$ and $\delta_1$, with respect to the input photons. For the cavity-included setup, according to Ref.~\cite{photon-cavity_int}, the transmission coefficient formula reads
\begin{equation}
    t_{\text{0,1}}^{\rm cav}\left(\delta\right)=\frac{\frac{2\kappa_{{\rm c}}}{2\kappa_{{\rm c}}+\kappa_{{\rm l}}}}{1+2\mathrm{i}\frac{\omega}{2\kappa_{{\rm c}}+\kappa_{{\rm l}}}+\frac{4C_{2}}{1+2\mathrm{i}\delta_{\text{0,1}}/\gamma}},
    \label{eq:car_cav_coef}
\end{equation}
For the waveguide-included setup, according to Ref.~\cite{photon-waveguide_int}, the transmission coefficient formula reads
\begin{equation}
    t_{\text{0,1}}^{\rm wg}\left(\delta\right)=\frac{1+2{\rm i}\delta_{\text{0,1}}/\gamma}{1+P+2{\rm i}\delta_{\text{0,1}}/\gamma}.
    \label{eq:car_wg_coef}
\end{equation}
Notably, the $t^{\rm wg}$ can be viewed as $t^{\rm cav}$ when taking the limit of $\kappa_{{\rm c}}\rightarrow \infty$.

The symbols in Eqn.~(\ref{eq:car_cav_coef}) and (\ref{eq:car_wg_coef}) are explained in Fig.~\ref{tab_para_car_cav} and \ref{tab_para_car_wg} in the main text.

To maximize the GHZ state fidelity, $t(\delta_0)$ should take the maximum modulus and $t(\delta_1)$ the minimum modulus. Besides, the two spin transitions have a fixed detuning $\Delta$ hence $\delta_0-\delta_1=\Delta$. Based on these, the optimum values of $\delta_1$ and $\omega$ can be determined analytically which gives $\delta_1=0$ for both the cavity and waveguide cases and $\omega=\frac{4C_2\Delta(2\kappa_a+\kappa_c)}{1+4\Delta^2}$ for the cavity case. As in the reflection scheme, $\delta_1$ and $\omega$ are also assumed to fluctuate following Gaussian distributions with the same deviations $\sigma$ around the optimal values. The density matrix of the resulting GHZ state is calculated as an average over $\delta_1$ and $\omega$.

The optimal number of scattering times $n_{\rm sc}$ is determined by looking for the lowest infidelity in the regime where the success rate is above $10^{-4}$. We found the optima by scanning the GHZ fidelity and success rates over the number of scattering times, as shown in Fig.~\ref{fig:scan_n_sc_single}.

\section{\label{app:car_coherent}Carving scheme with coherent light source}

\begin{figure*}
    \centering
    \includegraphics[width=\textwidth]{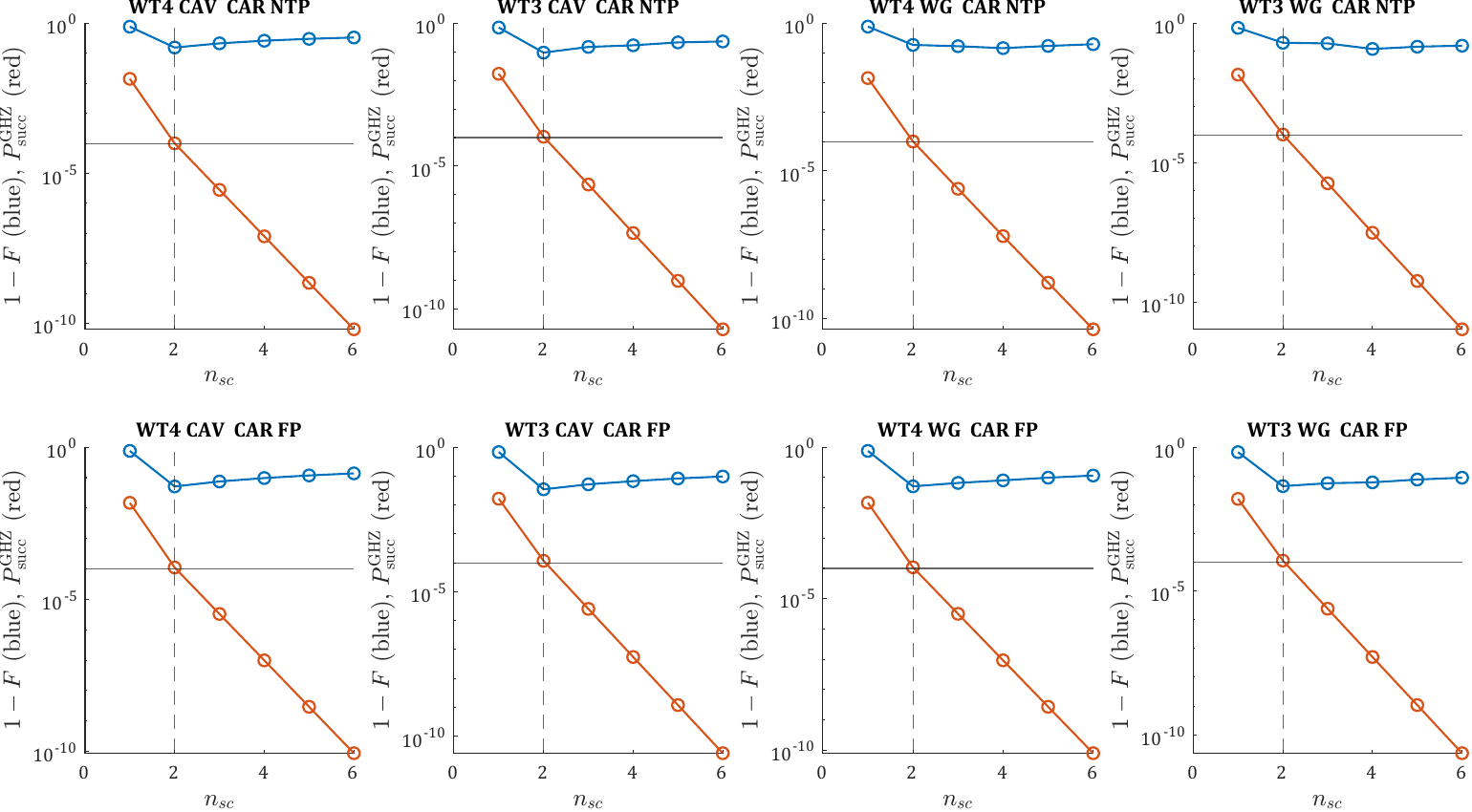}
    \caption{The scan of fidelity and success rate over the number of scattering times in the carving scheme with a single photon source. The title in each plot denotes different architectures, entangling schemes, and parameter sets. The red curves represent the success rate, and the blue curves represent the infidelity. The horizontal solid black lines denote the minimum requirement of the success rate of $10^{-4}$. The vertical dashed black lines denote the selected optimum numbers of scattering times. They are also listed in the table in Fig.~\ref{tab_para_car_cav} and \ref{tab_para_car_wg} in the main text.}
    \label{fig:scan_n_sc_coh}
\end{figure*}

Here, we describe the calculation method for the carving scheme with coherent light. We will use the following properties of the coherent light. (1) A beam splitter converts a coherent light beam into two independent beams of coherent light: $\ket{\alpha}\rightarrow\ket{\frac{\alpha}{\sqrt{2}}}_{+} \ket{\frac{\alpha}{\sqrt{2}}}_{-}$. (2) If there are two coherent light beams incident on both input ports of the beam splitter, the conversion of the beam splitter is: $\ket{\alpha}\ket{\beta}\rightarrow\ket{\frac{\alpha+\beta}{\sqrt{2}}}_{+}\ket{\frac{\alpha-\beta}{\sqrt{2}}}_{-}$. (3) A coherent light beam after scattering from a spin-system will be split into three independent coherent light beams in the linear regime considered here (i.e. the pulse is assumed long enough that only a single photon from the beam scatters from the spin at a time) : $\ket{\alpha}\rightarrow|r\alpha\rangle|d\alpha\rangle|t\alpha\rangle$, where $r$, $d$ and $t$ are the reflection, dissipation, and transmission coefficients of the spin. They have the relationships $|r|^2+|d|^2+|t|^2=1$.

We first take the case with two spins, one at the upper route and one at the lower routes, as an example to demonstrate the calculation method and then generalize to the case with $n_{\rm wt}$ spins. The spins are initially in a state $\ket{\psi_0}=\frac{1}{2}\left(1, 1, 1, 1\right)^\text{T}$. When the coherent light beam $\ket{\alpha}$ from the light source passes the first beam splitter, it turns into the state $\ket{\frac{\alpha}{\sqrt{2}}}_{\rm u}\ket{\frac{\alpha}{\sqrt{2}}}_{\rm d}$, where the subscripts denote the upper ("$_{\rm u}$") and lower ("$_{\rm d}$") routes. Next, when the beams have passed through the spin-systems, the state of the spins-systems and light field becomes:
\begin{equation}
    \ket{\psi_1^{\text{p}}}=\frac{1}{2}
    \begin{pmatrix}
        \ket{\frac{r_0\alpha}{\sqrt{2}}}_{\rm u}\ket{\frac{d_0\alpha}{\sqrt{2}}}_{\rm u}\ket{\frac{t_0\alpha}{\sqrt{2}}}_{\rm u} \ket{\frac{r_0\alpha}{\sqrt{2}}}_{\rm d}\ket{\frac{d_0\alpha}{\sqrt{2}}}_{\rm d}\ket{\frac{t_0\alpha}{\sqrt{2}}}_{\rm d}\\
        \ket{\frac{r_0\alpha}{\sqrt{2}}}_{\rm u}\ket{\frac{d_0\alpha}{\sqrt{2}}}_{\rm u}\ket{\frac{t_0\alpha}{\sqrt{2}}}_{\rm u} \ket{\frac{r_1\alpha}{\sqrt{2}}}_{\rm d}\ket{\frac{d_1\alpha}{\sqrt{2}}}_{\rm d}\ket{\frac{t_1\alpha}{\sqrt{2}}}_{\rm d}\\
        \ket{\frac{r_1\alpha}{\sqrt{2}}}_{\rm u}\ket{\frac{d_1\alpha}{\sqrt{2}}}_{\rm u}\ket{\frac{t_1\alpha}{\sqrt{2}}}_{\rm u} \ket{\frac{r_0\alpha}{\sqrt{2}}}_{\rm d}\ket{\frac{d_0\alpha}{\sqrt{2}}}_{\rm d}\ket{\frac{t_0\alpha}{\sqrt{2}}}_{\rm d}\\
        \ket{\frac{r_1\alpha}{\sqrt{2}}}_{\rm u}\ket{\frac{d_1\alpha}{\sqrt{2}}}_{\rm u}\ket{\frac{t_1\alpha}{\sqrt{2}}}_{\rm u} \ket{\frac{r_1\alpha}{\sqrt{2}}}_{\rm d}\ket{\frac{d_1\alpha}{\sqrt{2}}}_{\rm d}\ket{\frac{t_1\alpha}{\sqrt{2}}}_{\rm d}
    \end{pmatrix}
\end{equation}
where the subscriptions "$_0$" and "$_1$" of the reflection, dissipation, and transmission coefficients are regarding the $\ket{0}$ and $\ket{1}$ states of the spin, respectively.

When the beams are interfered at the second beam splitter, the state of the spin-systems and photonic field becomes:
    \begin{equation}
    \ket{\psi_1^{\text{b}}}=\frac{1}{2}
    \begin{pmatrix}
        \ket{\frac{r_0\alpha}{\sqrt{2}}}_{\rm u}\ket{\frac{d_0\alpha}{\sqrt{2}}}_{\rm u} \ket{\frac{r_0\alpha}{\sqrt{2}}}_{\rm d}\ket{\frac{d_0\alpha}{\sqrt{2}}}_{\rm d}\ket{\frac{2t_0\alpha}{\sqrt{2}}}_{+}\ket{0}_{-}\\
        \ket{\frac{r_0\alpha}{\sqrt{2}}}_{\rm u}\ket{\frac{d_0\alpha}{\sqrt{2}}}_{\rm u} \ket{\frac{r_1\alpha}{\sqrt{2}}}_{\rm d}\ket{\frac{d_1\alpha}{\sqrt{2}}}_{\rm d}\ket{\frac{t_0+t_1}{\sqrt{2}}\alpha}_{+}\ket{\frac{t_0-t_1}{\sqrt{2}}\alpha}_{-}\\
        \ket{\frac{r_1\alpha}{\sqrt{2}}}_{\rm u}\ket{\frac{d_1\alpha}{\sqrt{2}}}_{\rm u} \ket{\frac{r_0\alpha}{\sqrt{2}}}_{\rm d}\ket{\frac{d_0\alpha}{\sqrt{2}}}_{\rm d}\ket{\frac{t_1+t_0}{\sqrt{2}}\alpha}_{+}\ket{\frac{t_1-t_0}{\sqrt{2}}\alpha}_{-}\\
        \ket{\frac{r_1\alpha}{\sqrt{2}}}_{\rm u}\ket{\frac{d_1\alpha}{\sqrt{2}}}_{\rm u} \ket{\frac{r_1\alpha}{\sqrt{2}}}_{\rm d}\ket{\frac{d_1\alpha}{\sqrt{2}}}_{\rm d}\ket{\frac{2t_1\alpha}{\sqrt{2}}}_{+}\ket{0}_{-}
    \end{pmatrix}
    \label{eq:coh-spin}
\end{equation}
where the subscripts "$_+$" and "$_-$" denote the two sides of the second beam splitter.

We now perform partial traces over the reflection and dissipation modes and project the transmission modes onto the cases where only detector $d_+$ or $d_-$ records a photon. One may use the following formula to accomplish the calculation: for arbitrary coherent states $\ket{\beta}$ and $\ket{\gamma}$,
\begin{align}
    \sum_{l=0}^{\infty} \bra{l} \ket{\beta}\bra{\gamma} \ket{l}=\mathrm{e}^{-\frac{|\beta-\gamma|^{2}}{2}}\mathrm{e}^{\frac{\beta\gamma^{*}-\beta^{*}\gamma}{2}},
\end{align}
and
\begin{equation}
    \sum_{l=1}^{\infty} \bra{l} \ket{\beta}\bra{\gamma} \ket{l}=\sum_{l=0}^{\infty} \bra{l} \ket{\beta}\bra{\gamma} \ket{l}-\bra{0} \ket{\beta}\bra{\gamma} \ket{0}
    =\mathrm{e}^{-\frac{|\beta-\gamma|^{2}}{2}}\mathrm{e}^{\frac{\beta\gamma^{*}-\beta^{*}\gamma}{2}}-\mathrm{e}^{-\frac{|\beta|^{2}+|\gamma|^{2}}{2}}.
\end{equation}
where $\ket{l}$ are the Fock states.

After one traces out the reflection and dissipation modes and projects the transmission modes on the corresponding bases, one gets the density matrix of the spin systems after the first round of light scattering. One can then apply \textsc{not} gates on it and calculate the effect of another round of photon scattering on top of this similar to the procedure for a perfect single photon source described above.  

Based on the calculation above, one can see that the overall effect of photon scattering is that it multiplies each entry of the spin density matrix with a certain coefficient. Therefore, one can derive an iteration relationship of the density matrix of the spin as
\begin{align}
    \rho_{n_{\rm sc}}=\begin{cases}
        \mathbf{T}_{\rm coh}\odot \rho_{n_{\rm sc}-1},\quad n_{\rm sc} =1 \\
        \mathbf{T}_{\rm coh}\odot \left( O_{\rm not} \rho_{n_{\rm sc}-1} O_{\rm not} \right),\quad n_{\rm sc} \geq 2
    \end{cases}
\end{align}
where $\rho_{n_{\rm sc}}$ is the density matrix after scattering $n_{\rm sc}$th coherent states, and $O_{\rm not}$ is the operation of \textsc{not} gates on every spin. The coefficient matrix $\mathbf{T}_{\rm coh}$ is defined as
%\begin{widetext}
    \begin{equation}
    \mathbf{T}_{\rm coh}=
    \begin{cases}
        \left(\bigodot_{j=1}^{n_{\rm wt}} S^{\rm R}_j \odot S^{\rm D}_j \right)\odot S^+,\text{ detection on "$+$" side}\\
        \left(\bigodot_{j=1}^{n_{\rm wt}} S^{\rm R}_j \odot S^{\rm D}_j \right)\odot S^-,\text{ detection on "$-$" side}
    \end{cases}
\end{equation}
%\end{widetext}
where $n_{\rm wt}$ is the total number of spins,
\begin{align}
    S^{\rm R}_j = \sum_{l=0}^{\infty} \left(\left[\bra{l}\right] \odot \mathbf{R}_j \right) \left(\mathbf{R}^\dag_j \odot \left[\ket{l}\right]^\mathrm{T} \right),
    \label{eq:G_R}
\end{align}
\begin{align}
    S^{\rm D}_j= \sum_{l=0}^{\infty} \left( \left[\bra{l}\right] \odot \mathbf{D}_j \right) \left( \mathbf{D}^\dag_j  \odot \left[\ket{l}\right]^\mathrm{T}\right),
\end{align}
\begin{equation}
\begin{split}
    S^+=&\sum_{l=1}^{\infty} \left( \left[\bra{l}\right] \odot \mathbf{T}_+\right) \left(\mathbf{T}^\dag_+ \odot \left[\ket{l}\right]^\mathrm{T} \right) \odot\\
    &\left(\left[\bra{0}\right] \odot \mathbf{T}_- \right) \left(\mathbf{T}^\dag_- \odot \left[\ket{0}\right]^\text{T}\right),
\end{split}
\end{equation}
and
\begin{equation}
\begin{split}
    S^-=&\left(\left[\bra{0}\right] \odot \mathbf{T}_+ \right) \left(\mathbf{T}^\dag_+ \odot \left[\ket{0}\right]^\text{T}\right) \odot\\
    &\sum_{l=1}^{\infty} \left( \left[\bra{l}\right] \odot \mathbf{T}_-\right)  \left(\mathbf{T}^\dag_- \odot \left[\ket{l}\right]^\mathrm{T} \right),
\end{split}
\end{equation}
where $\left[\bra{l}\right]$ is a column vector whose elements are all $\bra{l}$. And, $\mathbf{R}_j$, $\mathbf{D}_j$, $\mathbf{T}_+$, and $\mathbf{T}_-$ are vectors whose entries are wavefunctions of the reflection, dissipation, and transmission coherent light beams. Similar to Eqn.~\ref{eq:coh-spin}, they form a relationship
\begin{align}
    \ket{\psi_{1}^{\rm b}}= \left(\bigodot_{j=1}^{n_{\rm wt}}\mathbf{R}_j\odot \mathbf{D}_j\right)\odot \mathbf{T}_+ \odot \mathbf{T}_- \odot \ket{\psi_0}.
\end{align}
Their entries are defined as
\begin{equation}
    \begin{split}
        \mathbf{R}_{j}[k]=&\ket{\vec{\alpha}^{\rm r}_{j}[k]}\\
        \mathbf{D}_{j}[k]=&\ket{\vec{\alpha}^{\rm d}_{j}[k]}\\
        \mathbf{T}_{+}[k]=&\ket{\vec{\alpha}_{\rm t+}[k]}\\
        \mathbf{T}_{-}[k]=&\ket{\vec{\alpha}_{\rm t-}[k]}
    \end{split}
\end{equation}
where $\vec{\alpha}^{\rm r}_{j}$ is defined as

\begin{align}
\vec{\alpha}^{\rm r}_{j}=
\begin{cases}
\frac{\alpha}{\sqrt{2}}
\left(\begin{array}{c}
t_0\\
t_1
\end{array}\right)^{\otimes (j-1)}
\otimes
\left(\begin{array}{c}
r_0\\
r_1
\end{array}\right)
\otimes
\left(\begin{array}{c}
1\\
1
\end{array}\right)^{\otimes (n_{\rm wt}-j)}\;,\; j\leq n_{\rm u}   \\
\frac{\alpha}{\sqrt{2}}
\left(\begin{array}{c}
1\\
1
\end{array}\right)^{\otimes (n_{\rm u})}
\otimes
\left(\begin{array}{c}
t_0\\
t_1
\end{array}\right)^{\otimes (j-n_{\rm u}-1)}
\otimes
\left(\begin{array}{c}
r_0\\
r_1
\end{array}\right)\otimes\left(\begin{array}{c}
1\\
1
\end{array}\right)^{\otimes (n_{\rm wt}-j)}\;,\; j\geq n_{\rm u}
\end{cases}.
\end{align}
The vector $\vec{\alpha}^{\rm d}_{j}$ follows a similar definition as $\vec{\alpha}^{\rm r}_{j}$, which can be derived by replacing $r_0$ and $r_1$ with $d_0$ and $d_1$. And, $\vec{\alpha}_{\rm t\pm}=\frac{1}{\sqrt{2}}\left(\vec{\alpha}_{\rm u}\pm \vec{\alpha}_{\rm d}\right)$ with
\begin{align}
\vec{\alpha}_{\rm u}=
\frac{\alpha}{\sqrt{2}}
\left(\begin{array}{c}
t_0\\
t_1
\end{array}\right)^{\otimes n_{\rm u}}
\otimes
\left(\begin{array}{c}
1\\
1
\end{array}\right)^{\otimes n_{\rm d}}  \\
\vec{\alpha}_{\rm d}=
\frac{\alpha}{\sqrt{2}}
\left(\begin{array}{c}
1\\
1
\end{array}\right)^{\otimes n_{\rm u}}
\otimes
\left(\begin{array}{c}
t_0\\
t_1
\end{array}\right)^{\otimes n_{\rm d}},
\end{align}
with $n_{\rm u}$ and $n_{\rm d}$ the number of spins in the upper and lower optical routes, respectively. By now, we get the unnormalized density matrix of the spin after $n_{\rm sc}$ rounds of light scattering. The trace of the density matrix gives the success rate for generating the GHZ state.

The optimum values of $\delta_1$ and $\omega$ are the same as in the carving scheme with a single-photon source as discussed in App.~\ref{app:car_single}. However, the optimal values of $\alpha$ and the number of scattering times $n_{\rm sc}$ must be determined. 

On the one hand, having more light-scattering rounds can potentially increase the GHZ fidelity, as in the carving scheme with a deterministic single-photon source. On the other hand, it also means more reflection and dissipation modes that will leak the quantum information of the system into the environment. We found that for all the cases in the carving scheme with coherent light, the optimum number of light-scattering rounds is $n_{\rm sc}=2$ because the fidelity gain from more scattering rounds doesn't outweigh the information leakage. 

Regarding the determination of $\alpha$, the larger the $\alpha$, the larger the success rate, but the lower the fidelity. For small $\alpha$, the coherent light resembles a probabilistic single-photon source. Hence, the fidelity of this scheme will approach the scheme with a deterministic single-photon source. However, the success rate will be much lower than it. On the other hand, when $\alpha$ is large, the success rate will approach the scheme with a deterministic single-photon source, but the fidelity will decrease. To get the maximum fidelity, we set the $\alpha$, letting the success rate take the minimum requirement value of $10^{-4}$ at $n_{\rm sc}=2$ as shown in Fig.~\ref{fig:scan_n_sc_coh}. However, the fidelities in that circumstance are still too low to realize modular quantum computing with surface code. Hence, the carving scheme with a coherent light source is not promising.

\section{Example: Diamond defect center implementation}
\label{app:diamond_center}
To demonstrate a particular physical implementation of our formalism, we give an example of diamond defect center-based hardware. We define a coherence times parameter set (called Set-D) inspired by nitrogen-vacancy (NV) centers in diamond~\cite{PhysRevApplied.15.024049,Bradley2022,PhysRevA.99.052330,Pompili2021,10.1145/3341302.3342070,Abobeih2022}. NV centers may, however, not be suited for the direct entangling schemes due to the difficulty of coupling these defect centers to nanophotonic cavities and waveguides.  Nonetheless, we anticipate that other defect centers, such as SnV or SiV, where strong coupling to nanophotonic cavities and waveguides has been experimentally demonstrated, can achieve similar coherence and operational times.

\begin{figure}[hbtp]
\centering
\includegraphics[width=0.48\textwidth]{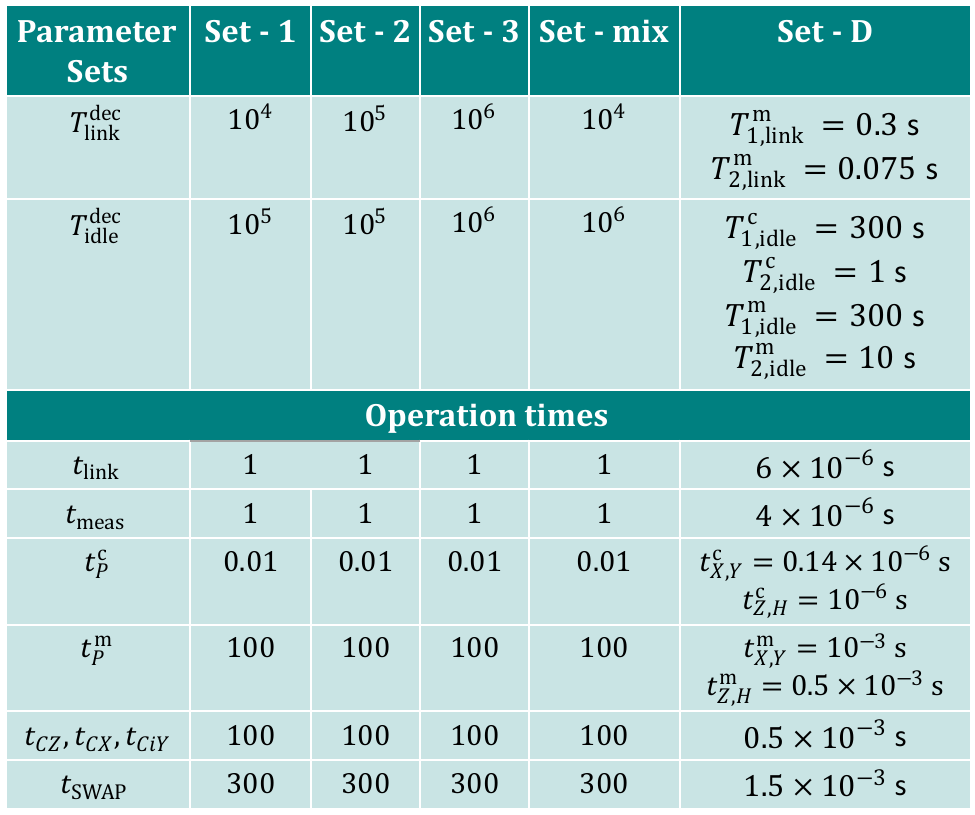}
\caption{Table describing the coherence and operation times, including two additional parameter sets, Set-mix and Set-D. All values except Set-D are expressed w.r.t. absolute value of $t_\text{link}=10^{-5}$ s, which is assumed to take one unit of time. Set-D is a special set with absolute values in seconds.}
\label{tab:coherence_parameters_all_appendix}
\end{figure}

%For the case of NV centers in diamond, an electron spin qubit (e) acts as the communication qubit and carbon nuclear qubits (n) are used as memory qubits, as indicated by the corresponding superscripts in Fig.~\ref{tab:coherence_parameters_all_appendix} for Set-NV. 
The coherence parameters' values for Set-D are inspired by Ref.~\cite{Pompili2021} and shown in Fig.~\ref{tab:coherence_parameters_all_appendix}. We consider separate $T_1$ and $T_2$ times for the generalized amplitude and phase damping channels inspired by the experiments. The values indicate a relatively strong dephasing noise (Z-biased). Set-D is also a special set for the inclusion of dynamical decoupling (DD) methods that reduce noise during gate pulse sequences~\cite{PhysRevX.9.031045}. In this model, gate pulses can only be applied during the focusing point between two DD pulses. This refocusing time is expressed as $t_\text{DD}=t_\text{pulse}+2n_\text{DD}t_\text{link}$, where $t_\text{pulse}$ is the time duration of a $\pi$-pulse for DD sequence. For a given value of $t_\text{pulse}$ inspired from the experiments, $n_\text{DD}$ was optimized in Ref.~\cite{10.1116/5.0200190}. We also use the same values of $t_\text{pulse}=10^{-3}$ s and $n_\text{DD}=18$ for Set-D. These parameters are not used for other sets in the simulations.

Another quantity for more insight into the distributed systems is link efficiency $\eta^*$, as also explored in Ref.~\cite{10.1116/5.0200190} in the context of NV centers using an EM scheme. We can express the entanglement generation efficiency (EGE) for EM or direct GHZ generation schemes as:
\begin{equation}
\label{eqn:EGE_eta}
\eta^*=\frac{2P_\text{succ}^\text{link/GHZ}}{t_\text{link/GHZ}\big( (T_\text{1,link/GHZ}^\text{dec})^{-1}+(T_\text{2,link/GHZ}^\text{dec})^{-1} \big)}
\end{equation}
The entanglement generation efficiency captures the average number of Bell pairs or GHZ states that can be generated within the coherence times of the memory qubits. We also calculate the link (entanglement generation) efficiency for all schemes in App.~\ref{app:threshold_details} via tables in Fig.~\ref{tab:weight_4_full_data} and Fig.~\ref{tab:weight_3_full_data}.

\section{Threshold simulation details}
\label{app:threshold_details}
\begin{figure}
    \centering
    \includegraphics[width=0.49\textwidth]{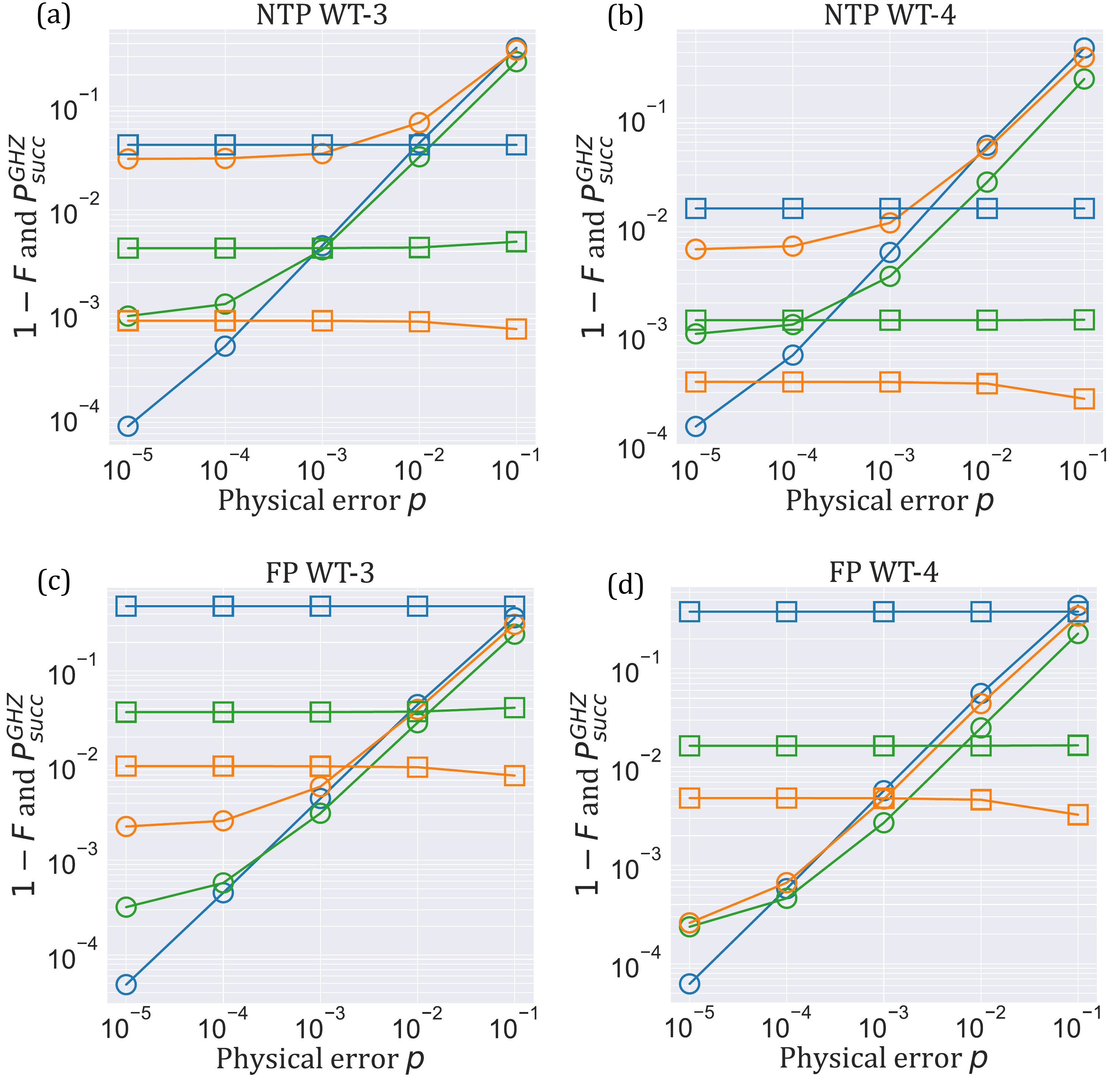}
    \caption{Success rates $P_\text{succ}^\text{GHZ}$ (squares) and GHZ state infidelity $1-F$ (circles) of RFL scheme (blue), CAV CAR (green) with SPS and WG CAR (orange) with SPS, as a function of physical error probability. The subtitles of the plots denote the parameter used (NTP/FP) and the surface code architecture (WT-3/4). Due to different state mixing in the protocol run via the depolarizing channel, there is a weak dependence of $P_\text{succ}^\text{GHZ}$ on the physical error. This is numerically evident in the plots.}
    \label{fig:GHZFidAndRate}
\end{figure}

In this appendix, we provide details on the numerical calculations of the distributed surface code thresholds. We also introduce two additional parameter sets for the coherence and operational times: Set-mix and Set-D. Set-mix is a set composed of $T_\text{link}^\text{dec}$ from Set-1 and $T_\text{idle}^\text{dec}$ from Set-3. The motivation is to analyze the performance of technologies that exhibit excellent quantum memory capabilities with long-lasting $T_\text{idle}^\text{dec}$ times. Set-D is inspired by nitrogen-vacancy (NV) center implementation of distributed surface code, explored in Ref.~\cite{10.1116/5.0200190}. See App. \ref{app:diamond_center} for more details on the physical model based on defect centers in diamond. Fig.~\ref{tab:coherence_parameters_all_appendix} presents a summary of all the five parameter sets together. For Set-D, we use the absolute values of the parameters and not relative numbers w.r.t. $t_\text{link}$ times. We use different $T_\text{1/2,idle/link}^\text{dec}$ times as shown, inspired via the experiments with dynamical decoupling (DD) incorporated gates. Before performing the full threshold calculations, we first plot the GHZ success probability and infidelity against the varied parameter, $p$, for circuit-level noise. This gives us estimate on which values of $p$ are the most relevant. Making use of the numerical bounds which we found on the success probability and fidelities. This makes it handy to calculate the other values and scan the parameter space where we would find a threshold for that setting. This is shown in Fig.~\ref{fig:GHZFidAndRate} for the direct GHZ generation schemes for both the WT-3/4 modular architectures.

\begin{figure}[hbtp]
\centering
\includegraphics[width=0.58\textwidth]{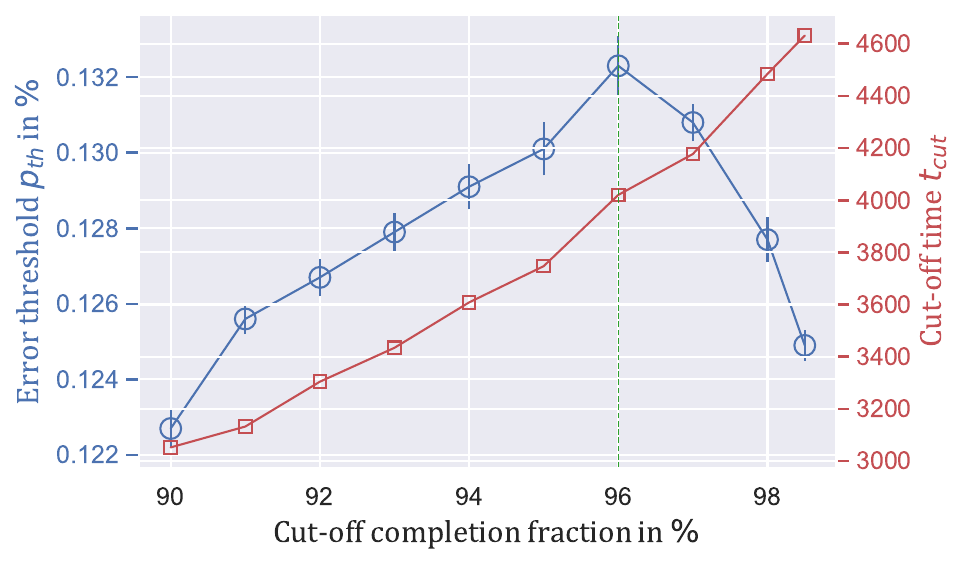}
\caption{Distributed surface code threshold $p_\text{th}$ (blue) and cut-off time $t_\text{cut}$ (red) w.r.t. GHZ cut-off completion fraction $x$. For each value of $x$, we find $t_\text{cut}^{x\%}$, which gives a certain threshold. A binary search for the highest value of $p_\text{th}$ results in an optimal value of $x$. The optimal value (highest) is shown in green. This plot was created for WT4 EM Set-3 FP.}
\label{fig:cutoff_em}
\end{figure}

\begin{figure}[hbtp]
\centering
\includegraphics[width=0.48\textwidth]{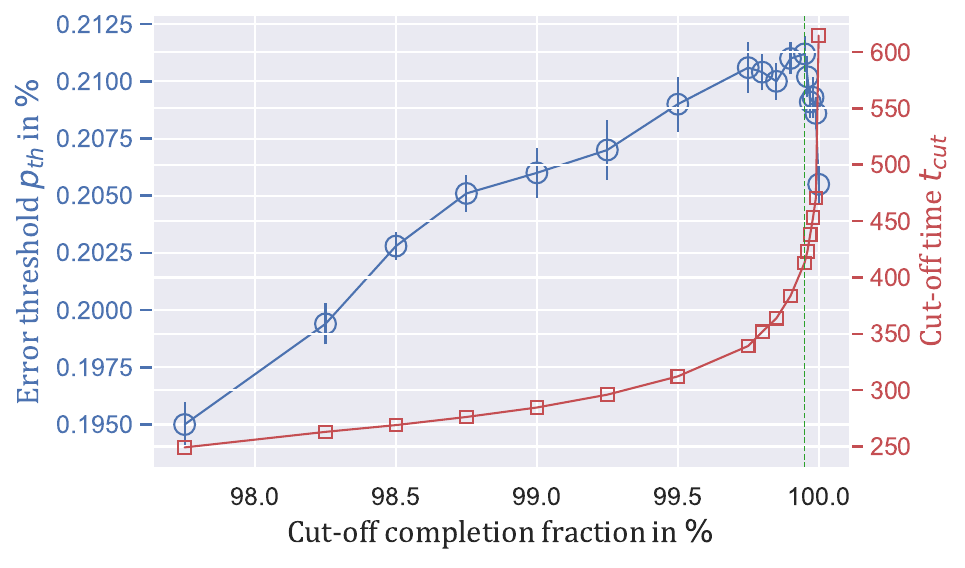}
\caption{Distributed surface code threshold $p_\text{th}$ (blue) and cut-off time $t_\text{cut}$ (red) w.r.t. GHZ cut-off completion fraction $x$ for WT4 RFL Set-mix NTP. Read Fig.~\ref{fig:cutoff_em}'s caption for details.}
\label{fig:cutoff_rfl}
\end{figure}

\subsection{\label{appsubsec:Cut-off time}Cut-off time optimization}
As described in the main text, the cut-off time $t_\text{cut}$ caps the maximum time allowed for GHZ generation for any scheme during the stabilizer measurement. Assuming $t_\text{cut}^{100\%}$ is the average time allowed for successful GHZ for a certain number of attempts, we can define some $t_\text{cut}^{x\%}$ where we get a GHZ success within $x\%$ attempts for entanglement generation using any scheme. $t_\text{cut}^{100\%}$ is computed in prior in our simulations. Then, we start with a trial value of $x$ (typically around $98\%-99\%$ for direct schemes or lower for EM schemes) to estimate logical error rates ($p_L$) in a wide spectrum of physical error rates $p$. We sweep $x$ in our noisy stabilizer simulator and deduce the $t_\text{cut}^{x\%}$ value. For each fixed value of $t_\text{cut}^{x\%}$, we perform the threshold simulations to get the logical error rates and estimate the threshold for the code. If the logical error rates are low enough (typically less than 0.2), we search ahead for a threshold. If the logical error rates are greater than 0.90, we abandon the search as there is a negligible chance of threshold within that regime, which we verified numerically. 

$t_\text{cut}^{x\%}$ increases exponentially as $x$ increases (see Fig. \ref{fig:cutoff_em} and Fig. \ref{fig:cutoff_rfl}). Starting from a low value of $x$, first, the code threshold $p_\text{th}$ increases with $x$, because more $t_\text{cut}$ allows for more error syndrome collection. However, if we keep increasing $x$, the code threshold starts to saturate to a maximum where more $t_\text{cut}$ does not assist in more syndrome data collection but starts contributing to more noise via decoherence. If we increase $x$ further, we see a rapid fall in the threshold near very high values of $x$. We perform a binary search over the variable $x$ to find the highest value of $p_\text{th}$, including the error bars. Using the respective entanglement generation scheme, we report the highest value as the code threshold $p_\text{th}$ for that architecture.

\subsection{\label{appsubsec:Threshold}Threshold estimation}
Now, we describe the threshold estimation in detail. A trial (guessed) value of $t_\text{cut}^{x\%}$ is chosen. Using a superoperator table for a certain input physical error probability $p$, we calculate the logical error probability $p_L$ (or logical success probability/rate $1-p_L$). A wide range of input $p$ values is considered such that a threshold lies within this range by numerically observing the behavior of $p_L$ with increasing lattice sizes (6,8,10,12). When a threshold exists, we narrow the search to a small range of $p$ values for the fitting procedure. We use the fitting function of the form \cite{WANG200331}:
\begin{equation}
    p_L=A+B\gamma+C\gamma^2
\end{equation}
where $\gamma=(p-p_\text{th})L^{1/\nu_0}$, with $L$ as different lattice sizes for different curves. The fitting parameters $A,B,C,p_\text{th},\nu_0$ are found from numerical optimization using the least-squares method. This is discussed in great detail in Ref.~\cite{10.1116/5.0200190}.

\begin{figure}[hbtp]
\centering
\includegraphics[width=0.48\textwidth]{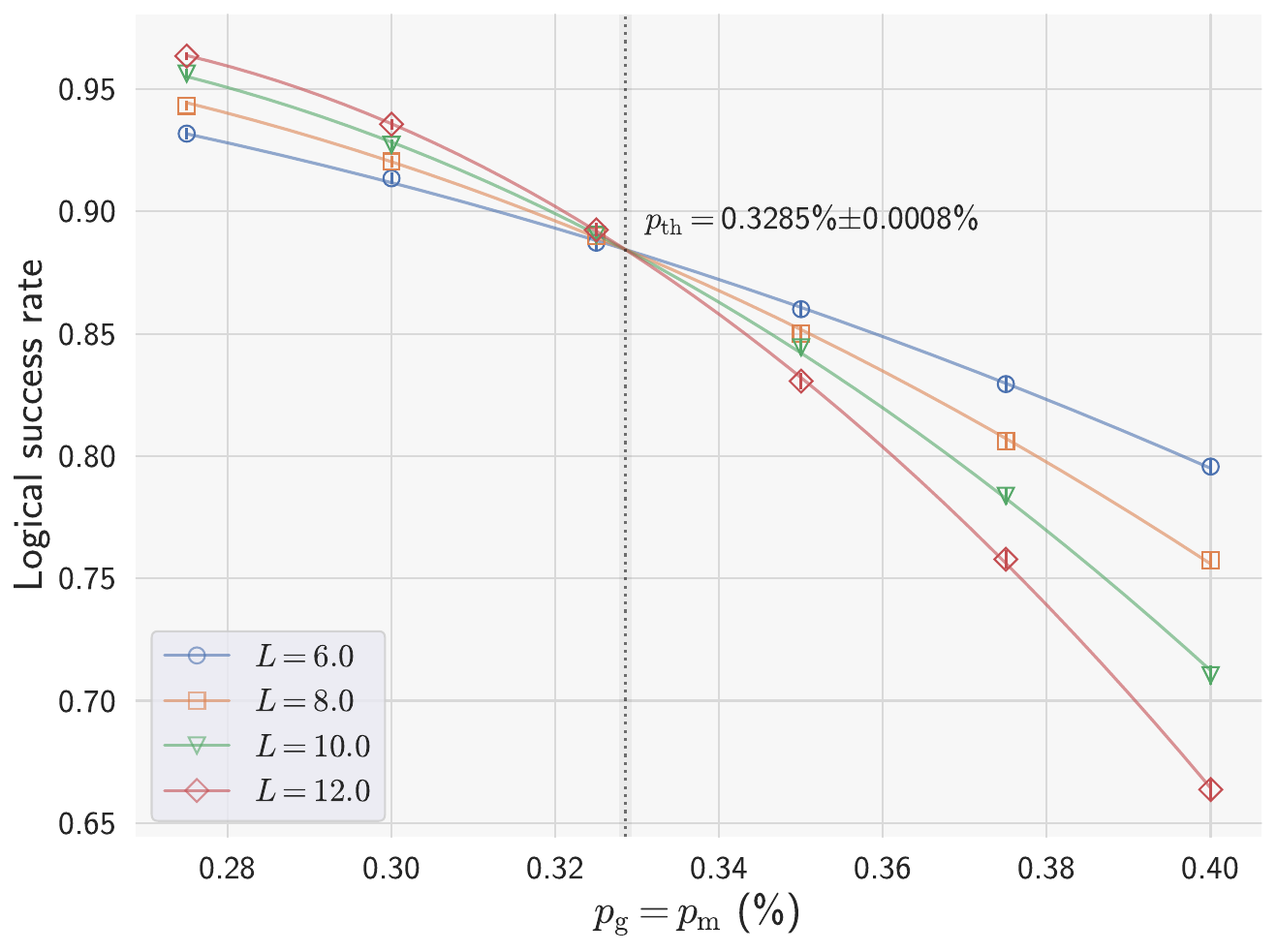}
\caption{Example threshold plot for WT4 RFL (Set-3) NTP. The plot shows the logical success rates $1-p_L$ against physical error rates $p=p_g=p_m$ for different even-valued distances for the distributed surface code. The fitting function reveals the code threshold within some error bar.}
\label{fig:example_threshold}
\end{figure}

Here, we show an example of threshold estimation in~\ref{fig:example_threshold} where we perform the fitting procedure for logical success curves for various even-sized code distances.

\subsection{\label{appsubsec:threshold_full_table}Detailed threshold simulation data}
\begin{figure*}
\centering
\includegraphics[width=0.89\textwidth]{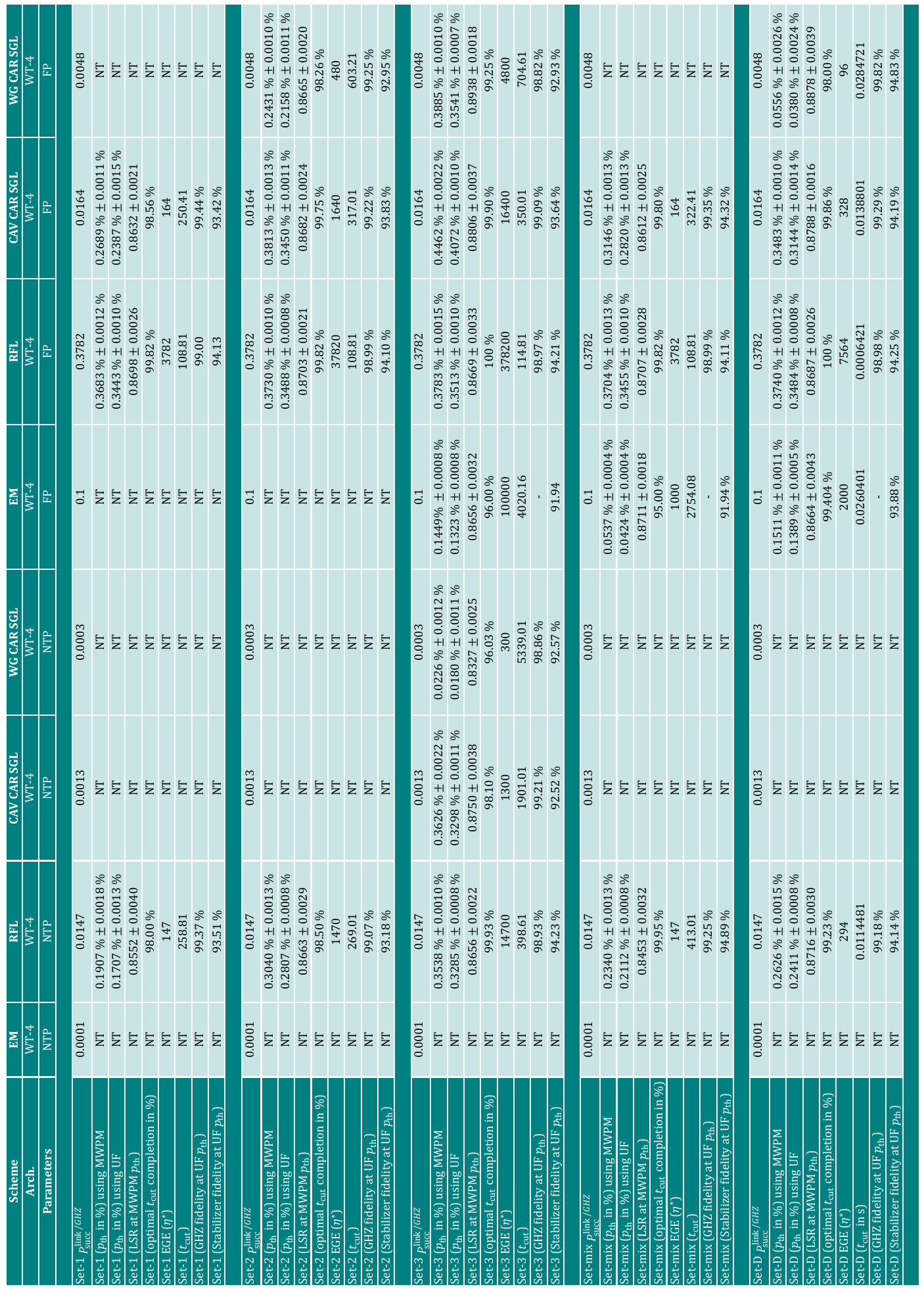}
\caption{Weight-4 full threshold simulations data. We give code thresholds for each architecture and scheme using two decoders, success probabilities,  entanglement generation efficiency, cut-off times, GHZ, and stabilizer fidelity. NT denotes no threshold.}
\label{tab:weight_4_full_data}
\end{figure*}

\begin{figure*}
\centering
\includegraphics[width=0.89\textwidth]{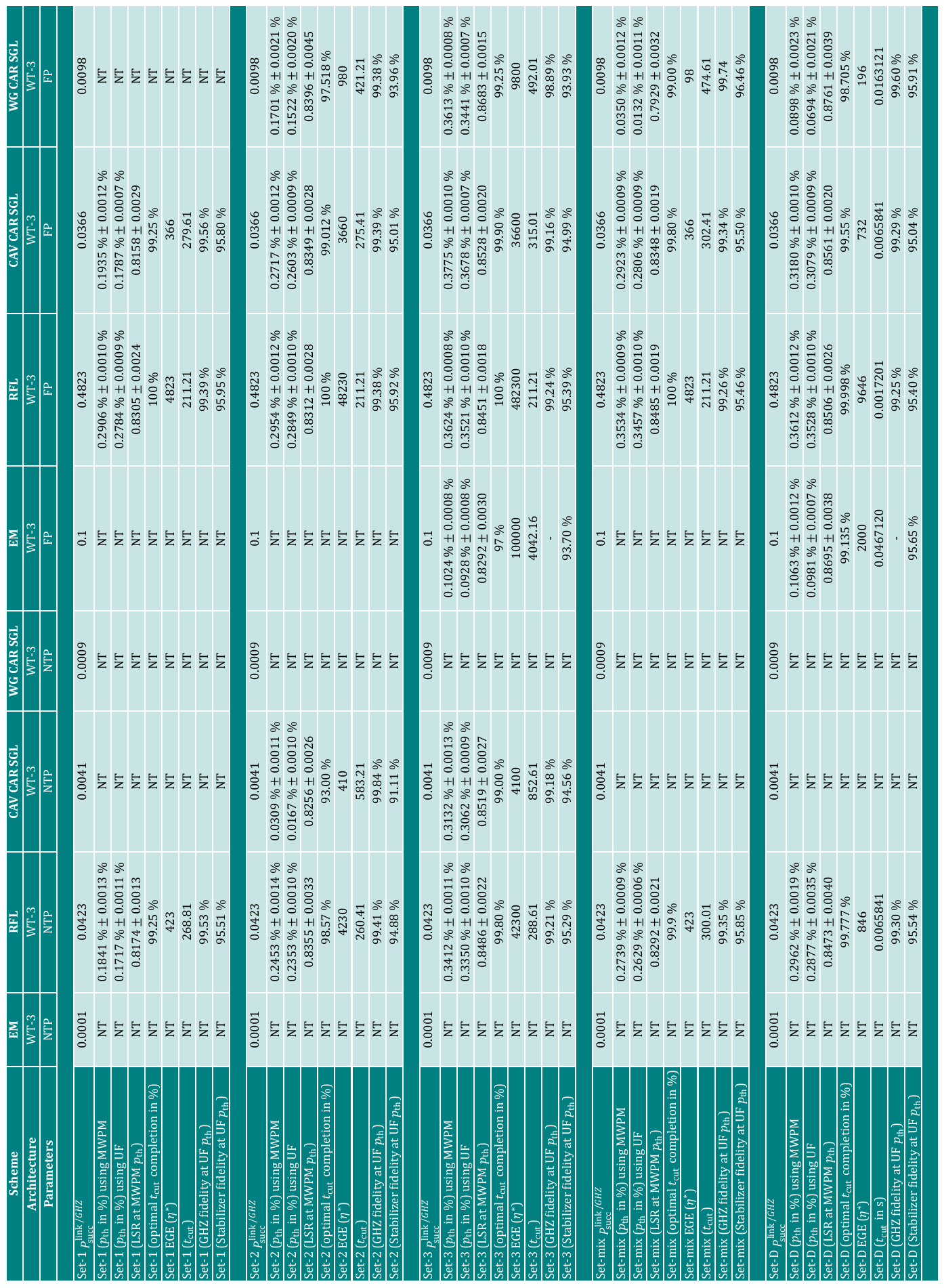}
\caption{Weight-3 full threshold simulations data. We give code thresholds for each architecture and scheme using two decoders, success probabilities,  entanglement generation efficiency, cut-off times, GHZ, and stabilizer fidelity. NT denotes no threshold.}
\label{tab:weight_3_full_data}
\end{figure*}

Finally, we provide all the simulation data for all the architectures and schemes in combination with all the coherence parameter sets, including the recently introduced Set-mix and Set-D. This is presented via full-page tables in Fig.~\ref{tab:weight_4_full_data} (WT4) and Fig.~\ref{tab:weight_3_full_data} (WT3). Each block of rows describes the simulation results for a particular coherence parameter set. The columns describe different combinations of architecture layout, scheme, and NTP/FP parameters.

The first row within each block describes the success probability $P_\text{succ}^\text{link/GHZ}$, which has been introduced earlier. While previously we reported thresholds found using the Union-Find (UF) decoder~\cite{Delfosse2021almostlineartime} by optimizing the binary search for threshold values, we report a much higher threshold ($p_\text{th}$ at MWPM) in this data using the minimum-weight perfect matching (MWPM) decoder~\cite{Edmonds_1965}. The UF threshold value is listed next, also the threshold value we reported in the main text. This was preferred because the UF decoder is fast with linear time complexity in the number of qubits. We find the logical success rate ($1-p_L$) at the threshold value using the MWPM decoder for more insight into the threshold performance. The optimal GHZ cut-off completion fraction ($x\%$) is also listed. This is the completion fraction at which the peak performance of the architecture was found using the respective scheme. The completion fraction is relatively less for EM schemes as these require fusion operations that take longer for GHZ generation. The decoherence vs. syndrome data rate trade-off is satisfied much earlier for EM schemes during QEC cycles. This can also be seen in Fig.~\ref{fig:cutoff_em} for the EM scheme and Fig.~\ref{fig:cutoff_rfl} for the RFL scheme. After that, we report the entanglement generation efficiency ($\eta^*$, described in Eqn.~\ref{eqn:EGE_eta}). The next number is the cut-off time $t_\text{cut}$ at the optimal GHZ cut-off completion rate. The cut-off time is crucial to define the logical clock speed of the quantum computer via the duration of the QEC cycle. It is decided by $t_\text{cut}$ or the two-qubit gate times $t^\text{m}_P$ (on the memory qubits), whichever is higher. GHZ fidelity at the threshold value using the UF decoder is also reported, which indicates the quality of the entanglement generation scheme used. The fidelity is calculated for only the direct schemes where the average fidelity was calculated using the average density matrix of the output GHZ state from the scheme. For the EM scheme, this was not possible due to the different protocols used for fusion. However, we can calculate the output Bell pair fidelity for any physical error rate using Eqn.~\ref{eqn:fid_single} and Eqn.~\ref{eqn:fid_double}, as the case may be. Finally, we report the stabilizer fidelity (at the UF threshold) for the entire stabilizer circuit protocol using any scheme considering the noisy two-qubit gates on the memory and noisy measurements. An improvement in two-qubit gate times or fidelity will drastically improve the stabilizer fidelity.

\clearpage
\twocolumngrid
\bibliographystyle{apsrev4-2}
\bibliography{bibliography}
\end{document}